\documentclass[useAMS, usenatbib, a4paper]{mnras}
\usepackage[spanish,es-minimal,english]{babel}
\usepackage[utf8]{inputenc}
\usepackage{graphicx}
\usepackage{xcolor}
\usepackage{fixltx2e}
\usepackage{hyperref}
\usepackage{savesym}
\savesymbol{tablenum}
\usepackage{siunitx}
\restoresymbol{SIX}{tablenum}
\usepackage{newtxtext}
\usepackage[varg,varvw,smallerops]{newtxmath}
\usepackage{xfrac} % for the \sfrac macro
\usepackage{booktabs}
\usepackage{longtable}
\usepackage{array}   % for \newcolumntype macro
\newcolumntype{L}{>{$}l<{$}} % math-mode version of lrc column types
\newcolumntype{R}{>{$}r<{$}} 
\newcolumntype{C}{>{$}c<{$}}
\hypersetup{colorlinks=True, linkcolor=blue!50!black, citecolor=black,
  urlcolor=blue!50!black}

\usepackage{etoolbox}
\robustify\bfseries
\robustify\itshape
\usepackage{enumerate}

\DeclareSIUnit\msun{\text{M\ensuremath{_\odot}}}
\DeclareSIUnit\lsun{\text{L\ensuremath{_\odot}}}
\DeclareSIUnit\zsun{\text{Z\ensuremath{_\odot}}}
\DeclareSIUnit\angstrom{\text{\AA}}

\newcounter{ionstage}
\renewcommand{\ion}[2]{\setcounter{ionstage}{#2}% 
  \ensuremath{\mathrm{#1\,\scriptstyle\Roman{ionstage}}}}
\newcommand\hii{\ion{H}{2}}

\definecolor{NEWcolor}{rgb}{0.7,0.2,0.1}
\newcommand\startNEW{\color{black}}
\newcommand\stopNEW{\color{black}}
\newcommand\NEW[1]{\startNEW #1\stopNEW\relax}

\newcommand\snrj{SNR~J\num{0059.4}\num{-7210}}

\newcommand*\chem[1]{\ensuremath{\mathrm{#1}}}
\newcommand\Config[1]{\ensuremath{\mathrm{#1}}}
\newcommand\Term[3]{\ensuremath{\mathrm{#1\ ^{#2}#3}}}
\newcommand\Level[4]{\ensuremath{\mathrm{#1\ ^{#2}#3_{#4}}}}

\newcommand\hmol{\chem{H_2}}

\newcommand\feni{Fe-Ni-Ca-Si}

\newcommand\ha{\ensuremath{\text{H}\alpha}}
\newcommand\hb{\ensuremath{\text{H}\beta}}

\newcommand\wav[1]{\ensuremath{\lambda #1}}
\newcommand\wavv[1]{\ensuremath{\lambda\lambda #1}}

\newcommand\avpdr{\ensuremath{A_V^{\text{PDR}}}}

\newcommand\Rat[2]{\ensuremath{\text{#1} / \text{#2}}}
\newcommand\RRat[2]{\ensuremath{\text{Zone #1} / \text{Zone #2}}}

\title[Optical molecular hydrogen lines from low-metallicity PDR]
{
  One hundred \NEW{optical emission lines of molecular hydrogen}\\
  from a low-metallicity photodissociation region
}

\author[Henney \& Valerdi]{
  William J. Henney\textsuperscript{1}\thanks{w.henney@irya.unam.mx}
  and Mabel Valerdi\textsuperscript{2}
  \\
  \textsuperscript{1}\foreignlanguage{spanish}{%
    Instituto de Radioastronomía y
    Astrofísica, Universidad Nacional Autónoma de México, Apartado
    Postal 3-72, 58090 Morelia, Michoacán, Mexico}\\
  \textsuperscript{2}\foreignlanguage{spanish}{%
    Instituto Nacional de Astrofísica, Óptica y Electrónica,
    (INAOE), Luis E. Erro No. 1, Sta. Ma. Tonantzintla, Puebla, C.P. 72840, México.}\\
}

\date{Accepted XXX. Received YYY; in original form ZZZ}

\pubyear{2023}

\begin{document}
\label{firstpage}
\pagerange{\pageref{firstpage}--\pageref{lastpage}}
\maketitle

\begin{abstract}
  We report the detection of a rich spectrum of more than one hundred
  optical emission lines
  {\NEW{of vibrationally hot molecular hydrogen (\chem{H_2})}}
  from the photodissociation region (PDR)
  around the mini-starburst cluster
  NGC~346 in the Small Magellanic Cloud.
  % We propose the term Deep Red Line (DRL) 
  % for these lines, which
  {\NEW{The lines}}
  are concentrated in the spectral range
  \SI{6000}{\angstrom} to \SI{9300}{\angstrom}
  and have observed brightnesses ranging from
  0.01\% to  0.4\% times that of the
  \hb{} \wav{4861} hydrogen recombination line.
  Analysis of the spatial distribution of the
  {\NEW{\chem{H_2} lines}}
  shows that they originate from a range of depths in the PDR,
  {\NEW{intermediate}}
  between the shallow layers probed by fluorescent lines
  of neutral nitrogen and oxygen,
  and the more shielded layers probed by
  neutral carbon recombination lines.
  Comparison with other PDRs shows that the relative strength of the
  {\NEW{\chem{H_2} lines}}
  with respect to the [\ion{C}{1}] \wav{8727} line increases rapidly with
  decreasing metallicity,
  {\startNEW
  being at least 40 times larger in NGC~346 than
  in the prototypical PDR of the Orion Bar.
  The internal PDR dust extinction is also found to be anomalously low in NGC~346.
  A separate result is the discovery of a high-ionization bow shock
  around the O2 star Walborn~3. 
  \stopNEW}
\end{abstract}

%\facilities{VLT:Yepun (MUSE); OANSPM:2.1m (Mezcal); Keck (HIRES)}
%\object{M42}

\section{Introduction}
\label{sec:introduction}

Photoionized nebulae around high mass stars emit a great variety
of optical emission lines, which dominate the total radiative
luminosity of these regions \citep{Osterbrock:2006a}.
The strong cooling lines, together with other weaker lines,
are used to diagnose the physical
conditions, chemical abundances, and kinematics of the gas
\citetext{for instance, in the archetypal Orion Nebula:
  \citealp{Baldwin:1991a, Mesa-Delgado:2008a, Garcia-Diaz:2008a, Mc-Leod:2016a}}.
The neutral and molecular
photodissociation regions (PDRs) that surround
these nebulae, on the other hand,
do not emit brightly in optical lines.
Instead, the important coolants in PDRs are
infrared and radio lines of molecules and atoms,
together with continuum radiation from dust grains
\citep{Hollenbach:1999a, Wolfire:2022r}.
Nonetheless, even weak optical lines could  provide useful diagnostics
of physical conditions in the outer layers of PDRs
where the visual extinction is low enough for optical photons to escape.
Despite rapid recent advances in infrared observations
from satellite and airborne observatories
\citep{Gardner:2006a,Pilbratt:2010a,Young:2012a},
optical observations still have some advantages:
namely, higher spatial resolution
and the availability of a more diverse variety
of instrumentation on ground-based telescopes.

Previous studies of optical emission lines from PDRs
\citep{Storzer:2000a} have concentrated on well-known
lines such as [\ion{O}{1}] \wav{6300} and
[\ion{S}{2}] \wavv{6716, 6731},
but these have the disadvantage that they are also
emitted by ionized or partially ionized gas
in the \hii{} region and ionization front and it is therefore
difficult to isolate the contribution from the PDR.
Other lines, such as \ion{O}{1} \wav{8446} \citep{Grandi:1975a}
and \ion{N}{1} \wav{5200} \citep{Ferland:2012a}
are typically emitted in a thin layer at the surface of the PDR
\citep{Henney:2021b}.
The only known \NEW{atomic lines} that form at any depth in the PDR
are the far-red [\ion{C}{1}] recombination lines at \wavv{8727, 9824, 9850}
\citep{Escalante:1991a}.
\startNEW
The strongest molecular hydrogen lines are detected
in the infrared, but weak transitions between
highly vibrationally excited states are predicted to be found at
optical wavelengths \citep{Black:1976a, Neufeld:1996a}.
These lines have been observed in reflection nebulae
\citep{Burton:1992a}
and Herbig-Haro bow shocks
\citep{Gredel:2007a,Giannini:2015a}
but to the best of our knowledge have not been previously reported
in the dense PDRs that surround massive star forming regions.
\stopNEW

In this paper, we present observations of
a very rich optical emission line spectrum
from neutral gas around
the massive star cluster NGC~346 in the Small Magellanic Cloud (SMC).
\NEW{In addition to lines of neutral metals},
over one hundred \NEW{rovibrational \hmol{}} lines are detected
in the red part of the spectrum,
with intensities that are typically one hundred times lower
than those of the strong optical lines from the \hii{} region.
% If these lines could be identified with known atomic or molecular transitions,
% they would provide a fascinating new window into the physical conditions
% in the region between the ionization front and the molecular cloud core. 

The Small Magellanic Cloud (SMC) is a dwarf irregular galaxy
that is a satellite of the Milky Way at a distance of \SI{62}{kpc} \citep{Graczyk:2020g}
and with a metallicity of roughly one-fifth solar
\citetext{\(Z \approx \SI{0.2}{\zsun}\), \citealp{1992ApJ...384..508R, Narloch:2021t}}.
The open cluster NGC~346
\citep{Walborn:1978k, Kudritzki:1989a, 2022A&A...666A.189R, Sabbi:2022a}
and its associated emission nebula N66
\citep{Henize:1956v, Tsamis:2003b, Lebouteiller:2008a, Whelan:2013d, Valerdi:2019a}
comprise the brightest star-forming region in the SMC
\citep{Geist:2022a}.
The cluster contains about 30 O-type stars with masses in the range of
\num{35}--\SI{100}{\msun} and age of \SI{3}{Myr} 
\citep{2006A&A...456..623E, 2019A&A...626A..50D},
which are responsible for exciting the N66 \hii{} region and surrounding PDR
\citep{Requena-Torres:2016g}. 
Star formation is ongoing in the region, as evidenced by the presence of
a large number of young stellar objects (YSOs)
\citep{Sabbi:2007h, Gouliermis:2008a},
including massive protostars \citep{Simon:2007r, Sewio:2013f}
and sub-solar mass objects \citep{Jones:2023a}.

The paper is organized as follows.
In section~\ref{sec:observations}, we describe the observations and data reduction.
In section~\ref{sec:prop-unid-lines} we describe the properties of the \NEW{\hmol{} lines},
their spatial distribution and relation with different varieties
of known emission lines.
% , together with statistical properties of their wavelength distribution.
% In section~\ref{sec:cand-line-ident} we critically evaluate potential identifications of the
% Deep Red Lines with known atomic and molecular transitions,
% although we find no convincing matches.
In section~\ref{sec:discussion} we discuss the uniqueness of the \NEW{molecular hydrogen}
spectrum from NGC~346
\startNEW
and its possible relation to the low metallicity of the region, while
\stopNEW
% , and possible next steps to further investigate the nature of the lines.
in section~\ref{sec:summary} we summarize our conclusions.
\startNEW
Supplementary figures and a table of all emission lines detected in the MUSE spectra
are given in appendices.
A further series of appendices describes in detail how
we ruled out other candidate identifications before making the positive
identification with molecular hydrogen.
\stopNEW

\section{Observations}
\label{sec:observations}
\begin{figure}
  \centering
  \includegraphics[width=\linewidth]{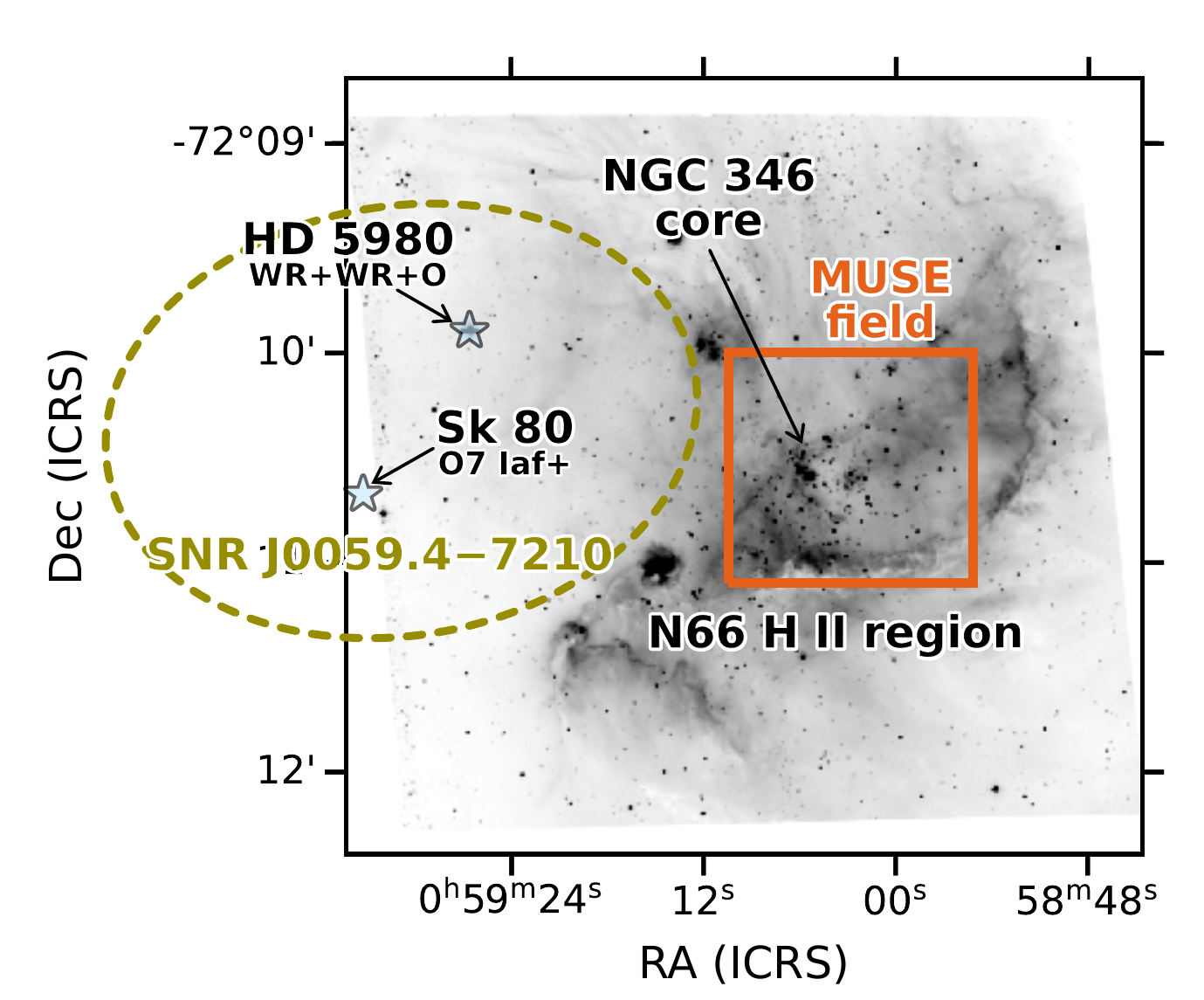}
  \caption{%
    Location of the \(1' \times 1'\) MUSE observed field within
    the NGC~346/N66 region.
    The field is centered on the dense ionized and
    molecular gas that surrounds the numerous luminous
    O~stars in the core of the star cluster.
    Additional luminous ionizing stars are located
    to the east, as marked. 
    The supernova remnant \snrj{} is also located nearby,
    but
    based on the X ray emission \citep{Maggi:2019q}
    does not overlap with the MUSE field.
  }
  \label{fig:finding}
\end{figure}

\begin{figure*}
  \centering
  \includegraphics[width=\linewidth]{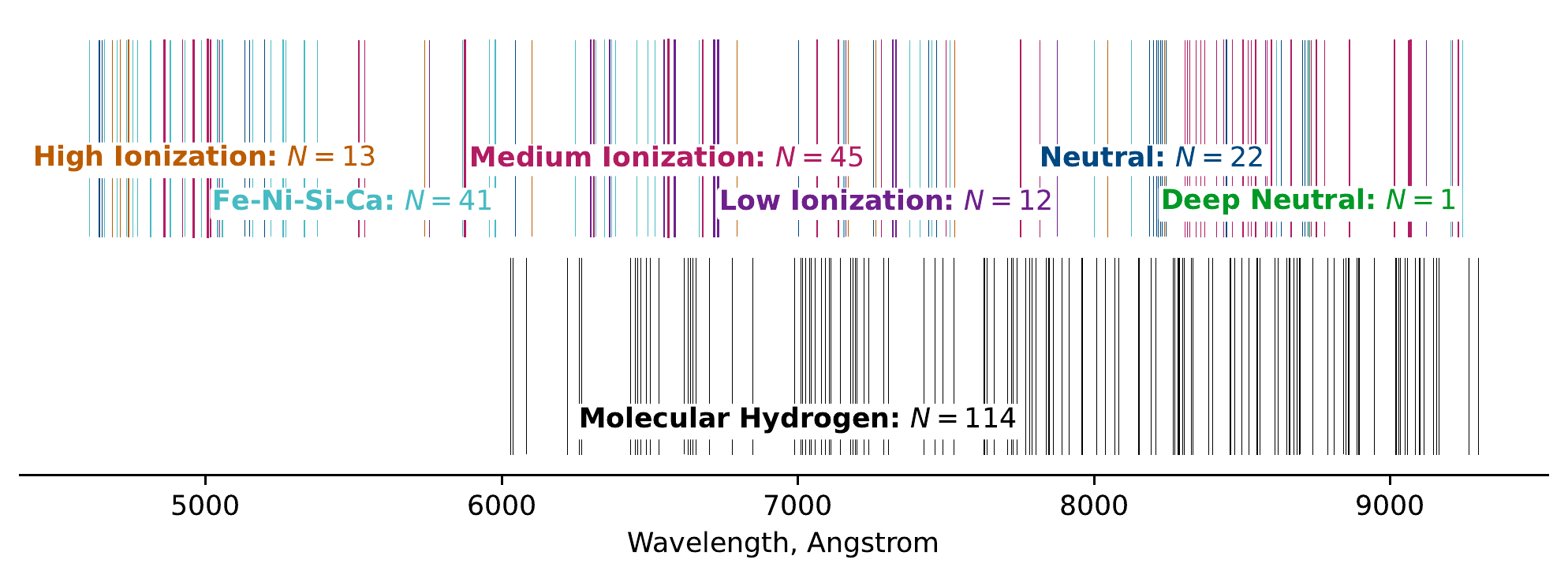}
  \caption{
    Wavelength distribution of \NEW{atomic} emission lines (color)
    and \NEW{molecular hydrogen} emission lines (black) from the MUSE spectrum of NGC~346.
    Lines are drawn with a width proportional to the logarithm
    of their brightness
    For the \NEW{atomic} lines,
    different colors correspond to the different classes of emission line
    that are listed in Table~\ref{tab:ion-class}.
    % There are clear systematic patterns in the distribution of the unidentified lines,
    % which are analyzed in detail in Appendix~\ref{sec:struct-wavel-distr}. 
  }
  \label{fig:bar-code}
\end{figure*}

We analyzed an integral field spectral cube of the central NGC~346 region
in the Small Magellanic Cloud (see Figure~\ref{fig:finding})
obtained as part of program 098.D-0211(A) (PI:  W.-R. Hamann)
using the MUSE spectrograph \citep{Bacon:2010a, Bacon:2014a}
on the VLT UT4 telescope.\@
The field of view is approximately
\SI{64}{arcsec} times \SI{60}{arcsec} with
spaxel size of \SI{0.2}{arcsec} and estimated seeing
full-width half maximum (FWHM) width of \SI{0.961}{arcsec}.
The wavelength range is \SI{4595}{\angstrom} to \SI{9366}{\angstrom}
sampled at \SI{1.25}{\angstrom.pix^{-1}} and the
spectral resolving power varies from \(R \approx 2000\) in the blue
to \(R \approx 4000\) in the red.
The pipeline-reduced data was downloaded from the
ESO Science Archive Facility.\footnote{
  \label{fn:1}
  Direct link to the data set:
  \url{https://archive.eso.org/dataset/ADP.2017-10-16T11:04:19.247}
}
The data has undergone the standard ESO pipeline processing
\citep{Weilbacher:2020a}
and as part of the MUSE-DEEP collection\footnote{
  \label{fn:2}
  \url{https://doi.org/10.18727/archive/42}
}
was co-added across multiple observations obtained on \mbox{2016-08-22}
with a total effective exposure time of \SI{12600}{s}.

Further processing was then carried out using
bespoke Python scripts as follows.
An initial continuum subtraction was performed using a two-pass median filter
along the wavelength axis, applied independently to each
spatial pixel (spaxel)
and incorporating a clipping of positive peaks
between the two median filter passes in order to mitigate
the over--subtracted dark halos on either side of bright emission lines.
This process has the effect of removing any spectral variation
on wavelength scales larger than the median filter width,
while leaving narrower features untouched.
However, it can produce artifacts when there is
significant spectral structure on scales comparable to the filter width,
so it is prudent to compare results from filters of different widths.
We have used widths in wavelength pixels of \(n = 7, 11, 31, 101, 1001\)
but, unless otherwise stated, all results shown here
are for \(n = 7\), corresponding to a filter width of \SI{9.8}{\angstrom}.
The larger widths are necessary for investigating the broad
Raman-scattered wings of the hydrogen lines, which will be reported
elsewhere.

\begin{table}
  \centering
  \caption{Adopted classification of nebular emission lines}
  \label{tab:ion-class}
  \begin{tabular}{l p{3cm} >{\raggedright\arraybackslash}p{2cm}}
    \toprule
    Line class
    & Description
    & Species \\
    \midrule
    High ionization
    & Collisional lines of triply-ionized metals,
      recombination/fluorescent lines of ionized metals,
      recombination lines of ionized helium 
    & [\ion{Ar}{4}], [\ion{Cl}{4}], [\ion{K}{4}], \ion{Si}{3}, \ion{O}{2}, \ion{He}{2}
    \\ \addlinespace
    Medium ionization
    & Collisional lines of doubly-ionized metals,
      recombination lines of neutral helium and hydrogen 
    & [\ion{Ar}{3}], [\ion{Cl}{3}], [\ion{O}{3}], [\ion{S}{3}], \ion{He}{1}, \ion{H}{1}
    \\ \addlinespace
    Low ionization
    & Collisional lines of neutral and singly-ionized metals
    & [\ion{O}{1}], [\ion{Cl}{2}], [\ion{N}{2}], [\ion{O}{2}], [\ion{P}{2}], [\ion{S}{2}]
    \\ \addlinespace
    Neutral
    & Fluorescent lines of neutral metals
    & [\ion{N}{1}], \ion{N}{1}, \ion{O}{1}
    \\ \addlinespace
    Deep neutral
    & Recombination lines of neutral metals with low ionization potential, \NEW{molecular lines}
    & [\ion{C}{1}], \NEW{\hmol}
    \\ \addlinespace
    \feni
    & Lines from refractory elements affected by gas-phase depletion
    & [\ion{Fe}{3}], [\ion{Fe}{2}], [\ion{Ni}{2}], \ion{Ca}{1}], \ion{Fe}{2}, \ion{Ni}{2}, \ion{Si}{2}
    \\
    \bottomrule
  \end{tabular}
\end{table}

A preliminary list of potential emission lines was prepared
using the function \texttt{find\_peaks()}
from the \texttt{scipy.signal} library \citep{Virtanen:2020a}
to automatically identify spectral peaks above a certain threshold.
In order to capture lines of different stages of ionization,
the function was applied to 1d spectra extracted from different spatial regions
of the nebula (\(20 \times 20\)~spaxels),
which yielded a list of 525 candidate lines.
The spatial emission distribution at the peak wavelength
of each candidate line was inspected and the approximate rest wavelength
was compared against lists of known nebular emission lines
\citep{Esteban:2004a, van-Hoof:2018a, Garcia-Rojas:2018a, Mendez-Delgado:2021d}
and night sky lines \citep{Osterbrock:1996a}.
From this comparison, we identified 135 known nebular emission lines,
which we classify according to the degree of ionization
of the emitting species, as detailed in Table~\ref{tab:ion-class}. 
These are shown by the colored vertical lines in Figure~\ref{fig:bar-code}.
A further 6 lines were found to come from stellar sources only
and are not considered further.
In addition, we identified 83 night sky emission lines of \chem{OH} and \chem{O_2}
plus 29 artifacts from over correction for telluric absorption,
while 160 of the candidate lines were judged to be false positives
due to noise or instrumental artifacts.
The remaining 113 lines show a spatial distribution suggestive of a nebular origin,
but have no clear identifications in known \NEW{atomic} line lists.
\startNEW
After exhaustively ruling out possible identifications with
neutral metals, diffuse interstellar bands, and Raman scattering
for these lines (see Appendices~\ref{sec:path-ident-molec} to \ref{sec:possible-id-ni}),
we found that they all correspond to ro-vibrational transitions
between vibrationally excited levels of molecular hydrogen up to \(v = 14\).
We obtained rest wavelengths for these lines by calculating a model with the
Cloudy plasma physics code \citep{Chatzikos:2023a}
using energy level data from \citep{Komasa:2011a}.
A table of all observed nebular and night sky lines is given
at the end of the paper (Table~\ref{tab:all-lines}).
\stopNEW

% These are the unidentified lines that are the main focus of this paper.
% As a reflection of both their spectral and spatial distribution
% (see next section),
% we propose the term Deep Red Line (DRL) for the unidentified lines. 

\section{\NEW{Emission} properties of the \NEW{optical molecular hydrogen} lines}
\label{sec:prop-unid-lines}

\begin{figure}
  \includegraphics[width=\linewidth]{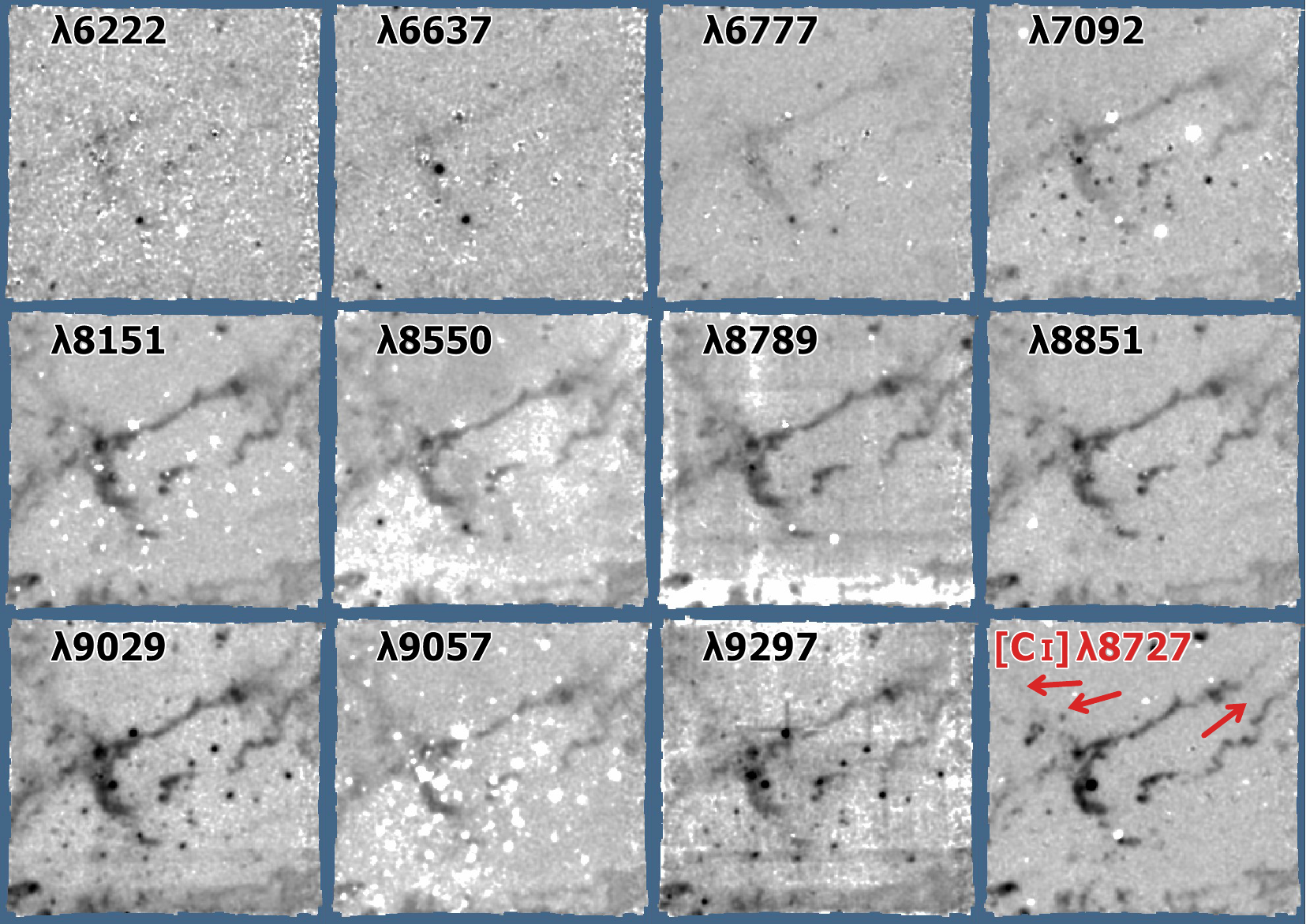}
  \caption{Emission maps of selected \NEW{optical \hmol{}} lines and the
    [\ion{C}{1}] \wav{8727} line.
    Red arrows on the [\ion{C}{1}] map indicate regions that show
    \NEW{\hmol{}} line emission but without any corresponding
    [\ion{C}{1}] emission. 
  }
  \label{fig:uil-stamps}
\end{figure}

\begin{table}
  \centering
  \caption{Spatial zones within the nebula}
  \label{tab:zones}
  \begin{tabular}{l l l l l}
    \toprule
    & Zone
    & Description
    & Key emission
    & Color\\
    \midrule
    Diffuse
    & 0 & Deep neutral & \NEW{\hmol{}} sum & Black \\
    & I & Neutral & [\ion{N}{1}] \wav{5199} & Blue \\
    & II & Low ionization & [\ion{S}{2}] \wav{6716} & Purple \\
    & III & Medium ionization & [\ion{S}{3}] \wav{9069} & Pink \\
    & IV & High ionization & [\ion{Ar}{4}] \wav{4740} & Orange \\
    \addlinespace
    Compact
    & MYSO & Massive YSOs & \ion{O}{1} \wav{8446} & Red \\
    & S & OB Stars & Continuum & Cyan \\
    \bottomrule
  \end{tabular}
\end{table}

The spectral distribution of the \NEW{\hmol{}} lines
over the observed wavelength range is shown by the black
vertical lines in Figure~\ref{fig:bar-code},
where it can be seen that they are concentrated in the red and far-red regions
(\(\lambda > \SI{6000}{\angstrom}\)).  In order to investigate their spatial distribution, 
images were created of all nebular emission lines
by summing 3 wavelength pixels
around each peak in the median-filtered cube.
Figure~\ref{fig:uil-stamps} shows images of 11 representative
\NEW{\hmol{}} lines that span the full wavelength range, together
with [\ion{C}{1}] \wav{8727} (lower right), which is the known line
whose emission distribution most closely matches that of the \NEW{\hmol{}} lines.
In high-mass star forming regions
the far-red [\ion{C}{1}] emission typically arises from recombination of \chem{C^+}
deep within neutral photodissociation regions \citep{Escalante:1991a}
\startNEW
The \hmol{} lines clearly trace the same filaments as [\ion{C}{1}]
but \hmol{} emission is also seen from some regions that do not emit [\ion{C}{1}],
as indicated by red arrows on the lower right panel of Figure~\ref{fig:uil-stamps}.
These seem to correspond to diffuse envelopes surrounding the denser filaments.
\stopNEW
% The similarity in spatial distribution suggests that the unidentified lines are emitted in the PDR rather than in the \hii{} region.
% Thus the word ``deep'' in the term Deep Red Line refers
% both to their origin from deeper into the neutral/molecular gas
% than is typical for optical lines
% and to their spectral distribution,
% which is towards the extreme red end of the visual range. 

\subsection{Division of the nebula into spatial zones}
\label{sec:division-nebula-into}

\begin{figure*}
  \centering
  \includegraphics[width=\linewidth]{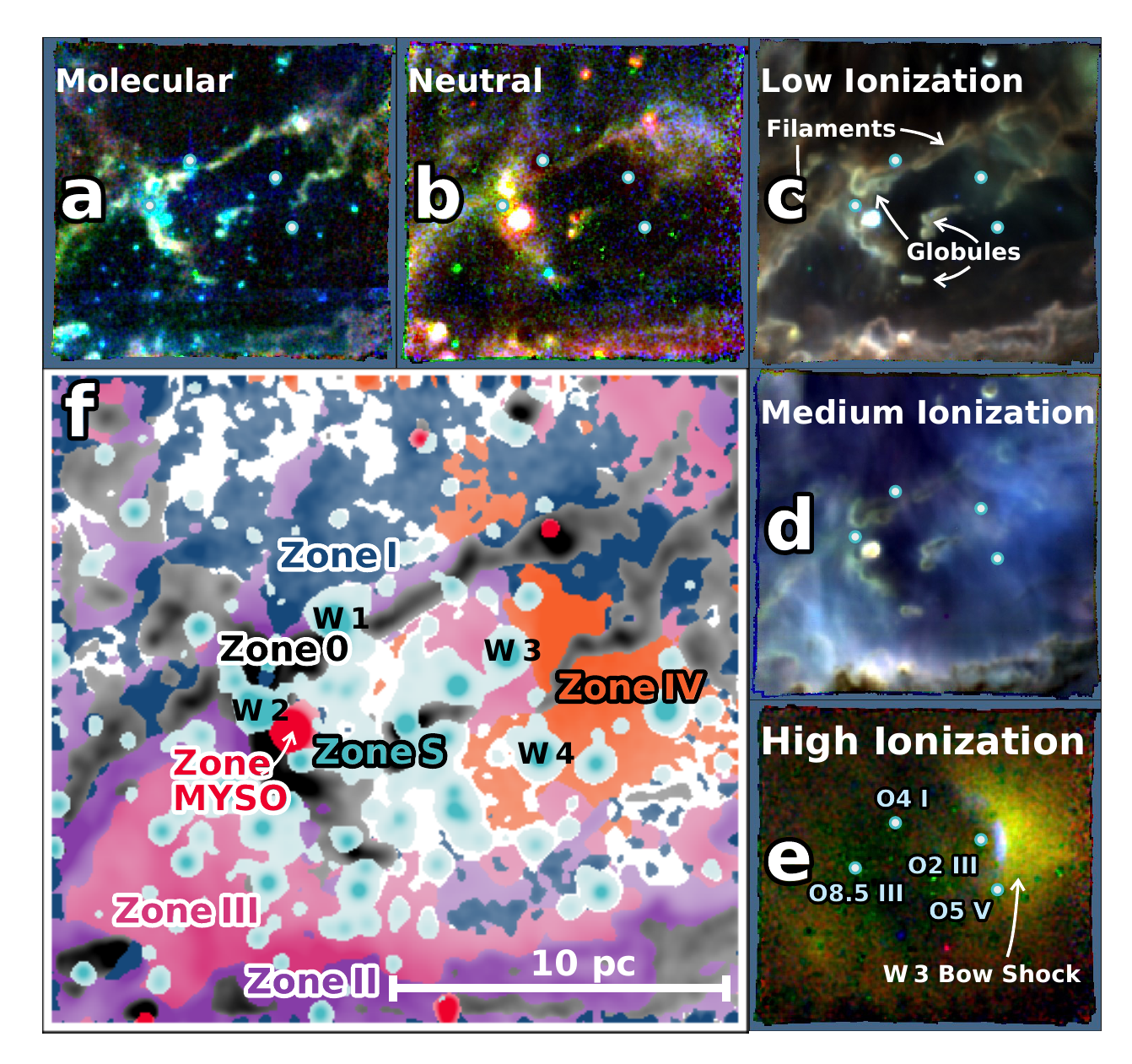}
  \caption{%
    (a--e) Three-channel RGB emission line images extracted from the
    MUSE data cube, showing different emission line types.
    (a)~\NEW{Optical \hmol{} lines}:
    \NEW{4-1 Q(2)} \wav{9057} (red), 
    \NEW{4-1 Q(1)} \wav{9029} (green), 
    \NEW{4-1 O(1)} \wav{9298} (blue).
    (b)~Neutral Fluorescent Lines:
    \ion{O}{1} \wav{8446} (red),
    [\ion{N}{1}] \wav{5199} (green),
    \ion{N}{1} \wav{8223} (blue).
    (c)~Low Ionization Lines:
    [\ion{S}{2}] \wav{6716} (red),
    [\ion{N}{2}] \wav{6583} (green),
    [\ion{O}{2}] \wav{7320} (blue).
    (d)~Medium Ionization Lines:
    [\ion{S}{3}] \wav{9069} (red),
    [\ion{Ar}{3}] \wav{7136} (green),
    [\ion{O}{3}] \wav{4959} (blue).
    (e)~High Ionization Lines:
    [\ion{Cl}{4}] \wav{7531} (red),
    [\ion{Ar}{4}] \wav{4740} (green),
    \ion{He}{2} \wav{4686} (blue).
    (f)~Division of the nebula into different zones,
    according to which type of emission is dominant
    (see text and Table~\ref{tab:zones} for details).
    The positions of the principal ionizing stars
    are shown as W~1 to W~4 in panel~a
    \citep{Walborn:1978k, Walborn:1986y}
    with corresponding spectral types indicated in panel~e
    \citep{Heydari-Malayeri:2010i}.
  }
  \label{fig:zones}
\end{figure*}

Images of the nebula in \NEW{various emission} lines are shown in Figure~\ref{fig:zones}.
Panels a-e show the emission distribution
for six different line classes listed in Table~\ref{tab:ion-class},
with each RGB image combining three emission lines with a broadly similar
degree of ionization (see caption for details).
The basic architecture of the nebula is similar in all but the highest ionization lines,\footnote{
  The highest ionization lines are instead dominated by a large bow-shock structure
that seems to be associated with the O2 star Walborn~3 and which will be discussed elsewhere.
}
consisting of a network of bright filaments and globules,
but differences between the ionization classes become apparent
on a small spatial scale.
In the Low Ionization and Medium Ionization lines, the globules and filaments have limb-brightened
edges and dark cores,
with the Medium Ionization lines showing additional diffuse emission between the filaments that
is absent in the Low Ionization lines.
The Neutral lines also trace the edges of the globules and filaments, but less distinctly than
the Low Ionization lines, and additionally show
weak diffuse emission that is not well correlated with
the diffuse emission of the Medium Ionization lines.
The Deep Neutral lines \NEW{and \hmol{} lines}
are concentrated towards the cores of the globules and filaments. 

The larger panel, Figure~\ref{fig:zones}f, shows
a division of the nebula into seven spatial zones according
to which type of emission predominates, as listed in Table~\ref{tab:zones}.
Normalized surface brightness maps were prepared in the
key emission process for each zone (see table)
and smoothed to a FWHM of \SI{1}{arcsec}.
Each spaxel was then assigned to the zone
whose key emission map has the greatest brightness,
with the map normalizations being adjusted by hand
to give roughly equal numbers of spaxels in each zone.
Spaxels that are below the tenth brightness percentile
in \emph{all} the key emission maps are not included in
any zone (shown as white in Figure~\ref{fig:zones}f).

Five of the zones correspond to diffuse emission of
progressively higher ionization (Zones~0, I, II, III, IV)
whereas the other two zones are dominated by compact sources:
massive Young Stellar Objects (Zone MYSO)
and OB stars (Zone~S).
It can be seen that there is no clear global trend
of ionization as a function of radius, except that
Zone~IV is concentrated near the star W~3.
On the other hand, there are many local regions
where one sees a systematic progression between Zones III-II-0
or, more rarely, III-II-I-0. 

\subsection{Emission line spectra of the different ionization zones}
\label{sec:emiss-line-spectra}

\begin{figure*}
  \centering
  \includegraphics[width=\linewidth]{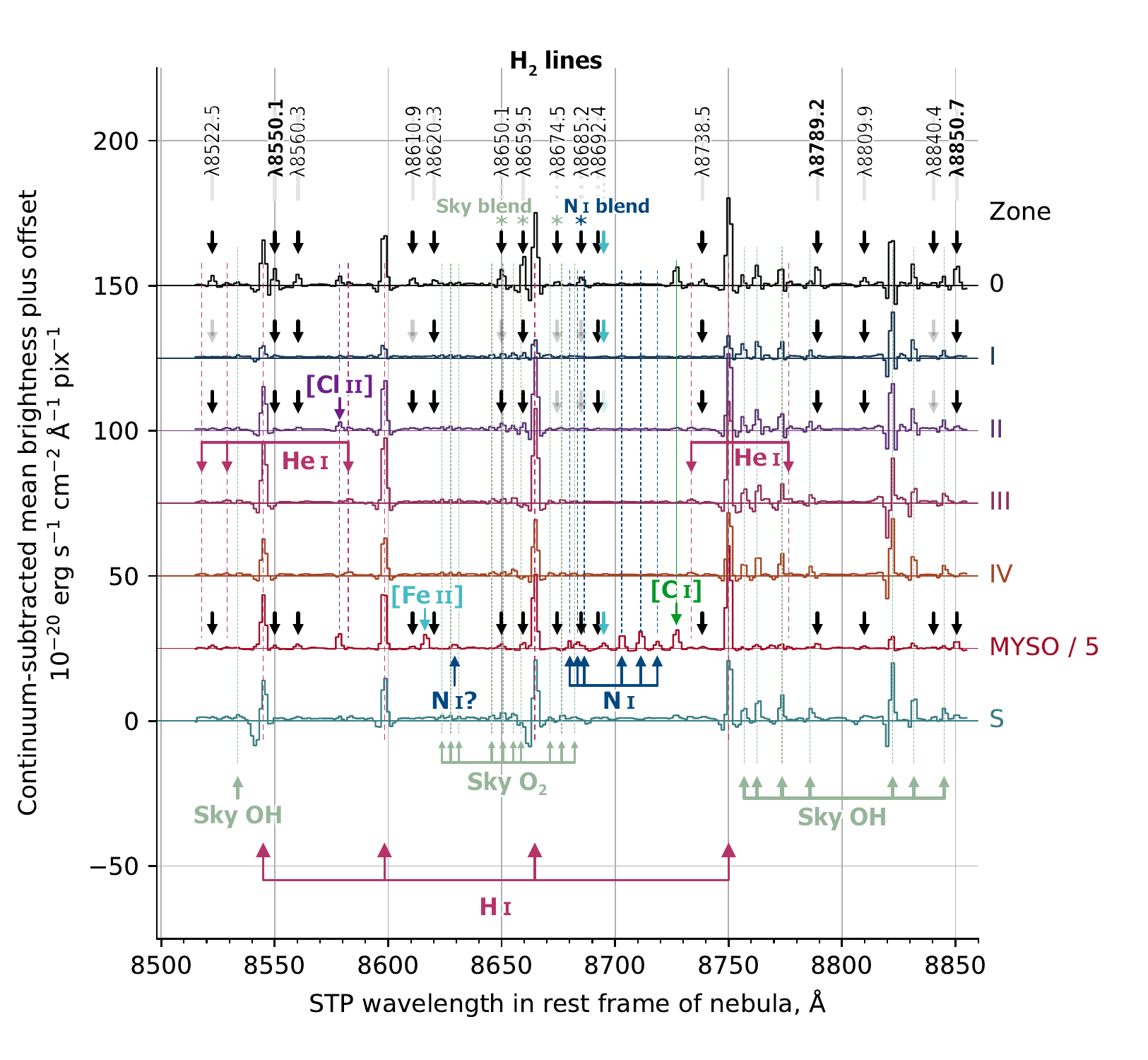}
  \caption{
    Example section of the spectrum at far-red wavelengths
    showing a small sample of the \NEW{\hmol} emission lines
    (downward pointing arrows).
    The most prominent \NEW{\hmol{} lines},
    whose emission is mapped in Figure~\ref{fig:uil-stamps},
    are highlighted in bold. 
    Each spectrum is for a different spatial zone of the nebula
    as labeled at right (see Figure~\ref{fig:zones}).
    All spectra are shown on the same scale but with regular vertical offsets,
    with the exception of Zone~MYSO which is divided by 5
    in order to make the \ion{H}{1} lines have a similar height to the other zones.
    Unidentified lines are detected primarily in Zone~0,
    but many are also found in Zones~I, II, and MYSO.
    Those that are not detected in a given zone are shown by faint gray arrows. 
    Wavelengths of \NEW{atomic} emission lines are marked with vertical dashed lines,
    with color indicating the type of line as in Figure~\ref{fig:bar-code},
    with the addition of gray-green for Terrestrial night sky lines.
    Asterisks mark \NEW{\hmol{}} lines that are affected by blending
    with known lines. 
    The wavelength scale is in the average frame of the nebula
    (\(V_\odot = \SI{+171.1}{kms^{-1}}\)), so nebular lines
    appear at their rest air wavelength,
    but sky lines appear shifted by roughly \SI{-5}{\angstrom}. 
  }
  \label{fig:spectrum-8xxx-B}
\end{figure*}

\begin{figure*}
  \centering
  \includegraphics[width=\linewidth]{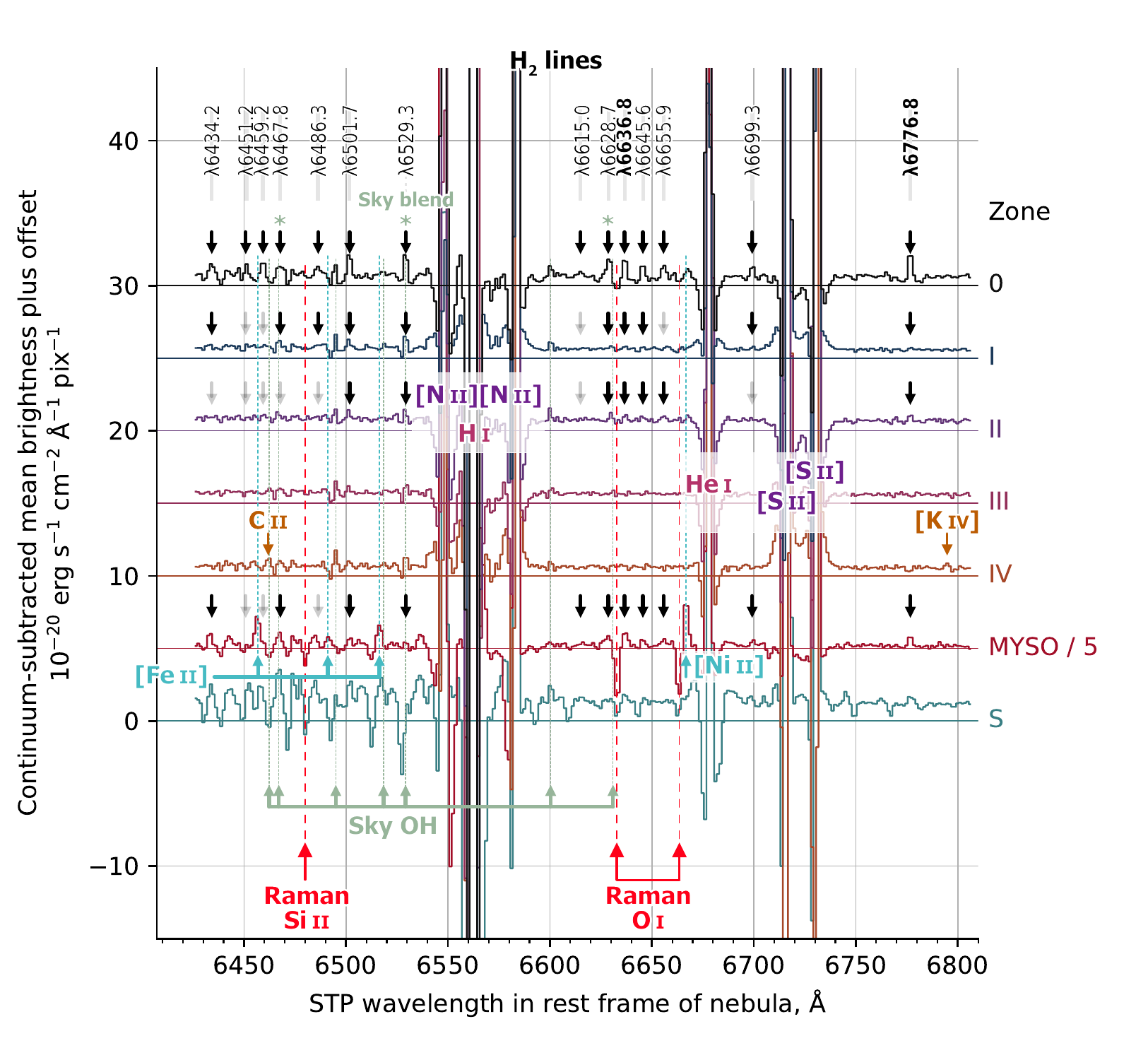}
  \caption{
    As Figure~\ref{fig:spectrum-8xxx-B} but for the region around
    the \ha{} hydrogen recombination line.
    In this spectral range, the well known nebular emission lines of
    \ha{} \wav{6563},
    [\ion{N}{2}] \wavv{6548, 6583},
    \ion{He}{1} \wav{6678}, and
    [\ion{S}{2}] \wavv{6716, 6731}
    emit strongly from all spatial zones.
    Some weak [\ion{Fe}{2}] and [\ion{Ni}{2}] emission lines
    are detected,
    but only from Zone~MYSO,
    while the weak high ionization line
    [\ion{K}{4}] \wav{6795} is detected from Zone~IV
    (the stronger \wav{6102} line of the same doublet is seen in
    Figure~\ref{fig:spectrum-6xxx-A})
    and possibly the \ion{C}{2} \wav{6462}
    recombination line from Zones~III and IV,
    although the latter is severely blended with
    a \chem{OH} night sky line,
    which makes the identification uncertain.
    In addition, Raman scattered absorption features
    of \ion{O}{1} and \ion{Si}{2} are seen very strongly
    in Zone~MYSO and more weakly in Zones~0 and I.
    The \NEW{\hmol{} lines}
    are again most prominent in Zone~0,
    but, in regions that are relatively uncontaminated by
    sky \chem{OH} emission, are also detected in Zones~MYSO, I and II.
  }
  \label{fig:spectrum-6xxx-B}
\end{figure*}

A mean spectrum is calculated for each spatial zone
by averaging over the corresponding spaxels.
Example spectral ranges that are particularly rich in
\NEW{\hmol{} lines} are illustrated in
Figures~\ref{fig:spectrum-8xxx-B} and~\ref{fig:spectrum-6xxx-B},
while additional spectral ranges are shown in Appendix~\ref{sec:additional-spectra}.
The \NEW{\hmol{}} lines are most clearly detected in the Zone~0 spectrum
(top row of the figures)
and are indicated by short black
downward-pointing arrows.
Many are also detected in Zone~MYSO and roughly half are
detected in one or both of Zones~I and II.
Non-detections in a given zone are indicated by light gray arrows.
None are detected from Zones~III or IV.
Some lines are contaminated by overlapping
night sky airglow lines of \chem{OH} or \chem{O_2},
which are marked with asterisks in the figures.
In addition, there are a handful of examples of blends
of \NEW{\hmol{} lines} with unrelated nebular lines:
\ion{N}{1} \wav{8686} in Figure~\ref{fig:spectrum-8xxx-B},
[\ion{Cl}{4}] \wav{8046} in Figure~\ref{fig:spectrum-7xxx-B},
and \ion{H}{1} \wavv{8299, 8306, 8334}  in Figure~\ref{fig:spectrum-8xxx-A}.
In most cases, these blends make only a minor contribution to the
spectrum in Zone~0 but make it impossible to measure the \NEW{\hmol{} line strength}
in the other spatial zones.
Another interesting case is the pair of \NEW{\hmol{} lines 5-1 S(5) and S(7)} at
\wavv{6628.7, 6636.8} (Figure~\ref{fig:spectrum-6xxx-B}),
which closely flank the absorption feature at \wav{6633}
that arises from Raman scattering of the FUV \ion{O}{1} \wav{1027}
line \citep{Henney:2021b}.
The Raman absorption feature can be seen to be
far stronger in Zone~MYSO
than in the other zones,\footnote{%
  Note that the broad Raman scattered wings of \ha{}
  are suppressed by the median filter that has been applied to
  the spectra (see section~\ref{sec:observations}).
}
which means that the flanking \NEW{\hmol{} lines}
in the Zone~0 spectrum are only marginally affected.
At wavelengths longer than \SI{9000}{\angstrom} (Figure~\ref{fig:spectrum-9xxx-A})
some of the \NEW{\hmol{} lines} are affected by
underlying telluric absorption. 

The mean line intensities and central wavelengths for each
spatial zone are determined by fitting Gaussian profiles\footnote{%
  Note that the wavelength pixel size \SI{1.25}{\angstrom}
  provides a relatively coarse sampling of the 
  instrumental profile \citep{Weilbacher:2015a},
  so it is important to integrate the model Gaussian profile
  over the bin width before fitting.
  To that end, we use \texttt{astropy.modeling} framework,
  defining a custom model using the \texttt{scipy} library function
  \texttt{scipy.stats.norm.cdf()}.
}
in 7~pixel windows around the peak wavelength of each line
after masking out any pixels affected by neighboring lines.
Additionally, for the \NEW{\hmol{}} lines where contamination
by sky lines is suspected, the observed spectrum
in Zone~IV is used as an estimate of the sky line profile,
which is subtracted from the spectra of the other zones
before fitting
(under the assumption that there is no \NEW{\hmol} emission from Zone~IV).
Results for \NEW{all nebular} lines are given in Table~\ref{tab:all-lines}.
The first \NEW{three} columns
give respectively
\NEW{the observed wavelength in the nebula reference frame,
  the atomic or molecular transition ID, and the}
rest wavelength.
\NEW{Transitions with the same label in column~4 are potentially
  blended in the observed spectra}.
\NEW{Columns 5--10 give the dereddened}
line intensity
(relative to \(\hb = 100\)) for
\NEW{each spatial zone.}
The uncertainty in the rest wavelength may arise from
both the signal-to-noise of the observations
and additional systematic effects.
This is investigated in detail in Appendix~\ref{sec:rest-wavel-accur},
where we find results consistent with signal-to-noise being
the dominant factor for lines with intensity less than 1\%
of \hb{}.
Comparison with \NEW{atomic} lines of comparable brightness
(for which the rest wavelength is known to high precision)
then allows us to estimate the accuracy in
the rest wavelengths of the unidentified lines.
This is of order \SI{\pm 0.2}{\angstrom} for the brightest
\NEW{\hmol{} lines}, but rising to \SI{\pm 1}{\angstrom} for the weaker lines or
those that are affected by blends (indicated \NEW{by labels in column~4 of} the table).

\startNEW

\subsection{Extinction and reddening}
\label{sec:extinction}
\begin{figure}
  \centering
  \includegraphics[width=\linewidth]{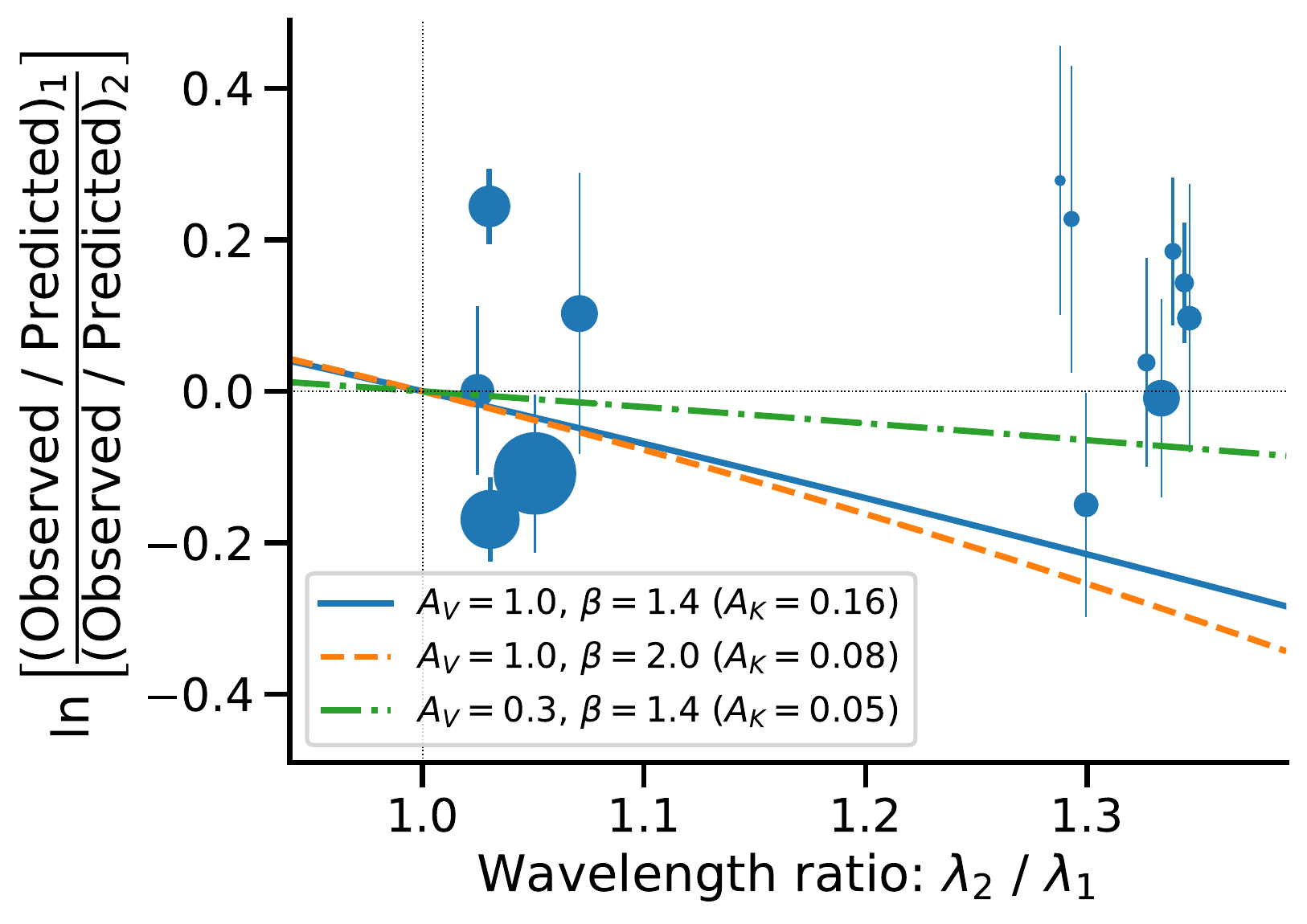}
  \startNEW
  \caption{
    Determination of additional reddening in the PDR. 
    Each point corresponds to a pair of \hmol{} lines (labeled 1 and 2)
    that arise from a common upper level, with \(\lambda_1 < \lambda_2\).
    For example, 5-1 S(7) \wav{6637} and 5-2 S(7) \wav{8851}.
    The predicted intrinsic intensity ratio is then independent of
    physical conditions, and given by the ratio of the
    transition probabilities multiplied by the photon energies. 
    The \(x\) axis shows \(\lambda_2/\lambda_1\), while the \(y\) axis shows
    the natural logarithm of the ratio of the observed intensities,
    after correcting for foreground extinction to the \hii{} region,
    normalized by the predicted ratios.
    Symbol size is proportional to the product of the intensities
    of the two lines.
    For a differential extinction law of the form \(\lambda^{-\beta}\),
    the predicted behavior is \(\ln(\text{ratio}) = \tau_2 [1 - (\lambda_2/\lambda_1)^\beta]\)
    where \(\tau_2 = (0.4 \ln 10) (\SI{5500}{\angstrom} / \lambda_2)^\beta \avpdr\)
    and \avpdr{} is the visual dust extinction in magnitudes
    between the ionization front and the region where the \hmol{} lines arise.
    This is indicated by colored lines for different values of \avpdr{} and \(\beta\),
    assuming \(\lambda_2 = \SI{8700}{\angstrom}\), which is the average value for
    the group of points on the right hand side of the figure.
    It can be seen that the observations are consistent with \(\avpdr \approx 0\)
    and that \(\avpdr > 1\) is strongly excluded.
  }
  \stopNEW
  \label{fig:reddening-h2}
\end{figure}

To account for differential extinction by dust,
the spectra are de-reddened using the observed hydrogen Balmer decrement in each zone.
An intrinsic Balmer decrement of \(I(\ha)/I(\hb) = 2.821\) is assumed,
corresponding to Case~B recombination at the average density
and temperature of the ionized nebula
\citetext{\(n = \SI{100}{cm^{-3}}\), \(T = \SI{12500}{K}\),
  \citealp{Valerdi:2019a}}.
We use an extinction law tailored to the SMC with
total-to-selective extinction ratio \(R_V = 2.74\)
\citep{Gordon:2003l}, which we calculate using PyNeb \citep{Luridiana:2015a}.
The derived visual extinction \(A_V\) for the different zones
% ranges from \(E(B-V) = 0.06\) in Zone~IV to \(E(B-V) = 0.22\)
ranges from \(A_V = \SI{0.16}{mag}\) in Zone~IV to \(A_V = 0.60\)
in Zones~II and III.\@
This reddening correction will be most accurate for lines
whose emission zones overlap with those of the Balmer lines,
which includes all lines emitted in the \hii{} region and
the ionization front.

The \hmol{} lines may have additional extinction due
to dust in the PDR between the ionization front and the dissociation front.
We can use the \hmol{} spectrum itself to investigate this, since there are several pairs
of lines with a common upper level.
After eliminating all lines that are potentially effected by blends,
we are left with five pairs of lines with closely spaced wavelengths (\(\lambda_2/\lambda_1 = 1.04 \pm 0.02\))
and eight pairs of lines with more widely separated wavelengths (\(\lambda_2/\lambda_1 = 1.32 \pm 0.02\)).
If there is any additional reddening in the PDR, then the latter group
should show intensity ratios (shorter wavelength line over longer wavelength line)
that are smaller than the predicted intrinsic values.  
Figure~\ref{fig:reddening-h2} shows that this is not the case:
both groups cluster around the zero line, indicating that any
additional reddening in the PDR must be small.
Assuming a power-law extinction \(A_\lambda \propto \lambda^{-\beta}\) with index \(\beta = 1.4\)
(appropriate for \(R_V = 2.74\)), then the formal result is \(\avpdr = (-0.4 \pm 0.6)\),
and \(\avpdr > 1\) is excluded at the 5\% significance level. 
\stopNEW

\subsection{Interzone diagnostic ratios}
\label{sec:interz-diagn-rati}
\begin{figure*}
  \centering \includegraphics[width=\linewidth]{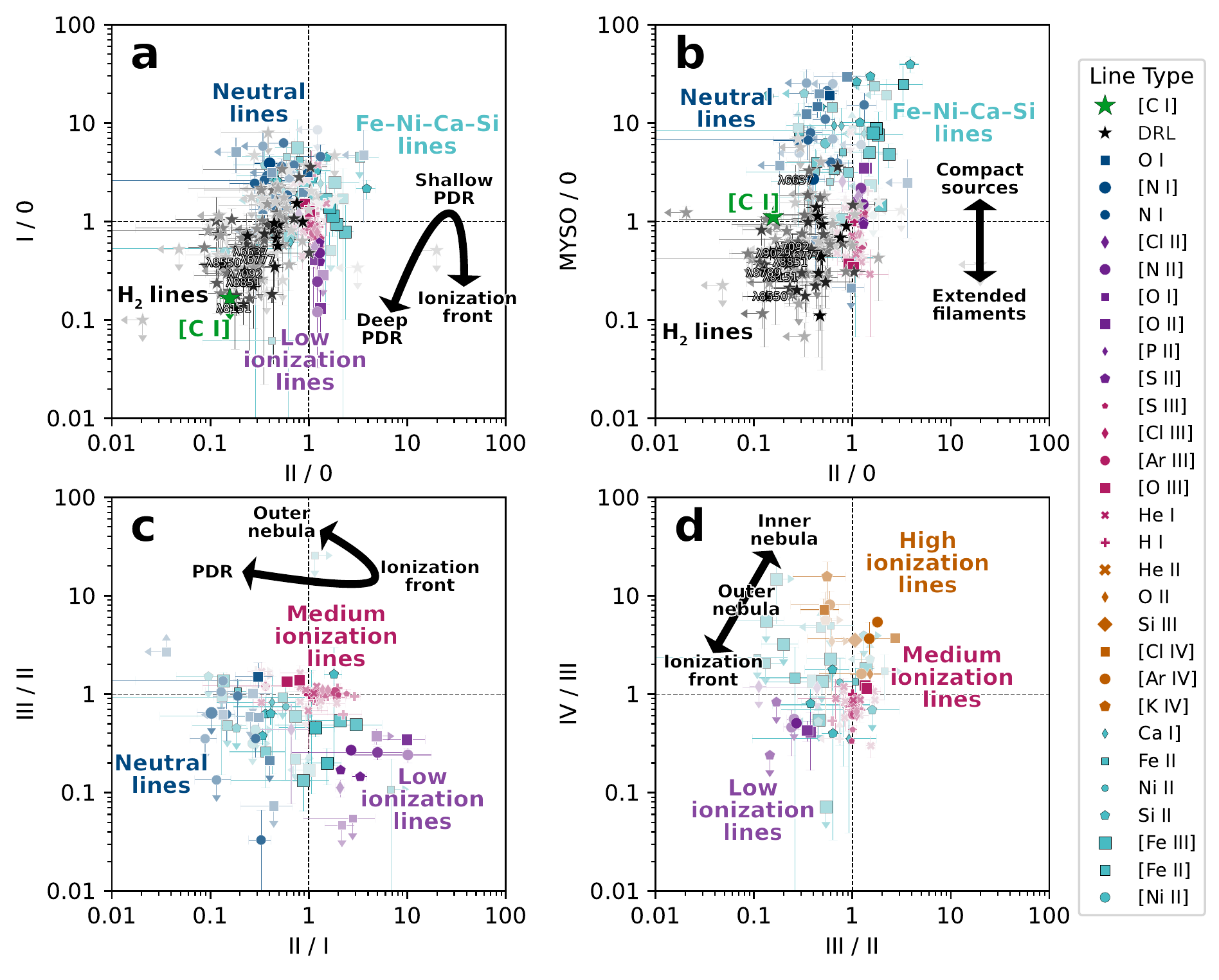}
  \caption{
    Ratios of mean line brightness between different spatial zones.
    (a)~\RRat{I}{0} versus \RRat{II}{0}. 
    (b)~\RRat{MYSO}{0} versus \RRat{II}{0}. 
    (c)~\RRat{III}{II} versus \RRat{II}{I}. 
    (d)~\RRat{IV}{III} versus \RRat{III}{II}. 
    High Ionization lines are omitted from panels a and b
    due to lack of detection from the relevant zones,
    while \NEW{\hmol{} lines}  are
    similarly omitted from panels c and d.
    Selected \NEW{\hmol{} lines} from Figure~\ref{fig:uil-stamps} are
    labeled with their wavelength on panels a and b.
    Symbol colors indicate the line classes (Table~\ref{tab:ion-class})
    as in Figure~\ref{fig:bar-code}, with
    color intensity that is proportional to
    the logarithm of the line strength.
    Symbol shapes indicate the specific ion species within each
    class, as given in the key at right.
    Error bars indicate one-sigma uncertainties,
    while arrows indicate three-sigma upper or lower
    limits in case of detection in only one of the zones
    that contribute to the ratio.
    Double-ended thick black arrows indicate suggested
    interpretations of trends seen in each diagram. 
  }
  \label{fig:ratios}
\end{figure*}

Ratios of relative line intensities between
different spatial zones are shown in Figure~\ref{fig:ratios}.
The most important result for \NEW{optical molecular hydrogen lines} is shown in
panel~a, which plots the ratio \Rat{I}{0} against \Rat{II}{0}.
It can be seen that the PDR lines
([\ion{C}{1}] and \hmol) are the only class of lines that
consistently show both \(\Rat{I}{0} < 1\) and \(\Rat{II}{0} < 1\).
They are thus well separated from the other classes of line
on this diagram.
The Low Ionization lines, which should arise from ionized gas
close to the outer edge of the \hii{} region
or from the ionization front,
show \(\Rat{I}{0} < 1\) but \(\Rat{II}{0} > 1\), presumably due to
limb brightening at the edge of the globules and filaments.
The Neutral lines on the other hand,
which should arise just beyond the ionization front,
show \(\Rat{II}{0} < 1\) but \(\Rat{I}{0} > 1\).
The \NEW{\hmol{} lines} seem to fall on a sequence between the
position of the [\ion{C}{1}] line
(\(\Rat{I}{0} < 0.17\), \(\Rat{II}{0} = 0.16\))
and the main clump of Neutral lines
(\(\Rat{I}{0} \approx 3\), \(\Rat{II}{0} \approx 0.5\)).
One possible explanation for this is that the \NEW{\hmol{} lines} arise
at a range of intermediate depths in the PDR
between the ionization front and the CO dissociation front.

Figure~\ref{fig:ratios}b plots the ratio
\Rat{MYSO}{0} against \Rat{II}{0}.
It is notable that the \NEW{\hmol{}} lines maintain almost
the same distribution as in panel~a, although other
classes of line change significantly.
In particular, the [\ion{C}{1}] line,
the Low Ionization lines,
and the Neutral lines all
show typically \(\Rat{MYSO}{0} \approx 3 \times \Rat{I}{0}\) 
(compare panel~b with panel~a).
This could be explained by a combination of a
higher gas density and softer illuminating spectrum in
the compact massive YSO sources.
The \feni{} lines show an even greater enhancement in the
MYSO zone, with typically \(\Rat{MYSO}{0} \approx 10 \times \Rat{I}{0}\),
which suggests that the gas-phase abundance of refractory elements
is less depleted in compact sources than in the extended nebula. 

Figure~\ref{fig:ratios}c and d show gradations within the
more ionized regions of the nebula and do not include the \NEW{\hmol{} lines} since
they are not detected from zones III and IV.
Panel~c shows the transition between the neutral and
ionized sides of the ionization front,
while panel~d shows the increasing degree of ionization
within the \hii{} region, which is driven by a combination
of local ionization stratification and the different effective temperatures of the principal ionizing stars
(see Figure~\ref{fig:zones}). 

\section{Discussion}
\label{sec:discussion}

\startNEW
Optical \hmol{} emission lines from a PDR were first detected in the reflection
nebula NGC~2023 \citep{Burton:1992a}, where about 30 lines were observed in the
wavelength range \SIrange{7600}{8800}{\angstrom}
(a subset of the lines that we detect in NGC~346).
However,
very few studies of the optical \hmol{} lines have been published in the intervening 30 years,
at the same time as observations of near-infrared (\SIrange{1}{2}{\micron})
vibrationally excited \hmol{} lines have become commonplace.
Notable exceptions have been studies of shock-excited molecular hydrogen
in the Herbig-Haro objects HH~91A \citep{Gredel:2007a} and HH~1 \citep{Giannini:2015a}.
It is therefore worth asking: is there something special about NGC~346 that makes these lines particularly easy to detect?
\stopNEW

\begin{table}
  \newcommand\tss{\textsuperscript}
  \setlength\tabcolsep{3pt}
  \caption{\NEW{Optical PDR lines} in NGC~346 and other \NEW{star forming regions}}
  \label{tab:region-comparison}
  \begin{tabular}{@{} l r r@{\({}={}\)}r S r@{}}\toprule
    &\multicolumn{4}{c}{Relative intensity (\(\hb = 100\))}
    & \\
    Region/zone
    & [\ion{C}{1}] \wav{8727}
    & \multicolumn{2}{r}{\(\max(\hmol)\)}
    & \(\sum\hmol\)
    & \(n(\hmol)\)
    \\
    {(1)} & {(2)} & \multicolumn{2}{r}{(3)} & {(4)} & {(5)} \\
    \midrule
    Orion S\tss{a} & 0.17 & \wav{9114}& 0.014 & 0.026 & 3 \\
    Orion Bar\tss{a} & 0.06 & \wav{9114}& 0.009 & 0.017 & 3 \\      
    \addlinespace
    M 17 A\tss{b} & 0.27 & \wav{8459}& 0.019 & 0.16 & 15 \\
    M 17 B\tss{b} & 0.73 & \wav{8894}& 0.080 & 0.63 & 16 \\
    \addlinespace
    30 Dor Globule\tss{c} & 0.12 & \wav{9029}& 0.022 & 0.16 & 18 \\
    30 Dor Clump\tss{c} & 0.06 & \wav{9029}& 0.086 & 0.71 & 20 \\
    \addlinespace
    NGC 346 MYSO\tss{d} & 0.24 & \wav{9297}& 0.200 & 5.93 & 80 \\
    NGC 346 Zone 0\tss{d} & 0.21 & \wav{9029}& 0.399 & 14.83 & 114 \\
    \bottomrule
    \addlinespace
    \multicolumn{6}{@{} p{0.95\linewidth} @{}}{%
    Columns:
    (1)~\NEW{Massive} star forming region and specific PDR zone (see Notes below). 
    (2)~[\ion{C}{1}] \wav{8727} reddening-corrected relative intensity
    as percentage of \hb{}.
    (3)~Wavelength and relative intensity of the brightest observed \NEW{optical \hmol{} line}.
    (4)~Sum of relative intensities of all observed \NEW{optical \hmol{} lines}.
    (5)~Number of observed \NEW{optical \hmol{} lines}.
    }\\
    \addlinespace
    \multicolumn{6}{@{} p{0.95\linewidth} @{}}{%
    Notes: \tss{a}Orion S (South) and Orion Bar zones are as depicted in Figure~2b of
    \citet{Henney:2021b}.
    \tss{b}M~17~A and M~17~B are two zones within Tile~03 of
    the MUSE mosaic of M~17, located \SI{2.5}{arcmin} north of the central star cluster.
    Zone~A is adjacent to a bright ionization front,
    while Zone~B is in a more shielded location.
    \tss{c}In 30~Doradus, the Globule zone is a bright compact emission knot with
    a cometary morphology located \SI{40}{arcsec} SSE of the central cluster R136,
    while the Clump zone is a more diffuse and shielded region
    located \SI{1}{arcmin} SE of the central cluster.
    \tss{d} See this paper.
    }
  \end{tabular}
\end{table}

\begin{figure}
  \centering
  \includegraphics[width=\linewidth]{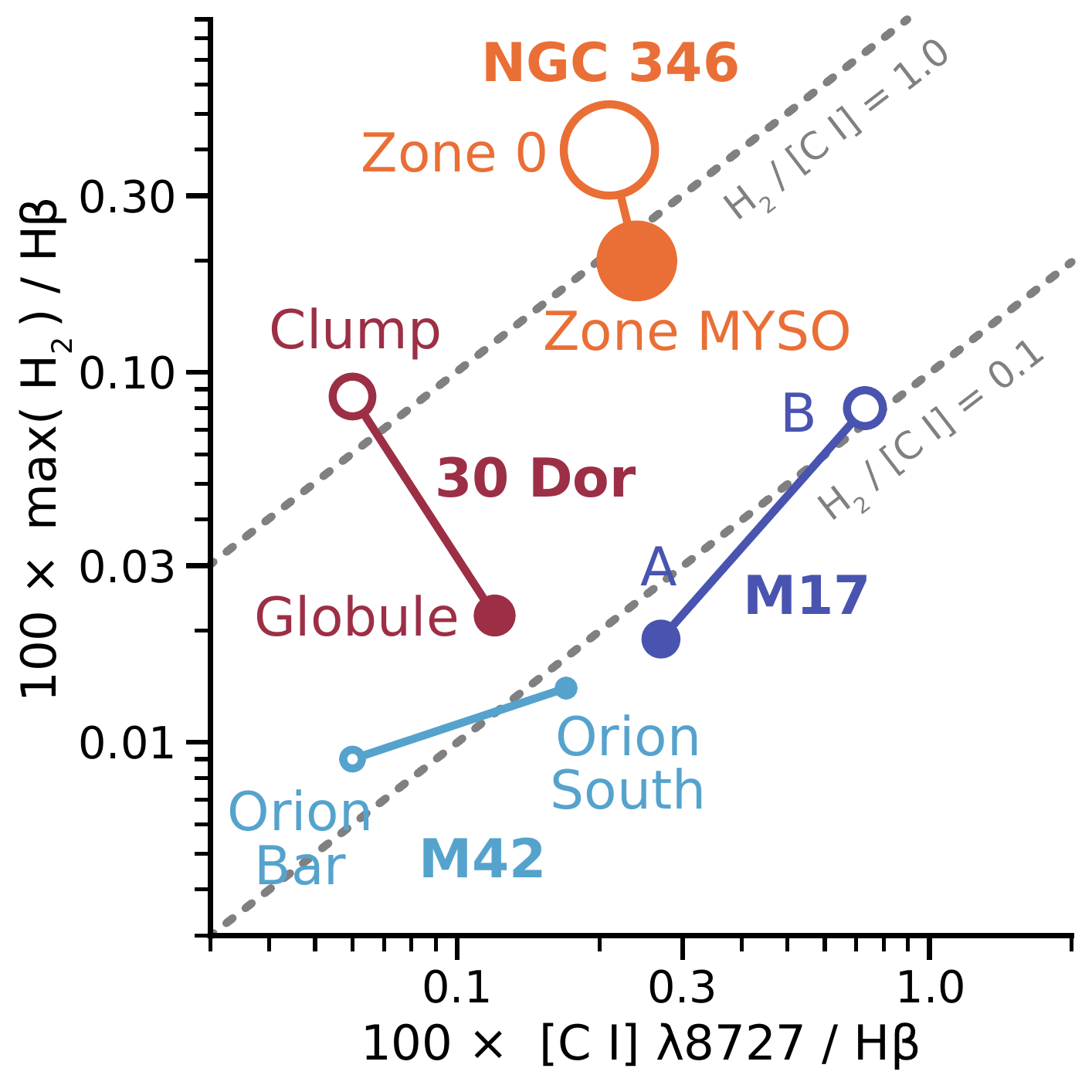}
  \caption{
    Relative intensity of the brightest observed \NEW{optical \hmol{} line} plotted
    against relative intensity of the [\ion{C}{1}] \wav{8727} line for the
    different star forming regions and zones listed in Table~\ref{tab:region-comparison}.
    The symbol size (area) is proportional to the number of observed \NEW{\hmol{} lines} in each zone.
    Galactic regions are shown in blue, while lower metallicity extragalactic
    regions are shown in orange/red.
    Two different PDR zones are shown for each region: the filled symbol
    indicates the zone that is closer to the illuminating stars.
    Observational uncertainties are generally less than 10\%, but
    as high as 50\% in the case of the Orion \NEW{\hmol{}} measurements.
  }
  \label{fig:different-regions}
\end{figure}

We have made a preliminary search for \NEW{optical \hmol{} lines} in MUSE data cubes of
other bright PDRs.
These include the Milky Way regions of the Orion Nebula \citep{Weilbacher:2015a},
and the Omega Nebula (M~17),\footnote{
  Unpublished MUSE observations under Programme ID 0103.C-0288(A), PI: M. Reiter,
  \url{http://archive.eso.org/wdb/wdb/eso/abstract/query?progid=0103.C-0288(A)}.
}
together with 30~Doradus in the Large Magellanic Cloud \citep{Castro:2018a}.
Results are shown in Table~\ref{tab:region-comparison} and
Figure~\ref{fig:different-regions} for two separate zones in each region,
which are compared with Zones~0 and MYSO in NGC~346.
For all regions, we selected zones that were bright in [\ion{C}{1}] \wav{8727}
emission as the most promising places to look for \NEW{the optical \hmol{} lines}
(see notes to Table~\ref{tab:region-comparison} for detailed locations).
Note that intensities are corrected for foreground reddening using the
observed Balmer decrement, but there may be additional extinction within the PDR,
which is not corrected for. 

We did detect \NEW{optical \hmol{} lines} in each one of these regions, but the number
of lines and their relative intensities are both far smaller than in NGC~346.
For instance, in the Orion Nebula, we detected only two \NEW{optical \hmol{} lines},
with relative intensities of order 0.01\% of \hb{}.%
\footnote{
  We also detect a third line at \wav{6814}, which is not found in NGC~346 or
  any of the other regions, but whose spatial distribution is
  \NEW{similar to the \hmol{} lines. As yet, we have found no identification for this line.}
}
The summed relative brightness of all the observed \NEW{optical \hmol{} lines}
in the prototypical Orion Bar PDR
is nearly 1000 times smaller than in NGC~346.
The other two regions, M~17 and 30~Doradus, show intermediate levels of \NEW{optical \hmol{}} emission:
tens of lines are detected, with individual relative brightness of
up to 0.09\% of \hb{}. The total \NEW{optical \hmol{} line} relative brightness is
10 to 50 times higher than in Orion,
but still 20 to 100 times lower than in NGC~346.

Figure~\ref{fig:different-regions} shows that despite the generally
excellent spatial correlation between the \NEW{\hmol{}} and [\ion{C}{1}] emission
that is seen \emph{within} each region,
the two types of emission show almost no correlation
in their strength between \emph{different} star forming regions.
However, if we divide the regions into Galactic and extragalactic groups,
a significant trend with metallicity emerges.
The Galactic regions (blue symbols), which both have metallicities
that are approximately solar (\(Z_\odot\)),
show \(\hmol/[\ion{C}{1}] \approx 0.1\),
whereas the extragalactic regions (red/orange symbols)
show higher values, reaching \(\hmol/[\ion{C}{1}] \ge  1\)
for the lowest metallicity region NGC~346 (\(Z \approx 0.2\,Z_\odot\)).
The intermediate metallicity region 30~Dor (\(Z \approx 0.5\,Z_\odot\))
shows intermediate values of \(\hmol/[\ion{C}{1}] \approx 0.2\) to \(1\).
In addition, for all four regions, \(\hmol/[\ion{C}{1}]\)
is consistently higher in the zone that is farther from the illuminating stars.
On the other hand, we have found no consistent trends with other
parameters of the regions such as luminosity, ionization parameter, density,
or surface brightness.

\startNEW
It is also noteworthy that we derive a very low extinction through the PDR
in NGC~346 (section~\ref{sec:extinction}).
In Galactic PDRs, a typical extinction in the \(K\) band (\SI{2.2}{\micron})
of \(A_K \sim 0.5\)
is derived from the near-infrared \hmol{} lines \citep{Kaplan:2021a}.
The visual extinction \(A_V\) will be 3 to 10 times larger, depending
on the details of the extinction law
\citep{Fitzpatrick:1999a, Fahrion:2023a},
but the foreground extinction needs to be subtracted to obtain the
internal extinction in the PDR. 
For different samples in the Orion Bar,
the result is \(\avpdr = 2\) to \(10\) \citetext{Table~1 of \citealp{Peeters:2023a}},
which is much larger than the value \(\avpdr = (-0.4 \pm 0.6)\)
that we derive for NGC~346.
It is possible that variations in geometry may explain part of this difference,
since the Orion Bar is viewed edge-on, which amplifies the path lengths of the line
of sight through the PDR.
However, the results of \citet{Kaplan:2021a} show a variation by only a
factor of 2 in the PDR extinction over several different regions with a variety of
geometries, suggesting that the low extinction in NGC~346 is intrinsic.
\stopNEW

\startNEW
In conclusion, we find that NGC~346 is indeed special in both
the large number and high relative brightness
of its optical emission lines of molecular hydrogen, \chem{H_2},
together with the relatively low extinction through the PDR.
\stopNEW
Although we do not know the exact reason for this, it is likely to
be related to the low metallicity of the region.

\section{Summary}
\label{sec:summary}

\NEW{The principal results of this paper are as follows:}

\begin{enumerate}[1.]
\item \NEW{We have presented a catalogue of more than 100 optical
    emission lines of vibrationally excited molecular hydrogen
    (\chem{H_2}),} that are detected in the MUSE optical spectrum of
  filamentary photodissociation regions (PDRs) around the
  mini-starburst cluster NGC~346 in the Small Magellanic Cloud.
  They are found at wavelengths \(> \SI{6000}{\angstrom}\)
  and up to the upper limit of the MUSE observations at
  \SI{9300}{\angstrom} (Figure~\ref{fig:bar-code}).
\item \NEW{Transitions as high as \(v = 13\) with excitation temperature
  \(> \SI{50000}{K}\) are detected (for example, 13-6 S(2)
  \wav{7199}, Figure~\ref{fig:spectrum-7xxx-A}).}
\item Their spatial distribution in the nebula is different from that
  of most \NEW{atomic and ionized lines}, but is closest to that of
  the [\ion{C}{1}] \wav{8727} line (Figure~\ref{fig:uil-stamps}).  The
  pattern of line ratio variations between different zones of the
  nebula (Figure~\ref{fig:ratios}a) is consistent with \NEW{\hmol{}
    lines} originating at a range of depths in the PDR between the
  ionization front and the fully molecular zone.
\item In addition to diffuse filamentary emission, the \NEW{optical
    \hmol{} lines} are also detected from the compact circumstellar
  environments of massive young stellar objects (MYSOs) in the region,
  but with a relative intensity that is typically lower than in the
  diffuse filaments (Figure~\ref{fig:ratios}b).  This contrasts with
  the case of known fluorescent lines of neutral atoms (\ion{N}{1},
  \ion{O}{1}) or lines from ions of refractory elements
  ([\ion{Fe}{2}], [\ion{Fe}{3}], \ion{Si}{2}), which are all strongly
  enhanced in the MYSOs.
\item A preliminary study of MUSE observations of other bright
  PDRs (Orion Nebula, M17, 30~Dor) shows that
  the \NEW{optical \hmol{} lines} are also present in those regions,
  but with a relative intensity that is much lower than in NGC~346.
  \startNEW
  The summed relative brightness of optical \chem{H_2} lines
  with respect to \hb{} in NGC~346 is 800 times larger than
  in the prototypical PDR of the Orion Bar.
  \stopNEW
  We find that the intensity ratio \(\hmol/\text{[\ion{C}{1}]}\)
  is a strongly decreasing function of the metallicity.
\item \NEW{The internal extinction in the PDR to the \hmol{} emission
    zone is found to be \(\avpdr < 1\) (section~\ref{sec:extinction}),
    which is significantly smaller than is found in the Orion Bar
    and other PDRs in our Galaxy.}
\item A separate result of this study is the detection of a highly
  ionized bow shock around the O2 star Walborn~3
  (Figure~\ref{fig:zones}e) via the emission of [\ion{Ar}{4}] and
  \ion{He}{2} lines, which will be presented in more detail elsewhere.
\end{enumerate}

\section*{Acknowledgements}

Based on observations collected at the European Southern Observatory
under ESO program 098.D-0211(A).
We thank P. Zeigler for discussions on MUSE data reduction
and V. Escalante for introducing us to EMILI.
\startNEW
We are grateful to M. Richer and A. López for discussions that led
to the identification of the molecular hydrogen lines.
We thank the anonymous referee for a useful report.
\stopNEW
We gratefully acknowledge financial support provided by
\foreignlanguage{spanish}{%
  Dirección General de Asuntos del Personal Académico,
  Universidad Nacional Autónoma de México},
through grant
``\foreignlanguage{spanish}{%
  Programa de Apoyo a Proyectos de Investigación
  e Inovación Tecnológica IN109823}''.
MV acknowledges support from the Mexican
\foreignlanguage{spanish}{Consejo Nacional de Ciencia y Tecnología}
from the program
``\foreignlanguage{spanish}{Estancias Posdoctorales por México 2022}''.

Scientific software and databases used in this work include
SAOImage~DS9\footnote{\url{https://sites.google.com/cfa.harvard.edu/saoimageds9}} \citep{Joye:2003a},
the Atomic Line List\footnote{\url{https://www.pa.uky.edu/~peter/newpage/}} \citep{van-Hoof:2018a},
\startNEW
Cloudy\footnote{\url{https://gitlab.nublado.org/}} \citep{Chatzikos:2023a},
PyNeb\footnote{\url{http://research.iac.es/proyecto/PyNeb//}} \citep{Luridiana:2015a},
\stopNEW
EMILI\footnote{\url{https://web.pa.msu.edu/astro/software/emili/}}
\citep{Sharpee:2003a},
and the following 3rd-party Python packages:
numpy, scipy, pandas, \NEW{statsmodels}, astropy, astroquery, regions, mpdaf, matplotlib, seaborn, cmasher.

\section*{Data Availability}
All observational datasets analyzed in this work are available from the
ESO Science Archive Facility
(\url{http://archive.eso.org/eso/eso_archive_main.html}),
see footnotes \ref{fn:1} and \ref{fn:2} in section~\ref{sec:observations} for details.
All data analysis scripts,\footnote{
  \url{https://github.com/will-henney/muse-hii-regions/tree/main/scripts}
}
extracted spectra and other derived data products,\footnote{
  \url{https://github.com/will-henney/muse-hii-regions/tree/main/data/n346-lines}
}
and documentation of the analysis steps\footnote{
  \url{https://github.com/will-henney/muse-hii-regions/blob/main/docs/ngc-346.org}
}
are freely available.

\bibliography{smc-deep-neutral-refs}
\clearpage
\onecolumn

% \section{Table of all observed emission lines}
% \label{sec:known-lines-plus-h2}

\startNEW
{
  \setlength{\tabcolsep}{2pt}
  \begin{longtable}{RrRr RRRRRR}
  \caption{Wavelengths and intensities of observed emission lines in NGC~346}
  \label{tab:all-lines}
  \\
  \multicolumn{10}{p{\linewidth}}{
    Table columns:

\noindent (1)
Observed wavelength (air) in \si{\angstrom} in the reference frame of the nebula
\citetext{\(V_\odot = \SI{171}{km.s^{-1}}\), \citealp{Zeidler:2022a}}.
This is derived from a Gaussian fit to the mean profile from the dominant
spatial zone (Table~\ref{tab:zones}) of the respective line class,
with total uncertainty estimated as in Appendix~\ref{sec:rest-wavel-accur}.
Observed wavelengths of terrestrial night sky lines (indicated by italics)
are also given in the nebula frame to facilitate identification of blends with
nebular lines. As a result, these are systematically shorter than the
corresponding rest wavelengths.

\noindent (2) Atomic, ionic, or molecular species.
For molecular hydrogen lines, all transitions are within
the ground electronic state, with 
the spectroscopic designation given as
vibrational quantum numbers of upper and lower levels: \(v'\)-\(v\),
followed by rotational branch label:
O (\(\Delta J = +2\)), Q (\(\Delta J = 0\)), or S (\(\Delta J = -2\))
with lower level rotational quantum number in parentheses (\(J\)).
For instance, \hmol{} 12-5 S(1) is the transition from \(v' = 12\), \(J' = 3\)
to \(v = 5\), \(J = 1\).

\noindent (3) Rest wavelength (air) of the transition
\citep{Komasa:2011a, van-Hoof:2018a, Chatzikos:2023a}.

\noindent (4) Blend label. When two or more transitions lie at closely similar wavelengths,
they may potentially be blended in the MUSE spectra.
Each potential blend is assigned a sequential label of one or two letters.
In general we include all cases with rest wavelength separation
\(\delta \lambda < \SI{2.5}{\angstrom}\) (2 pixels),
but for blends between a nebular line and a sky line the Doppler shift
between the respective emission frames is accounted for. 

\noindent (5-10) Relative line intensity as percentage of \hb{} for each spatial zone.
These values are corrected for reddening based on the Balmer decrement
}
\\ \\
\toprule
 & & & & \multicolumn{6}{c}{Line intensity, \(I(\hb = 100)\)} \\
\lambda(\text{obs}) & Species & \lambda(\text{rest}) & Blend & \text{Zone 0} & \text{Zone I} & \text{Zone II} & \text{Zone III} & \text{Zone IV} & \text{Zone MYSO} \\
(1) & (2) & (3) & (4) & (5) & (6) & (7) & (8) & (9) & (10) \\
\midrule
\endfirsthead
\toprule
 & & & & \multicolumn{6}{c}{Line intensity, \(I(\hb = 100)\)} \\
\lambda(\text{obs}) & Species & \lambda(\text{rest}) & Blend & \text{Zone 0} & \text{Zone I} & \text{Zone II} & \text{Zone III} & \text{Zone IV} & \text{Zone MYSO} \\
(1) & (2) & (3) & (4) & (5) & (6) & (7) & (8) & (9) & (10) \\
\midrule
\endhead
\midrule
\multicolumn{10}{r}{Continued on next page} \\
\midrule
\endfoot
\bottomrule
\endlastfoot
4606.49 \pm 0.26 & [\ion{Fe}{3}] & 4607.12 & a & 0.03 \pm 0.01 & 0.05 \pm 0.02 &  & 0.01 \pm 0.01 &  & 0.07 \pm 0.01 \\
 & \ion{N}{2} & 4607.16 & a &  &  &  &  &  &  \\
 & \ion{O}{2} & 4638.86 & b &  &  &  &  &  &  \\
 & \ion{N}{3} & 4640.64 & b &  &  &  &  &  &  \\
4639.12 \pm 0.57 & \ion{O}{2} & 4641.81 & b &  &  &  & 0.02 \pm 0.02 & 0.06 \pm 0.03 &  \\
 & \ion{N}{2} & 4643.06 & b &  &  &  &  &  &  \\
4649.69 \pm 0.13 & \ion{O}{2} & 4649.13 & c &  &  &  & 0.10 \pm 0.02 & 0.16 \pm 0.02 &  \\
 & \ion{O}{2} & 4650.84 & c &  &  &  &  &  &  \\
4657.78 \pm 0.17 & [\ion{Fe}{3}] & 4658.10 &  & 0.11 \pm 0.02 & 0.14 \pm 0.08 & 0.16 \pm 0.01 & 0.07 \pm 0.02 &  & 0.79 \pm 0.10 \\
4685.42 \pm 0.10 & \ion{He}{2} & 4685.68 &  &  &  &  &  & 0.23 \pm 0.02 &  \\
4701.34 \pm 0.11 & [\ion{Fe}{3}] & 4701.62 &  & 0.03 \pm 0.03 &  & 0.06 \pm 0.02 & 0.04 \pm 0.02 &  & 0.26 \pm 0.02 \\
4711.44 \pm 0.13 & [\ion{Ar}{4}] & 4711.37 & d &  &  &  & 0.46 \pm 0.22 & 1.63 \pm 0.16 &  \\
 & \ion{He}{1} & 4713.14 & d &  &  &  &  &  &  \\
4733.31 \pm 0.64 & [\ion{Fe}{3}] & 4734.00 &  &  &  &  &  &  & 0.07 \pm 0.04 \\
4740.05 \pm 0.08 & [\ion{Ar}{4}] & 4740.17 &  &  &  &  & 0.28 \pm 0.03 & 1.47 \pm 0.08 &  \\
4754.45 \pm 0.12 & [\ion{Fe}{3}] & 4754.81 &  & 0.04 \pm 0.02 & 0.06 \pm 0.05 & 0.05 \pm 0.01 & 0.01 \pm 0.01 &  & 0.18 \pm 0.02 \\
4769.10 \pm 0.31 & [\ion{Fe}{3}] & 4769.53 &  &  &  &  & 0.02 \pm 0.01 &  & 0.09 \pm 0.02 \\
4814.26 \pm 0.10 & [\ion{Fe}{2}] & 4814.53 &  & 0.01 \pm 0.01 &  & 0.05 \pm 0.01 &  &  & 0.35 \pm 0.02 \\
4861.23 \pm 0.05 & \ion{H}{1} & 4861.32 &  & 100.00 & 100.00 & 100.00 & 100.00 & 100.00 & 100.00 \\
4880.73 \pm 0.10 & [\ion{Fe}{3}] & 4881.07 &  & 0.05 \pm 0.01 & 0.06 \pm 0.03 & 0.09 \pm 0.02 & 0.02 \pm 0.01 &  & 0.43 \pm 0.02 \\
4921.86 \pm 0.08 & \ion{He}{1} & 4921.93 &  & 0.90 \pm 0.04 & 0.83 \pm 0.04 & 0.97 \pm 0.04 & 1.16 \pm 0.06 & 1.02 \pm 0.04 & 0.41 \pm 0.08 \\
4930.61 \pm 0.62 & [\ion{Fe}{3}] & 4930.64 &  & 0.05 \pm 0.02 & 0.30 \pm 0.26 & 0.04 \pm 0.01 & 0.05 \pm 0.01 &  & 0.06 \pm 0.03 \\
4958.83 \pm 0.05 & [\ion{O}{3}] & 4958.91 &  & 119.37 \pm 3.59 & 180.32 \pm 5.91 & 107.73 \pm 3.43 & 144.04 \pm 5.35 & 180.12 \pm 6.67 & 44.93 \pm 1.36 \\
 & [\ion{Fe}{3}] & 4985.90 & e &  &  &  &  &  &  \\
4986.16 \pm 0.16 & [\ion{Fe}{3}] & 4987.20 & e & 0.16 \pm 0.02 & 0.15 \pm 0.10 & 0.30 \pm 0.02 & 0.16 \pm 0.01 &  & 0.23 \pm 0.03 \\
5006.75 \pm 0.05 & [\ion{O}{3}] & 5006.84 &  & 331.88 \pm 6.87 & 431.47 \pm 5.02 & 335.06 \pm 8.90 & 462.85 \pm 16.81 & 548.67 \pm 13.37 & 117.06 \pm 1.49 \\
5015.65 \pm 0.09 & \ion{He}{1} & 5015.68 &  & 1.92 \pm 0.08 & 1.72 \pm 0.15 & 2.20 \pm 0.12 & 2.37 \pm 0.15 & 1.67 \pm 0.06 & 1.03 \pm 0.09 \\
5040.89 \pm 0.46 & \ion{Si}{2} & 5041.03 &  & 0.07 \pm 0.02 & 0.23 \pm 0.04 & 0.08 \pm 0.03 & 0.05 \pm 0.02 & 0.10 \pm 0.02 & 0.71 \pm 0.32 \\
5047.68 \pm 0.11 & \ion{He}{1} & 5047.74 &  & 0.15 \pm 0.01 & 0.19 \pm 0.03 & 0.15 \pm 0.02 & 0.18 \pm 0.01 & 0.16 \pm 0.02 & 0.17 \pm 0.01 \\
5055.49 \pm 0.46 & \ion{Si}{2} & 5055.98 &  & 0.05 \pm 0.04 &  & 0.01 \pm 0.01 & 0.02 \pm 0.01 &  & 0.95 \pm 0.05 \\
5146.98 \pm 0.29 & \ion{O}{1} & 5146.61 &  & 0.02 \pm 0.01 & 0.11 \pm 0.03 &  &  &  & 0.08 \pm 0.06 \\
5158.36 \pm 0.12 & [\ion{Fe}{2}] & 5158.81 &  & 0.04 \pm 0.03 & 0.02 \pm 0.03 &  &  &  & 0.51 \pm 0.04 \\
5191.61 \pm 0.39 & [\ion{Ar}{3}] & 5191.82 &  & 0.06 \pm 0.02 &  & 0.07 \pm 0.01 & 0.07 \pm 0.02 & 0.08 \pm 0.02 & 0.08 \pm 0.05 \\
5198.76 \pm 0.19 & [\ion{N}{1}] & 5197.98 & f & 0.18 \pm 0.04 & 0.71 \pm 0.10 & 0.07 \pm 0.02 &  &  & 0.48 \pm 0.21 \\
 & [\ion{N}{1}] & 5200.26 & f &  &  &  &  &  &  \\
5261.31 \pm 0.12 & [\ion{Fe}{2}] & 5261.61 &  &  &  & 0.02 \pm 0.01 &  &  & 0.47 \pm 0.04 \\
5270.45 \pm 0.24 & [\ion{Fe}{3}] & 5270.40 & g & 0.04 \pm 0.01 & 0.04 \pm 0.03 & 0.10 \pm 0.01 & 0.05 \pm 0.01 &  & 0.21 \pm 0.04 \\
 & [\ion{Fe}{2}] & 5273.38 & g &  &  &  &  &  &  \\
5333.35 \pm 0.16 & [\ion{Fe}{2}] & 5333.65 &  & 0.04 \pm 0.02 & 0.13 \pm 0.04 & 0.02 \pm 0.01 &  &  & 0.18 \pm 0.02 \\
5376.11 \pm 0.30 & [\ion{Fe}{2}] & 5376.45 &  & 0.03 \pm 0.01 & 0.08 \pm 0.03 & 0.03 \pm 0.02 & 0.01 \pm 0.01 &  & 0.10 \pm 0.02 \\
5517.64 \pm 0.12 & [\ion{Cl}{3}] & 5517.71 &  & 0.29 \pm 0.02 & 0.24 \pm 0.02 & 0.34 \pm 0.02 & 0.38 \pm 0.03 & 0.28 \pm 0.01 & 0.21 \pm 0.02 \\
5537.76 \pm 0.11 & [\ion{Cl}{3}] & 5537.88 &  & 0.23 \pm 0.01 & 0.20 \pm 0.02 & 0.26 \pm 0.01 & 0.28 \pm 0.02 & 0.19 \pm 0.01 & 0.17 \pm 0.02 \\
5739.53 \pm 0.30 & \ion{Si}{3} & 5739.73 &  &  &  &  & 0.02 \pm 0.01 & 0.06 \pm 0.01 &  \\
5754.65 \pm 0.13 & [\ion{N}{2}] & 5755.08 &  & 0.19 \pm 0.02 & 0.02 \pm 0.02 & 0.22 \pm 0.02 & 0.05 \pm 0.01 &  & 0.40 \pm 0.03 \\
5866.57 \pm 0.80 & \ion{Ni}{2} & 5867.99 &  & 0.37 \pm 0.21 & 0.71 \pm 0.30 & 0.38 \pm 0.22 & 0.29 \pm 0.18 &  & 0.16 \pm 0.11 \\
5875.63 \pm 0.05 & \ion{He}{1} & 5875.62 &  & 10.84 \pm 0.42 & 10.56 \pm 0.32 & 10.92 \pm 0.38 & 11.14 \pm 0.42 & 10.98 \pm 0.43 & 6.97 \pm 0.27 \\
5978.94 \pm 0.25 & \ion{Si}{2} & 5978.93 &  & 0.02 \pm 0.01 & 0.10 \pm 0.02 & 0.03 \pm 0.01 & 0.01 \pm 0.01 & 0.01 \pm 0.01 & 0.61 \pm 0.03 \\
6029.50 \pm 0.34 & \chem{H_2} 7-2 S(7) & 6029.57 &  & 0.03 \pm 0.01 & 0.02 \pm 0.01 &  &  &  & 0.01 \pm 0.01 \\
6037.21 \pm 0.32 & \chem{H_2} 7-2 S(1) & 6037.46 &  & 0.04 \pm 0.01 &  &  &  &  & 0.02 \pm 0.01 \\
6046.19 \pm 0.11 & \ion{O}{1} & 6046.23 &  & 0.02 \pm 0.01 & 0.04 \pm 0.01 & 0.02 \pm 0.01 &  &  & 0.56 \pm 0.03 \\
6083.53 \pm 0.32 & \chem{H_2} 7-2 S(0) & 6083.08 &  & 0.01 \pm 0.01 &  &  &  &  & 0.01 \pm 0.01 \\
6101.49 \pm 0.32 & [\ion{K}{4}] & 6101.79 &  &  &  &  & 0.01 \pm 0.01 & 0.10 \pm 0.01 &  \\
6221.80 \pm 0.32 & \chem{H_2} 4-0 S(5) & 6221.63 &  & 0.04 \pm 0.01 &  &  &  &  & 0.02 \pm 0.01 \\
6248.26 \pm 0.35 & \ion{Fe}{2} & 6248.91 &  & 0.02 \pm 0.01 & 0.09 \pm 0.02 & 0.01 \pm 0.01 & 0.01 \pm 0.01 &  & 0.05 \pm 0.01 \\
6262.12 \pm 0.33 & \chem{H_2} 12-5 S(1) & 6261.73 &  & 0.02 \pm 0.01 & 0.05 \pm 0.01 & 0.01 \pm 0.01 &  &  & 0.01 \pm 0.01 \\
6270.85 \pm 0.39 & \chem{H_2} 4-0 S(3) & 6270.24 &  & 0.08 \pm 0.02 & 0.07 \pm 0.04 & 0.04 \pm 0.02 &  &  & 0.09 \pm 0.03 \\
6300.39 \pm 0.08 & [\ion{O}{1}] & 6300.30 &  & 1.99 \pm 0.08 & 0.84 \pm 0.58 & 2.12 \pm 0.08 &  &  & 3.24 \pm 0.12 \\
6311.98 \pm 0.11 & [\ion{S}{3}] & 6312.06 &  & 1.18 \pm 0.07 & 0.71 \pm 0.05 & 1.50 \pm 0.11 & 1.52 \pm 0.12 & 0.78 \pm 0.05 & 1.34 \pm 0.05 \\
6318.30 \pm 0.30 & \ion{Fe}{2} & 6317.99 &  &  & 0.19 \pm 0.04 &  &  &  & 0.09 \pm 0.03 \\
6347.56 \pm 0.30 & \ion{Si}{2} & 6347.11 &  & 0.02 \pm 0.01 & 0.07 \pm 0.02 & 0.03 \pm 0.01 & 0.02 \pm 0.01 & 0.01 \pm 0.01 & 0.64 \pm 0.10 \\
6363.84 \pm 0.08 & [\ion{O}{1}] & 6363.78 &  & 0.68 \pm 0.03 & 0.37 \pm 0.13 & 0.74 \pm 0.03 &  &  & 1.13 \pm 0.06 \\
6365.15 \pm 1.09 & \ion{Si}{2} & 6371.36 &  & 0.02 \pm 0.01 & 0.04 \pm 0.01 & 0.07 \pm 0.01 & 0.11 \pm 0.10 &  & 0.73 \pm 0.05 \\
6384.24 \pm 0.35 & \ion{Fe}{2} & 6383.73 & h &  &  &  & 0.01 \pm 0.01 &  & 0.02 \pm 0.01 \\
 & \ion{Fe}{2} & 6385.46 & h &  &  &  &  &  &  \\
6433.75 \pm 0.34 & \chem{H_2} 4-0 S(12) & 6434.04 & i & 0.04 \pm 0.01 & 0.04 \pm 0.02 & 0.00 \pm 0.01 &  &  & 0.04 \pm 0.01 \\
 & \chem{H_2} 4-0 S(0) & 6434.99 & i &  &  &  &  &  &  \\
6451.20 \pm 0.54 & \chem{H_2} 8-3 S(3) & 6450.83 &  & 0.02 \pm 0.01 &  &  &  &  &  \\
6456.60 \pm 0.34 & \ion{Fe}{2} & 6455.84 &  &  &  & 0.01 \pm 0.01 &  &  & 0.08 \pm 0.01 \\
6457.89 \pm 0.34 & \chem{H_2} 8-3 S(5) & 6459.07 &  & 0.04 \pm 0.01 &  &  &  &  & 0.08 \pm 0.06 \\
6467.57 \pm 0.31 & \chem{H_2} 8-3 S(2) & 6469.13 &  & 0.05 \pm 0.02 & 0.12 \pm 0.03 & 0.02 \pm 0.01 &  &  & 0.06 \pm 0.03 \\
6485.33 \pm 0.37 & \chem{H_2} 8-3 S(6) & 6486.30 &  & 0.02 \pm 0.01 & 0.04 \pm 0.03 &  &  &  & 0.01 \pm 0.02 \\
 & \ion{Fe}{2} & 6491.25 & j &  &  &  &  &  &  \\
6491.39 \pm 0.47 & \ion{Fe}{2} & 6493.04 & j &  &  &  &  &  & 0.02 \pm 0.01 \\
6501.82 \pm 0.33 & \chem{H_2} 8-3 S(1) & 6502.09 &  & 0.05 \pm 0.01 & 0.06 \pm 0.01 & 0.02 \pm 0.01 &  &  & 0.01 \pm 0.01 \\
6516.45 \pm 0.24 & \ion{Fe}{2} & 6517.03 &  & 0.03 \pm 0.03 &  & 0.01 \pm 0.01 & 0.00 \pm 0.01 &  & 0.10 \pm 0.02 \\
6529.09 \pm 0.12 & \chem{H_2} 8-3 S(7) & 6529.62 &  & 0.05 \pm 0.01 & 0.11 \pm 0.02 & 0.02 \pm 0.01 &  &  & 0.01 \pm 0.01 \\
6548.06 \pm 0.09 & [\ion{N}{2}] & 6548.05 &  & 2.49 \pm 0.13 & 0.65 \pm 0.57 & 2.89 \pm 0.16 & 0.74 \pm 0.11 &  & 5.45 \pm 0.29 \\
6562.73 \pm 0.05 & \ion{H}{1} & 6562.79 &  & 282.11 \pm 10.66 & 282.09 \pm 8.17 & 282.12 \pm 9.36 & 282.12 \pm 10.16 & 282.07 \pm 10.76 & 282.11 \pm 11.51 \\
6583.42 \pm 0.05 & [\ion{N}{2}] & 6583.45 &  & 8.63 \pm 0.26 & 4.37 \pm 0.35 & 10.45 \pm 0.39 & 2.83 \pm 0.11 &  & 13.17 \pm 0.23 \\
6615.52 \pm 0.35 & \chem{H_2} 4-0 Q(3) & 6615.09 &  & 0.01 \pm 0.01 & 0.01 \pm 0.01 & 0.01 \pm 0.01 &  &  & 0.01 \pm 0.01 \\
6628.10 \pm 0.34 & \chem{H_2} 5-1 S(5) & 6629.04 & k & 0.06 \pm 0.02 & 0.04 \pm 0.05 & 0.02 \pm 0.02 &  &  & 0.12 \pm 0.06 \\
\mathit{6630.22} & Sky OH & 6634.00 & k &  &  &  &  &  &  \\
6637.16 \pm 0.19 & \chem{H_2} 5-1 S(7) & 6636.69 &  & 0.06 \pm 0.01 & 0.03 \pm 0.01 & 0.01 \pm 0.01 &  &  & 0.16 \pm 0.07 \\
6645.67 \pm 0.34 & \chem{H_2} 5-1 S(4) & 6645.58 &  & 0.04 \pm 0.01 & 0.03 \pm 0.01 & 0.01 \pm 0.01 &  &  & 0.02 \pm 0.01 \\
6655.86 \pm 0.35 & \chem{H_2} 8-3 Q(1) & 6656.12 &  & 0.03 \pm 0.01 & 0.02 \pm 0.02 & 0.02 \pm 0.01 &  &  & 0.01 \pm 0.01 \\
6667.06 \pm 0.24 & [\ion{Ni}{2}] & 6666.80 &  & 0.03 \pm 0.01 &  & 0.01 \pm 0.01 &  &  & 0.17 \pm 0.03 \\
6678.08 \pm 0.13 & \ion{He}{1} & 6678.15 &  & 2.50 \pm 0.13 & 2.53 \pm 0.08 & 2.55 \pm 0.22 & 2.64 \pm 0.24 & 2.58 \pm 0.20 & 1.56 \pm 0.05 \\
6699.44 \pm 0.35 & \chem{H_2} 5-1 S(9) & 6699.11 &  & 0.03 \pm 0.01 & 0.02 \pm 0.01 & 0.00 \pm 0.01 &  &  & 0.01 \pm 0.01 \\
6716.43 \pm 0.04 & [\ion{S}{2}] & 6716.44 &  & 14.40 \pm 0.46 & 5.98 \pm 1.02 & 17.68 \pm 0.54 & 2.56 \pm 0.15 &  & 13.50 \pm 0.49 \\
6730.81 \pm 0.05 & [\ion{S}{2}] & 6730.82 &  & 11.03 \pm 0.32 & 7.11 \pm 0.51 & 13.23 \pm 0.51 & 2.25 \pm 0.10 &  & 11.15 \pm 0.19 \\
6776.73 \pm 0.36 & \chem{H_2} 5-1 S(1) & 6776.75 &  & 0.06 \pm 0.01 & 0.02 \pm 0.01 & 0.02 \pm 0.01 &  &  & 0.03 \pm 0.01 \\
6794.61 \pm 0.36 & [\ion{K}{4}] & 6795.10 &  &  &  &  & 0.00 \pm 0.01 & 0.02 \pm 0.01 &  \\
6847.54 \pm 0.36 & \chem{H_2} 5-1 S(0) & 6847.81 &  & 0.04 \pm 0.01 & 0.02 \pm 0.01 & 0.01 \pm 0.01 &  &  & 0.02 \pm 0.01 \\
6987.20 \pm 0.35 & \chem{H_2} 5-1 Q(1) & 6987.18 &  & 0.06 \pm 0.01 & 0.02 \pm 0.02 & 0.02 \pm 0.01 &  &  & 0.03 \pm 0.01 \\
7002.39 \pm 0.12 & \ion{O}{1} & 7001.92 &  & 0.03 \pm 0.01 & 0.05 \pm 0.01 & 0.01 \pm 0.01 &  &  & 0.38 \pm 0.02 \\
7009.74 \pm 0.37 & \chem{H_2} 5-1 Q(2) & 7009.46 & l & 0.05 \pm 0.01 & 0.04 \pm 0.02 & 0.02 \pm 0.01 &  &  & 0.04 \pm 0.01 \\
 & \chem{H_2} 9-4 S(3) & 7011.24 & l &  &  &  &  &  &  \\
7015.73 \pm 0.38 & \chem{H_2} 9-4 S(4) & 7015.36 &  & 0.02 \pm 0.01 &  & 0.02 \pm 0.01 &  &  & 0.03 \pm 0.01 \\
7026.54 \pm 0.39 & \chem{H_2} 9-4 S(2) & 7025.14 &  & 0.03 \pm 0.01 & 0.05 \pm 0.02 & 0.01 \pm 0.01 &  &  & 0.05 \pm 0.03 \\
7038.26 \pm 0.36 & \chem{H_2} 9-4 S(5) & 7038.03 &  & 0.04 \pm 0.01 & 0.04 \pm 0.01 & 0.02 \pm 0.01 &  &  & 0.03 \pm 0.01 \\
7043.27 \pm 0.36 & \chem{H_2} 5-1 Q(3) & 7042.98 &  & 0.04 \pm 0.01 & 0.03 \pm 0.02 & 0.01 \pm 0.01 &  &  & 0.02 \pm 0.01 \\
7056.84 \pm 0.42 & \chem{H_2} 9-4 S(1) & 7056.69 &  & 0.08 \pm 0.02 &  & 0.04 \pm 0.02 &  &  & 0.06 \pm 0.02 \\
7065.19 \pm 0.09 & \ion{He}{1} & 7065.28 &  & 1.98 \pm 0.08 & 1.86 \pm 0.06 & 1.87 \pm 0.07 & 1.87 \pm 0.08 & 1.94 \pm 0.09 & 1.43 \pm 0.06 \\
7078.86 \pm 0.40 & \chem{H_2} 9-4 S(6) & 7080.02 &  & 0.02 \pm 0.01 & 0.01 \pm 0.01 & 0.01 \pm 0.01 &  &  & 0.02 \pm 0.01 \\
7092.31 \pm 0.36 & \chem{H_2} 6-2 S(5) & 7091.93 &  & 0.09 \pm 0.01 & 0.03 \pm 0.01 & 0.02 \pm 0.01 &  &  & 0.05 \pm 0.01 \\
7105.35 \pm 0.36 & \chem{H_2} 6-2 S(4) & 7105.43 & m & 0.08 \pm 0.01 & 0.02 \pm 0.01 & 0.02 \pm 0.01 &  &  & 0.02 \pm 0.01 \\
 & \chem{H_2} 9-4 S(0) & 7105.68 & m &  &  &  &  &  &  \\
7111.31 \pm 0.36 & \chem{H_2} 6-2 S(7) & 7111.45 &  & 0.06 \pm 0.01 &  & 0.02 \pm 0.01 &  &  & 0.03 \pm 0.01 \\
7135.70 \pm 0.08 & [\ion{Ar}{3}] & 7135.78 &  & 6.94 \pm 0.20 & 5.24 \pm 0.19 & 8.17 \pm 0.25 & 8.61 \pm 0.29 & 6.73 \pm 0.20 & 4.03 \pm 0.12 \\
 & \chem{H_2} 5-1 O(2) & 7143.90 & n &  &  &  &  &  &  \\
7143.76 \pm 0.44 & \chem{H_2} 5-1 Q(5) & 7144.42 & n & 0.06 \pm 0.02 &  & 0.03 \pm 0.01 &  &  & 0.03 \pm 0.01 \\
 & \chem{H_2} 6-2 S(8) & 7144.85 & n &  &  &  &  &  &  \\
7154.79 \pm 0.14 & [\ion{Fe}{2}] & 7155.14 &  & 0.01 \pm 0.01 &  &  &  &  & 0.22 \pm 0.02 \\
7160.76 \pm 0.38 & \ion{He}{1} & 7160.13 &  & 0.01 \pm 0.01 & 0.04 \pm 0.03 & 0.01 \pm 0.01 & 0.02 \pm 0.01 & 0.01 \pm 0.01 &  \\
7170.26 \pm 0.37 & [\ion{Ar}{4}] & 7170.50 &  &  &  &  & 0.01 \pm 0.01 & 0.07 \pm 0.01 &  \\
7178.33 \pm 0.36 & \chem{H_2} 6-2 S(2) & 7178.58 &  & 0.07 \pm 0.01 & 0.01 \pm 0.01 & 0.02 \pm 0.01 &  &  & 0.03 \pm 0.01 \\
7184.60 \pm 0.13 & \chem{H_2} 11-5 O(3) & 7186.19 &  & 0.06 \pm 0.01 & 0.08 \pm 0.01 & 0.03 \pm 0.01 &  &  & 0.10 \pm 0.01 \\
7194.08 \pm 0.37 & \chem{H_2} 6-2 S(9) & 7194.43 &  & 0.03 \pm 0.01 &  & 0.01 \pm 0.01 &  &  &  \\
7198.70 \pm 0.37 & \chem{H_2} 13-6 S(2) & 7199.27 &  & 0.02 \pm 0.01 & 0.04 \pm 0.01 & 0.01 \pm 0.01 &  &  & 0.04 \pm 0.01 \\
7222.62 \pm 0.43 & \chem{H_2} 9-4 Q(1) & 7221.48 &  & 0.03 \pm 0.01 &  & 0.01 \pm 0.01 &  &  & 0.04 \pm 0.02 \\
 & [\ion{Ar}{4}] & 7237.40 & o &  &  &  &  &  &  \\
7237.76 \pm 0.48 & \chem{H_2} 6-2 S(1) & 7238.21 & o & 0.07 \pm 0.02 &  &  &  &  & 0.02 \pm 0.02 \\
7254.33 \pm 0.13 & \ion{O}{1} & 7254.15 & p & 0.05 \pm 0.02 & 0.08 \pm 0.01 & 0.02 \pm 0.01 &  &  & 0.73 \pm 0.03 \\
 & \ion{O}{1} & 7254.45 & p &  &  &  &  &  &  \\
 & \ion{O}{1} & 7254.53 & p &  &  &  &  &  &  \\
 & [\ion{Cl}{4}] & 7261.40 & q &  &  &  &  &  &  \\
7263.12 \pm 0.39 & [\ion{Ar}{4}] & 7262.70 & q &  &  &  & 0.03 \pm 0.02 & 0.07 \pm 0.01 &  \\
7281.25 \pm 0.14 & \ion{He}{1} & 7281.35 &  & 0.68 \pm 0.03 & 0.79 \pm 0.04 & 0.62 \pm 0.03 & 0.69 \pm 0.04 & 0.79 \pm 0.04 & 0.47 \pm 0.03 \\
7289.14 \pm 0.40 & \chem{H_2} 5-1 Q(7) & 7293.34 &  & 0.09 \pm 0.05 & 0.05 \pm 0.04 & 0.09 \pm 0.05 & 0.04 \pm 0.02 &  & 0.14 \pm 0.05 \\
7306.22 \pm 0.31 & \chem{H_2} 13-6 Q(2) & 7304.73 &  & 0.08 \pm 0.02 & 0.23 \pm 0.08 & 0.06 \pm 0.02 &  &  & 0.05 \pm 0.02 \\
 & [\ion{O}{2}] & 7318.39 & r &  &  &  &  &  &  \\
7319.84 \pm 0.09 & [\ion{O}{2}] & 7319.99 & r & 2.70 \pm 0.12 & 0.39 \pm 0.21 & 3.30 \pm 0.14 & 1.14 \pm 0.07 &  & 9.44 \pm 0.44 \\
 & [\ion{O}{2}] & 7329.66 & s &  &  &  &  &  &  \\
7330.19 \pm 0.15 & [\ion{O}{2}] & 7330.73 & s & 1.62 \pm 0.12 &  & 2.11 \pm 0.22 & 0.79 \pm 0.12 &  & 5.71 \pm 0.22 \\
 & [\ion{Ar}{4}] & 7331.40 & s &  &  &  &  &  &  \\
7377.57 \pm 0.13 & [\ion{Ni}{2}] & 7377.83 &  & 0.02 \pm 0.01 &  & 0.00 \pm 0.01 &  &  & 0.29 \pm 0.01 \\
7411.37 \pm 0.13 & [\ion{Ni}{2}] & 7411.61 &  &  &  &  &  &  & 0.12 \pm 0.01 \\
7423.56 \pm 0.43 & \chem{H_2} 4-0 O(7) & 7426.27 &  & 0.02 \pm 0.01 & 0.04 \pm 0.01 & 0.01 \pm 0.01 &  &  & 0.07 \pm 0.02 \\
7442.42 \pm 0.13 & \ion{N}{1} & 7442.30 &  & 0.01 \pm 0.01 & 0.07 \pm 0.01 & 0.01 \pm 0.01 &  &  & 0.22 \pm 0.02 \\
7452.18 \pm 0.38 & [\ion{Fe}{2}] & 7452.54 &  & 0.00 \pm 0.01 & 0.01 \pm 0.01 &  &  &  & 0.07 \pm 0.01 \\
7462.59 \pm 0.38 & \chem{H_2} 6-2 Q(1) & 7462.84 &  & 0.08 \pm 0.01 & 0.04 \pm 0.01 & 0.02 \pm 0.01 &  &  & 0.02 \pm 0.01 \\
7468.20 \pm 0.14 & \ion{N}{1} & 7468.31 & t & 0.05 \pm 0.02 & 0.14 \pm 0.07 & 0.02 \pm 0.02 &  &  & 0.34 \pm 0.02 \\
\mathit{7469.44} & Sky OH & 7473.70 & t &  &  &  &  &  &  \\
7488.35 \pm 0.38 & \chem{H_2} 6-2 Q(2) & 7488.21 &  & 0.06 \pm 0.01 & 0.03 \pm 0.02 & 0.01 \pm 0.01 &  &  & 0.01 \pm 0.01 \\
7499.83 \pm 0.38 & \ion{He}{1} & 7499.85 &  & 0.04 \pm 0.01 & 0.04 \pm 0.01 & 0.04 \pm 0.01 & 0.04 \pm 0.01 & 0.04 \pm 0.01 & 0.02 \pm 0.01 \\
7513.16 \pm 0.14 & \ion{Fe}{2} & 7513.18 &  & 0.03 \pm 0.01 & 0.13 \pm 0.03 & 0.02 \pm 0.01 & 0.02 \pm 0.01 &  & 0.14 \pm 0.01 \\
7526.35 \pm 0.13 & \chem{H_2} 6-2 Q(3) & 7526.43 & u & 0.07 \pm 0.01 & 0.12 \pm 0.02 & 0.03 \pm 0.01 &  &  & 0.03 \pm 0.01 \\
 & [\ion{Cl}{4}] & 7530.80 & u &  &  &  &  &  &  \\
7530.23 \pm 0.15 & [\ion{Cl}{4}] & 7530.80 &  &  &  &  & 0.03 \pm 0.01 & 0.13 \pm 0.01 &  \\
7628.76 \pm 0.16 & \chem{H_2} 7-3 S(5) & 7628.56 &  & 0.23 \pm 0.02 & 0.24 \pm 0.07 & 0.10 \pm 0.02 &  &  & 0.26 \pm 0.07 \\
7637.61 \pm 0.18 & \chem{H_2} 7-3 S(4) & 7636.91 & v & 0.21 \pm 0.02 & 0.20 \pm 0.07 & 0.09 \pm 0.02 &  &  & 0.32 \pm 0.07 \\
 & \chem{H_2} 7-3 S(6) & 7638.24 & v &  &  &  &  &  &  \\
7663.19 \pm 0.16 & \chem{H_2} 7-3 S(3) & 7663.09 &  & 0.14 \pm 0.02 & 0.11 \pm 0.02 & 0.05 \pm 0.01 &  &  & 0.14 \pm 0.02 \\
7706.79 \pm 0.31 & \chem{H_2} 7-3 S(2) & 7706.96 & w & 0.16 \pm 0.04 & 0.17 \pm 0.13 & 0.08 \pm 0.04 &  &  &  \\
 & \chem{H_2} 10-5 S(3) & 7707.12 & w &  &  &  &  &  &  \\
\mathit{7707.60} & Sky OH & 7712.00 & w &  &  &  &  &  &  \\
 & \chem{H_2} 6-2 Q(6) & 7720.70 & x &  &  &  &  &  &  \\
7720.99 \pm 0.55 & \chem{H_2} 12-6 S(2) & 7721.51 & x & 0.02 \pm 0.02 &  &  &  &  & 0.01 \pm 0.01 \\
\mathit{7721.60} & Sky OH & 7726.00 & x &  &  &  &  &  &  \\
7724.22 \pm 0.39 & \chem{H_2} 10-5 S(4) & 7724.23 &  & 0.01 \pm 0.01 &  &  &  &  & 0.01 \pm 0.01 \\
7739.74 \pm 0.41 & \chem{H_2} 10-5 S(1) & 7740.69 &  & 0.10 \pm 0.04 & 0.37 \pm 0.15 & 0.09 \pm 0.04 &  &  & 0.03 \pm 0.01 \\
7751.02 \pm 0.09 & [\ion{Ar}{3}] & 7751.10 &  & 1.61 \pm 0.07 & 1.37 \pm 0.23 & 1.68 \pm 0.08 & 1.71 \pm 0.07 & 1.37 \pm 0.08 & 0.96 \pm 0.04 \\
7768.64 \pm 0.16 & \chem{H_2} 7-3 S(1) & 7768.49 &  & 0.11 \pm 0.01 & 0.07 \pm 0.04 & 0.02 \pm 0.01 &  &  & 0.02 \pm 0.01 \\
7781.97 \pm 0.41 & \chem{H_2} 3-0 S(7) & 7782.22 &  & 0.09 \pm 0.02 & 0.10 \pm 0.04 & 0.03 \pm 0.01 &  &  & 0.05 \pm 0.01 \\
\mathit{7789.56} & Sky OH & 7794.00 & y &  &  &  &  &  &  \\
7790.34 \pm 0.42 & \chem{H_2} 10-5 S(0) & 7790.25 & y & 0.10 \pm 0.03 &  &  &  &  & 0.06 \pm 0.01 \\
 & \chem{H_2} 3-0 S(9) & 7790.99 & y &  &  &  &  &  &  \\
7803.31 \pm 0.24 & \chem{H_2} 4-0 O(9) & 7806.24 &  & 0.17 \pm 0.12 &  &  &  &  & 0.05 \pm 0.01 \\
7817.34 \pm 0.27 & \ion{He}{1} & 7816.13 & z & 0.16 \pm 0.03 & 0.55 \pm 0.08 & 0.16 \pm 0.02 & 0.13 \pm 0.03 & 0.20 \pm 0.02 & 0.05 \pm 0.01 \\
\mathit{7817.54} & Sky OH & 7822.00 & z &  &  &  &  &  &  \\
7837.56 \pm 0.15 & \chem{H_2} 3-0 S(5) & 7837.75 &  & 0.20 \pm 0.02 & 0.17 \pm 0.03 & 0.05 \pm 0.01 &  &  & 0.11 \pm 0.01 \\
7847.60 \pm 0.40 & \chem{H_2} 7-3 S(0) & 7847.68 & aa & 0.16 \pm 0.06 &  & 0.08 \pm 0.06 &  &  & 0.04 \pm 0.02 \\
\mathit{7848.52} & Sky OH & 7853.00 & aa &  &  &  &  &  &  \\
7862.55 \pm 0.59 & \chem{H_2} 3-0 S(11) & 7862.67 & ab & 0.05 \pm 0.03 &  & 0.03 \pm 0.03 &  &  & 0.02 \pm 0.01 \\
\mathit{7863.52} & Sky OH & 7868.00 & ab &  &  &  &  &  &  \\
7890.10 \pm 0.15 & \chem{H_2} 3-0 S(4) & 7890.29 &  & 0.10 \pm 0.01 & 0.05 \pm 0.02 & 0.03 \pm 0.01 &  &  & 0.03 \pm 0.01 \\
\mathit{7916.48} & Sky OH & 7921.00 & ac &  &  &  &  &  &  \\
7916.67 \pm 0.46 & \chem{H_2} 11-5 Q(9) & 7917.04 & ac & 0.05 \pm 0.02 &  &  &  &  & 0.03 \pm 0.01 \\
7959.48 \pm 0.21 & \chem{H_2} 3-0 S(3) & 7959.75 & ad & 0.19 \pm 0.06 &  &  &  &  & 0.10 \pm 0.02 \\
\mathit{7960.11} & Sky OH & 7964.65 & ad &  &  &  &  &  &  \\
7999.79 \pm 0.15 & [\ion{Fe}{2}] & 7997.03 & ae & 0.02 \pm 0.03 & 0.01 \pm 0.01 & 0.01 \pm 0.01 &  &  & 0.21 \pm 0.01 \\
 & [\ion{Fe}{2}] & 7999.47 & ae &  &  &  &  &  &  \\
 & [\ion{Cr}{2}] & 8000.08 & ae &  &  &  &  &  &  \\
8008.86 \pm 0.28 & \chem{H_2} 7-3 Q(1) & 8008.96 & af & 0.12 \pm 0.03 &  & 0.04 \pm 0.03 &  &  & 0.02 \pm 0.01 \\
\mathit{8009.53} & Sky OH & 8014.10 & af &  &  &  &  &  &  \\
8037.27 \pm 0.43 & \chem{H_2} 7-3 Q(2) & 8038.44 &  & 0.07 \pm 0.01 & 0.02 \pm 0.02 &  &  &  & 0.02 \pm 0.01 \\
8045.56 \pm 0.15 & [\ion{Cl}{4}] & 8045.62 &  &  &  &  & 0.03 \pm 0.01 & 0.25 \pm 0.02 &  \\
8069.13 \pm 0.41 & \chem{H_2} 10-5 O(2) & 8068.77 &  & 0.02 \pm 0.01 &  & 0.00 \pm 0.01 &  &  & 0.01 \pm 0.01 \\
8083.08 \pm 0.41 & \chem{H_2} 7-3 Q(3) & 8082.91 &  & 0.06 \pm 0.01 & 0.03 \pm 0.01 & 0.02 \pm 0.01 &  &  & 0.04 \pm 0.01 \\
8126.42 \pm 0.32 & \ion{Ca}{1}] & 8125.31 &  & 0.02 \pm 0.01 & 0.03 \pm 0.01 & 0.01 \pm 0.01 &  &  & 0.15 \pm 0.08 \\
8150.81 \pm 0.14 & \chem{H_2} 3-0 S(1) & 8150.65 &  & 0.22 \pm 0.01 & 0.03 \pm 0.02 & 0.04 \pm 0.01 &  &  & 0.07 \pm 0.01 \\
8187.57 \pm 0.14 & \ion{N}{1} & 8188.01 &  & 0.02 \pm 0.01 & 0.07 \pm 0.01 & 0.02 \pm 0.01 &  &  & 0.22 \pm 0.01 \\
8191.20 \pm 0.42 & \chem{H_2} 7-3 O(2) & 8192.31 &  & 0.07 \pm 0.01 & 0.03 \pm 0.02 & 0.02 \pm 0.02 &  &  & 0.04 \pm 0.01 \\
8200.13 \pm 0.43 & \ion{N}{1} & 8200.36 &  & 0.01 \pm 0.01 & 0.04 \pm 0.01 &  &  &  & 0.10 \pm 0.01 \\
8206.92 \pm 0.42 & \chem{H_2} 10-5 S(9) & 8206.90 &  & 0.03 \pm 0.01 & 0.01 \pm 0.01 & 0.00 \pm 0.01 &  &  & 0.03 \pm 0.01 \\
8210.11 \pm 0.17 & \ion{N}{1} & 8210.72 &  & 0.05 \pm 0.01 & 0.10 \pm 0.01 & 0.03 \pm 0.01 &  &  & 0.19 \pm 0.02 \\
8216.55 \pm 0.16 & \ion{N}{1} & 8216.34 &  & 0.06 \pm 0.02 & 0.15 \pm 0.01 & 0.01 \pm 0.01 &  &  & 0.37 \pm 0.03 \\
 & \ion{O}{1} & 8221.82 & ag &  &  &  &  &  &  \\
8223.36 \pm 0.15 & \ion{N}{1} & 8223.14 & ag & 0.02 \pm 0.01 & 0.12 \pm 0.03 &  &  &  & 0.49 \pm 0.03 \\
8229.21 \pm 0.16 & \ion{O}{1} & 8230.00 &  & 0.02 \pm 0.01 & 0.03 \pm 0.01 & 0.01 \pm 0.01 &  &  & 0.30 \pm 0.02 \\
8236.09 \pm 0.61 & \ion{He}{2} & 8236.78 &  &  &  &  &  & 0.01 \pm 0.01 &  \\
8242.58 \pm 0.16 & \ion{N}{1} & 8242.39 &  & 0.03 \pm 0.01 & 0.10 \pm 0.01 & 0.01 \pm 0.01 &  &  & 0.25 \pm 0.02 \\
 & \chem{H_2} 8-4 S(4) & 8266.30 & ah &  &  &  &  &  &  \\
8266.78 \pm 0.18 & \chem{H_2} 8-4 S(5) & 8266.55 & ah & 0.17 \pm 0.02 & 0.06 \pm 0.02 & 0.04 \pm 0.01 &  &  & 0.06 \pm 0.01 \\
8272.18 \pm 0.14 & \chem{H_2} 3-0 S(0) & 8272.65 &  & 0.17 \pm 0.01 & 0.04 \pm 0.03 & 0.04 \pm 0.01 &  &  & 0.03 \pm 0.02 \\
8283.33 \pm 0.22 & \chem{H_2} 4-1 S(7) & 8283.51 & ai & 0.13 \pm 0.02 & 0.12 \pm 0.06 &  &  &  &  \\
\mathit{8283.88} & Sky OH & 8288.60 & ai &  &  &  &  &  &  \\
8287.16 \pm 0.16 & \chem{H_2} 8-4 S(3) & 8287.27 &  & 0.12 \pm 0.03 & 0.14 \pm 0.03 & 0.08 \pm 0.01 &  &  & 0.44 \pm 0.03 \\
 & \ion{H}{1} & 8298.83 & aj &  &  &  &  &  &  \\
8298.99 \pm 0.16 & \chem{H_2} 4-1 S(6) & 8300.03 & aj & 0.22 \pm 0.03 & 0.24 \pm 0.05 & 0.17 \pm 0.02 &  &  & 0.20 \pm 0.02 \\
8304.89 \pm 0.33 & \chem{H_2} 4-1 S(9) & 8304.07 &  & 0.17 \pm 0.05 &  &  &  &  & 0.05 \pm 0.02 \\
8306.17 \pm 0.15 & \ion{H}{1} & 8306.11 &  & 0.13 \pm 0.02 & 0.16 \pm 0.02 & 0.14 \pm 0.01 & 0.15 \pm 0.01 & 0.14 \pm 0.01 & 0.16 \pm 0.07 \\
8314.22 \pm 0.15 & \ion{H}{1} & 8314.26 &  & 0.14 \pm 0.01 & 0.12 \pm 0.01 & 0.14 \pm 0.01 & 0.14 \pm 0.01 & 0.13 \pm 0.01 & 0.21 \pm 0.02 \\
8323.34 \pm 0.18 & \ion{H}{1} & 8323.42 &  & 0.14 \pm 0.02 & 0.13 \pm 0.02 & 0.13 \pm 0.02 & 0.13 \pm 0.02 & 0.12 \pm 0.02 & 0.20 \pm 0.02 \\
8328.99 \pm 0.43 & \chem{H_2} 8-4 S(2) & 8329.17 &  & 0.07 \pm 0.01 & 0.08 \pm 0.03 & 0.01 \pm 0.01 &  &  &  \\
 & \ion{H}{1} & 8333.78 & ak &  &  &  &  &  &  \\
8333.85 \pm 0.17 & \chem{H_2} 4-1 S(5) & 8334.66 & ak & 0.52 \pm 0.07 & 0.87 \pm 0.27 & 0.36 \pm 0.07 &  &  & 0.36 \pm 0.03 \\
8344.97 \pm 0.14 & \ion{H}{1} & 8345.55 &  & 0.17 \pm 0.02 & 0.10 \pm 0.06 & 0.19 \pm 0.01 & 0.12 \pm 0.01 & 0.09 \pm 0.01 & 0.23 \pm 0.02 \\
8359.46 \pm 0.22 & \ion{H}{1} & 8359.00 & al & 0.19 \pm 0.02 & 0.18 \pm 0.02 & 0.19 \pm 0.02 & 0.16 \pm 0.03 & 0.16 \pm 0.03 & 0.26 \pm 0.02 \\
 & \ion{He}{1} & 8361.73 & al &  &  &  &  &  &  \\
8374.27 \pm 0.16 & \ion{H}{1} & 8374.48 &  & 0.19 \pm 0.02 & 0.08 \pm 0.02 & 0.20 \pm 0.02 & 0.19 \pm 0.02 & 0.16 \pm 0.01 & 0.28 \pm 0.02 \\
8387.03 \pm 0.42 & \chem{H_2} 4-1 S(4) & 8387.65 &  & 0.08 \pm 0.01 & 0.03 \pm 0.02 &  &  &  & 0.02 \pm 0.01 \\
8399.92 \pm 0.47 & \chem{H_2} 8-4 S(8) & 8400.71 &  & 0.02 \pm 0.01 &  & 0.01 \pm 0.01 &  &  & 0.01 \pm 0.01 \\
\mathit{8410.40} & Sky OH & 8415.20 & am &  &  &  &  &  &  \\
8413.09 \pm 0.18 & \ion{H}{1} & 8413.32 & am & 0.21 \pm 0.02 & 0.68 \pm 0.53 & 0.21 \pm 0.02 & 0.25 \pm 0.03 & 0.23 \pm 0.03 & 0.30 \pm 0.02 \\
8437.88 \pm 0.15 & \ion{H}{1} & 8437.96 &  & 0.34 \pm 0.02 & 0.26 \pm 0.02 & 0.34 \pm 0.02 & 0.32 \pm 0.02 & 0.31 \pm 0.02 & 0.49 \pm 0.05 \\
8446.61 \pm 0.09 & \ion{O}{1} & 8446.48 &  & 0.35 \pm 0.02 & 0.59 \pm 0.02 & 0.19 \pm 0.01 &  &  & 6.86 \pm 0.26 \\
8459.24 \pm 0.21 & \chem{H_2} 4-1 S(3) & 8459.29 & an & 0.47 \pm 0.07 & 0.29 \pm 0.19 & 0.21 \pm 0.06 &  &  & 0.21 \pm 0.03 \\
\mathit{8460.37} & Sky OH & 8465.20 & an &  &  &  &  &  &  \\
8467.16 \pm 0.16 & \ion{H}{1} & 8467.26 &  & 0.37 \pm 0.03 & 0.24 \pm 0.06 & 0.38 \pm 0.02 & 0.36 \pm 0.03 & 0.34 \pm 0.02 & 0.41 \pm 0.02 \\
8475.33 \pm 0.18 & \chem{H_2} 8-4 S(0) & 8475.17 &  & 0.12 \pm 0.02 & 0.02 \pm 0.01 & 0.04 \pm 0.02 &  &  & 0.03 \pm 0.01 \\
8497.75 \pm 0.18 & \chem{H_2} 3-0 Q(1) & 8497.45 &  & 0.10 \pm 0.01 &  &  &  &  & 0.12 \pm 0.02 \\
8502.30 \pm 0.17 & \ion{H}{1} & 8502.49 &  & 0.37 \pm 0.02 & 0.25 \pm 0.02 & 0.40 \pm 0.02 & 0.47 \pm 0.04 & 0.45 \pm 0.04 & 0.43 \pm 0.02 \\
8518.10 \pm 0.45 & \ion{He}{1} & 8518.04 &  & 0.01 \pm 0.01 & 0.04 \pm 0.01 & 0.01 \pm 0.01 & 0.02 \pm 0.01 & 0.02 \pm 0.01 & 0.02 \pm 0.01 \\
8522.55 \pm 0.15 & \chem{H_2} 3-0 Q(2) & 8522.57 &  & 0.11 \pm 0.01 & 0.02 \pm 0.01 & 0.02 \pm 0.01 &  &  & 0.04 \pm 0.01 \\
8528.90 \pm 0.44 & \ion{He}{1} & 8528.95 &  & 0.04 \pm 0.01 & 0.04 \pm 0.01 & 0.03 \pm 0.01 & 0.02 \pm 0.01 & 0.02 \pm 0.01 & 0.02 \pm 0.01 \\
\mathit{8543.83} & Sky OH & 8548.70 & ao &  &  &  &  &  &  \\
8545.27 \pm 0.15 & \ion{H}{1} & 8545.38 & ao & 0.54 \pm 0.03 & 0.71 \pm 0.06 & 0.56 \pm 0.03 & 0.59 \pm 0.03 & 0.58 \pm 0.04 & 0.53 \pm 0.03 \\
8550.07 \pm 0.17 & \chem{H_2} 4-1 S(2) & 8549.88 &  & 0.21 \pm 0.02 & 0.08 \pm 0.02 & 0.03 \pm 0.02 &  &  & 0.04 \pm 0.01 \\
8560.25 \pm 0.15 & \chem{H_2} 4-1 S(13) & 8560.02 & ap & 0.12 \pm 0.01 & 0.03 \pm 0.01 & 0.02 \pm 0.01 &  &  & 0.05 \pm 0.01 \\
 & \chem{H_2} 3-0 Q(3) & 8560.35 & ap &  &  &  &  &  &  \\
8578.68 \pm 0.15 & [\ion{Cl}{2}] & 8578.69 &  & 0.09 \pm 0.01 & 0.06 \pm 0.01 & 0.09 \pm 0.01 & 0.01 \pm 0.01 &  & 0.16 \pm 0.01 \\
8582.88 \pm 0.44 & \ion{He}{1} & 8582.61 &  & 0.02 \pm 0.01 & 0.03 \pm 0.02 & 0.02 \pm 0.01 & 0.04 \pm 0.01 & 0.05 \pm 0.01 & 0.01 \pm 0.01 \\
8598.32 \pm 0.23 & \ion{H}{1} & 8598.39 &  & 0.54 \pm 0.06 & 0.49 \pm 0.02 & 0.56 \pm 0.08 & 0.56 \pm 0.08 & 0.53 \pm 0.07 & 0.59 \pm 0.03 \\
8610.90 \pm 0.44 & \chem{H_2} 11-6 S(3) & 8610.91 &  & 0.08 \pm 0.01 & 0.02 \pm 0.01 & 0.01 \pm 0.01 &  &  &  \\
8616.83 \pm 0.23 & [\ion{Fe}{2}] & 8616.95 &  & 0.03 \pm 0.01 & 0.03 \pm 0.01 & 0.02 \pm 0.01 & 0.00 \pm 0.01 &  & 0.11 \pm 0.02 \\
8620.53 \pm 0.52 & \chem{H_2} 11-6 S(1) & 8620.81 &  & 0.05 \pm 0.02 &  &  &  &  & 0.00 \pm 0.01 \\
8630.26 \pm 0.45 & \ion{N}{1} & 8629.24 &  & 0.01 \pm 0.01 & 0.06 \pm 0.01 & 0.01 \pm 0.01 &  &  & 0.05 \pm 0.01 \\
\mathit{8646.35} & Sky \chem{O_2} & 8651.28 & aq &  &  &  &  &  &  \\
8649.86 \pm 0.22 & \chem{H_2} 8-4 Q(1) & 8650.16 & aq & 0.18 \pm 0.03 & 0.07 \pm 0.03 & 0.08 \pm 0.02 &  &  & 0.05 \pm 0.01 \\
8659.47 \pm 0.17 & \chem{H_2} 4-1 S(1) & 8659.72 &  & 0.38 \pm 0.03 & 0.28 \pm 0.02 & 0.16 \pm 0.02 &  &  & 0.04 \pm 0.02 \\
8664.97 \pm 0.15 & \ion{H}{1} & 8665.02 &  & 0.79 \pm 0.04 & 0.78 \pm 0.04 & 0.90 \pm 0.04 & 0.86 \pm 0.04 & 0.85 \pm 0.04 & 0.75 \pm 0.03 \\
\mathit{8671.15} & Sky \chem{O_2} & 8676.10 & ar &  &  &  &  &  &  \\
8673.85 \pm 0.72 & \chem{H_2} 3-0 Q(5) & 8674.35 & ar & 0.05 \pm 0.03 &  &  &  &  & 0.03 \pm 0.02 \\
8680.39 \pm 0.61 & \ion{N}{1} & 8680.28 &  & 0.01 \pm 0.01 &  &  &  &  & 0.04 \pm 0.02 \\
 & \ion{N}{1} & 8683.40 & as &  &  &  &  &  &  \\
8684.39 \pm 0.44 & \chem{H_2} 8-4 Q(2) & 8685.29 & as & 0.10 \pm 0.01 & 0.08 \pm 0.06 & 0.01 \pm 0.01 &  &  & 0.08 \pm 0.03 \\
 & \ion{N}{1} & 8686.15 & as &  &  &  &  &  &  \\
 & \chem{H_2} 7-3 Q(9) & 8691.59 & at &  &  &  &  &  &  \\
8694.21 \pm 0.28 & \chem{H_2} 14-7 Q(2) & 8692.04 & at & 0.05 \pm 0.01 & 0.04 \pm 0.02 & 0.02 \pm 0.01 &  &  & 0.16 \pm 0.13 \\
8703.05 \pm 0.19 & \ion{N}{1} & 8703.25 &  & 0.01 \pm 0.01 & 0.06 \pm 0.01 & 0.01 \pm 0.01 &  &  & 0.16 \pm 0.02 \\
8711.64 \pm 0.16 & \ion{N}{1} & 8711.70 &  & 0.01 \pm 0.01 & 0.03 \pm 0.01 &  &  &  & 0.21 \pm 0.02 \\
8718.69 \pm 0.44 & \ion{N}{1} & 8718.84 &  & 0.02 \pm 0.01 & 0.02 \pm 0.02 &  &  &  & 0.09 \pm 0.01 \\
8727.05 \pm 0.16 & [\ion{C}{1}] & 8727.13 &  & 0.19 \pm 0.02 &  & 0.03 \pm 0.01 &  &  & 0.22 \pm 0.02 \\
8733.67 \pm 0.44 & \ion{He}{1} & 8733.43 &  & 0.04 \pm 0.01 & 0.02 \pm 0.01 & 0.03 \pm 0.01 & 0.03 \pm 0.01 & 0.04 \pm 0.01 & 0.02 \pm 0.01 \\
8738.57 \pm 0.45 & \chem{H_2} 8-4 Q(3) & 8738.38 &  & 0.06 \pm 0.01 & 0.02 \pm 0.01 & 0.01 \pm 0.01 &  &  & 0.03 \pm 0.01 \\
8750.36 \pm 0.10 & \ion{H}{1} & 8750.47 &  & 0.95 \pm 0.05 & 0.86 \pm 0.05 & 0.92 \pm 0.04 & 0.93 \pm 0.04 & 0.96 \pm 0.05 & 1.15 \pm 0.05 \\
8776.82 \pm 0.46 & \ion{He}{1} & 8776.71 &  & 0.02 \pm 0.01 &  & 0.01 \pm 0.01 & 0.02 \pm 0.01 & 0.01 \pm 0.01 & 0.02 \pm 0.01 \\
8789.12 \pm 0.16 & \chem{H_2} 4-1 S(0) & 8789.12 &  & 0.19 \pm 0.01 &  & 0.02 \pm 0.01 &  &  & 0.06 \pm 0.01 \\
 & \chem{H_2} 11-6 Q(1) & 8809.76 & au &  &  &  &  &  &  \\
8809.90 \pm 0.45 & \chem{H_2} 8-4 Q(4) & 8809.90 & au & 0.08 \pm 0.01 & 0.09 \pm 0.07 & 0.01 \pm 0.01 &  &  & 0.03 \pm 0.01 \\
8840.38 \pm 0.98 & \chem{H_2} 3-0 Q(7) & 8840.89 &  & 0.01 \pm 0.01 &  &  &  &  &  \\
8850.57 \pm 0.17 & \chem{H_2} 5-2 S(7) & 8850.66 &  & 0.23 \pm 0.02 & 0.06 \pm 0.03 & 0.05 \pm 0.01 &  &  & 0.09 \pm 0.01 \\
8862.77 \pm 0.10 & \ion{H}{1} & 8862.79 &  & 1.44 \pm 0.06 & 1.51 \pm 0.07 & 1.38 \pm 0.05 & 1.21 \pm 0.05 & 1.33 \pm 0.06 & 1.59 \pm 0.09 \\
8888.67 \pm 0.47 & \chem{H_2} 5-2 S(9) & 8888.85 &  & 0.09 \pm 0.02 &  &  &  &  & 0.02 \pm 0.01 \\
8893.50 \pm 0.20 & \chem{H_2} 5-2 S(5) & 8893.81 &  & 0.20 \pm 0.05 &  &  &  &  & 0.11 \pm 0.02 \\
8946.50 \pm 0.27 & \chem{H_2} 5-2 S(4) & 8946.15 &  & 0.15 \pm 0.03 & 0.13 \pm 0.05 & 0.07 \pm 0.02 &  &  & 0.07 \pm 0.02 \\
9014.88 \pm 0.12 & \ion{H}{1} & 9014.91 &  & 1.63 \pm 0.08 & 1.21 \pm 0.02 & 1.80 \pm 0.11 & 1.78 \pm 0.11 & 1.83 \pm 0.11 & 1.92 \pm 0.10 \\
9019.85 \pm 0.19 & \chem{H_2} 5-2 S(3) & 9019.30 &  & 0.23 \pm 0.07 &  &  &  &  & 0.03 \pm 0.01 \\
9029.22 \pm 0.17 & \chem{H_2} 4-1 Q(1) & 9029.26 &  & 0.29 \pm 0.03 &  & 0.04 \pm 0.02 &  &  & 0.14 \pm 0.02 \\
9033.99 \pm 0.49 & \chem{H_2} 9-5 S(4) & 9034.88 &  & 0.03 \pm 0.01 &  &  &  &  & 0.04 \pm 0.02 \\
 & \chem{H_2} 11-6 Q(4) & 9049.54 & av &  &  &  &  &  &  \\
9050.47 \pm 0.47 & \chem{H_2} 9-5 S(5) & 9049.69 & av & 0.10 \pm 0.01 &  & 0.01 \pm 0.01 &  &  & 0.04 \pm 0.04 \\
9057.18 \pm 0.16 & \chem{H_2} 4-1 Q(2) & 9057.20 &  & 0.20 \pm 0.01 &  & 0.02 \pm 0.01 &  &  &  \\
9061.42 \pm 0.55 & \ion{He}{1} & 9063.29 &  & 0.33 \pm 0.12 & 0.56 \pm 0.23 & 0.45 \pm 0.19 & 0.31 \pm 0.13 & 0.38 \pm 0.15 & 0.10 \pm 0.07 \\
9068.80 \pm 0.06 & [\ion{S}{3}] & 9068.90 &  & 10.18 \pm 0.36 & 7.68 \pm 0.24 & 12.53 \pm 0.40 & 12.32 \pm 0.50 & 5.62 \pm 0.28 & 12.49 \pm 0.51 \\
9084.49 \pm 0.50 & \chem{H_2} 9-5 S(2) & 9083.40 &  & 0.06 \pm 0.02 &  & 0.03 \pm 0.03 &  &  & 0.05 \pm 0.06 \\
9096.92 \pm 0.18 & \chem{H_2} 4-1 Q(3) & 9099.25 &  & 0.21 \pm 0.02 & 0.21 \pm 0.02 & 0.09 \pm 0.02 &  &  & 0.20 \pm 0.15 \\
9113.60 \pm 0.46 & \chem{H_2} 5-2 S(2) & 9113.54 &  & 0.07 \pm 0.01 &  &  &  &  &  \\
9121.27 \pm 0.39 & [\ion{Cl}{2}] & 9123.60 &  & 0.04 \pm 0.01 & 0.06 \pm 0.03 & 0.03 \pm 0.01 & 0.01 \pm 0.02 &  & 0.11 \pm 0.02 \\
9144.06 \pm 0.22 & \chem{H_2} 9-5 S(1) & 9145.66 &  & 0.12 \pm 0.02 &  & 0.03 \pm 0.02 &  &  & 0.06 \pm 0.04 \\
9155.28 \pm 0.52 & \chem{H_2} 4-1 Q(4) & 9155.56 &  & 0.07 \pm 0.02 & 0.02 \pm 0.01 & 0.02 \pm 0.01 &  &  &  \\
9162.90 \pm 0.55 & \chem{H_2} 9-5 S(7) & 9162.58 &  & 0.02 \pm 0.01 &  &  &  &  &  \\
9204.10 \pm 0.17 & \ion{Ca}{1}] & 9204.09 &  & 0.02 \pm 0.01 & 0.02 \pm 0.01 & 0.01 \pm 0.01 &  &  & 0.24 \pm 0.02 \\
9210.46 \pm 0.51 & \ion{He}{1} & 9210.28 &  & 0.05 \pm 0.01 & 0.05 \pm 0.01 & 0.05 \pm 0.01 & 0.04 \pm 0.01 & 0.05 \pm 0.03 &  \\
9217.91 \pm 0.21 & \ion{Fe}{1}] & 9218.47 &  & 0.03 \pm 0.02 & 0.03 \pm 0.01 & 0.02 \pm 0.01 &  &  & 0.24 \pm 0.03 \\
9228.92 \pm 0.11 & \ion{H}{1} & 9229.01 &  & 2.24 \pm 0.10 & 2.06 \pm 0.09 & 1.94 \pm 0.08 & 1.79 \pm 0.09 & 2.03 \pm 0.12 & 2.48 \pm 0.12 \\
9243.78 \pm 0.24 & \ion{Ca}{1}] & 9244.31 &  & 0.02 \pm 0.01 & 0.05 \pm 0.02 &  &  &  & 0.15 \pm 0.03 \\
9263.62 \pm 0.51 & \chem{H_2} 9-5 S(8) & 9264.11 &  & 0.03 \pm 0.01 & 0.02 \pm 0.02 & 0.01 \pm 0.01 &  &  & 0.02 \pm 0.01 \\
9297.28 \pm 0.16 & \chem{H_2} 4-1 O(2) & 9297.59 &  & 0.18 \pm 0.01 &  &  &  &  & 0.14 \pm 0.02 \\
\end{longtable}

}
\stopNEW
\twocolumn

\appendix
\section{Additional spectra}
\label{sec:additional-spectra}
\begin{figure*}
  \centering
  \includegraphics[width=\linewidth]{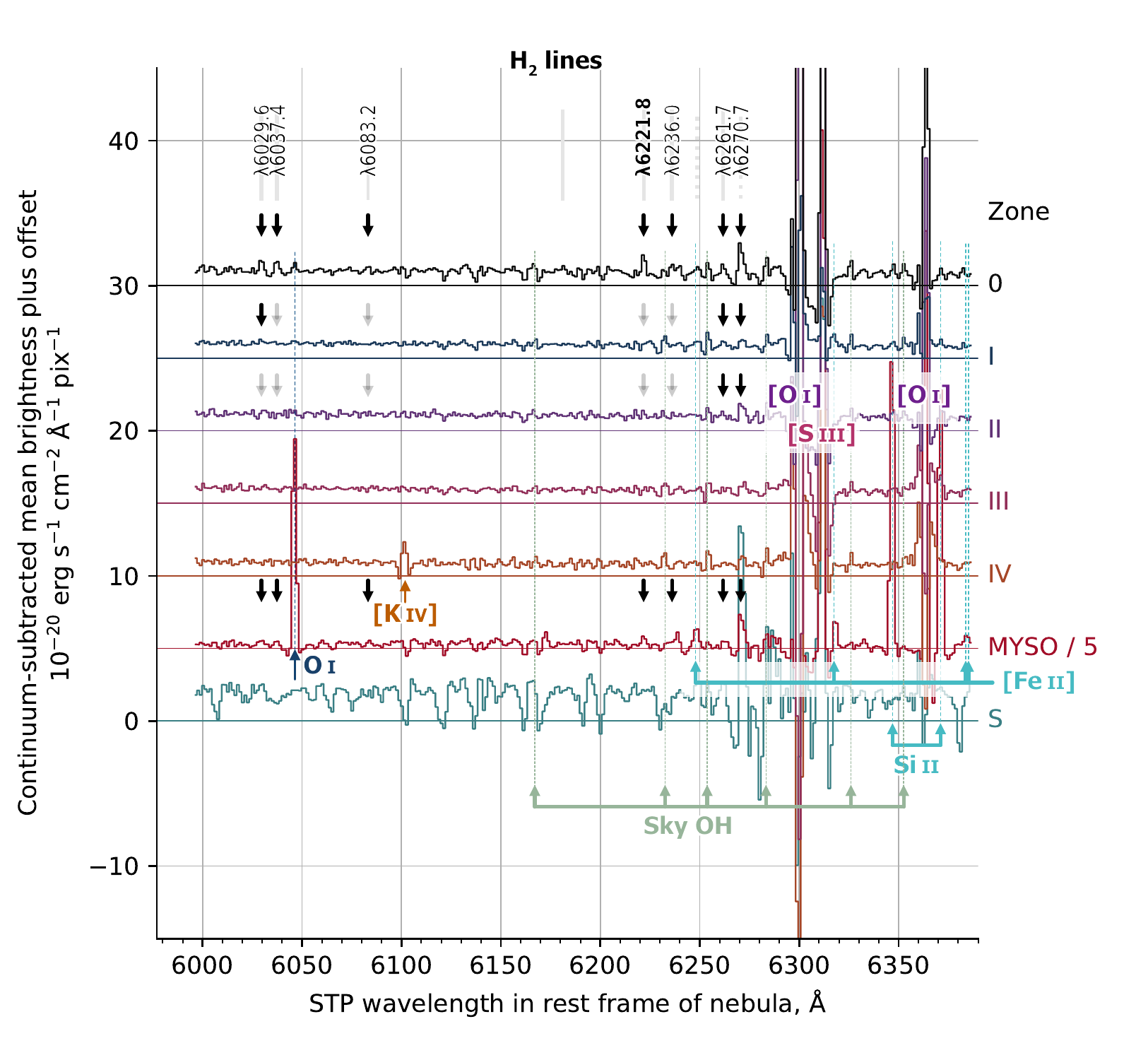}
  \caption{
    As Figure~\ref{fig:spectrum-8xxx-B} but for the region around
    the red [\ion{O}{1}] nebular lines and [\ion{S}{3}] auroral line.
  }
  \label{fig:spectrum-6xxx-A}
\end{figure*}

\begin{figure*}
  \centering
  \includegraphics[width=\linewidth]{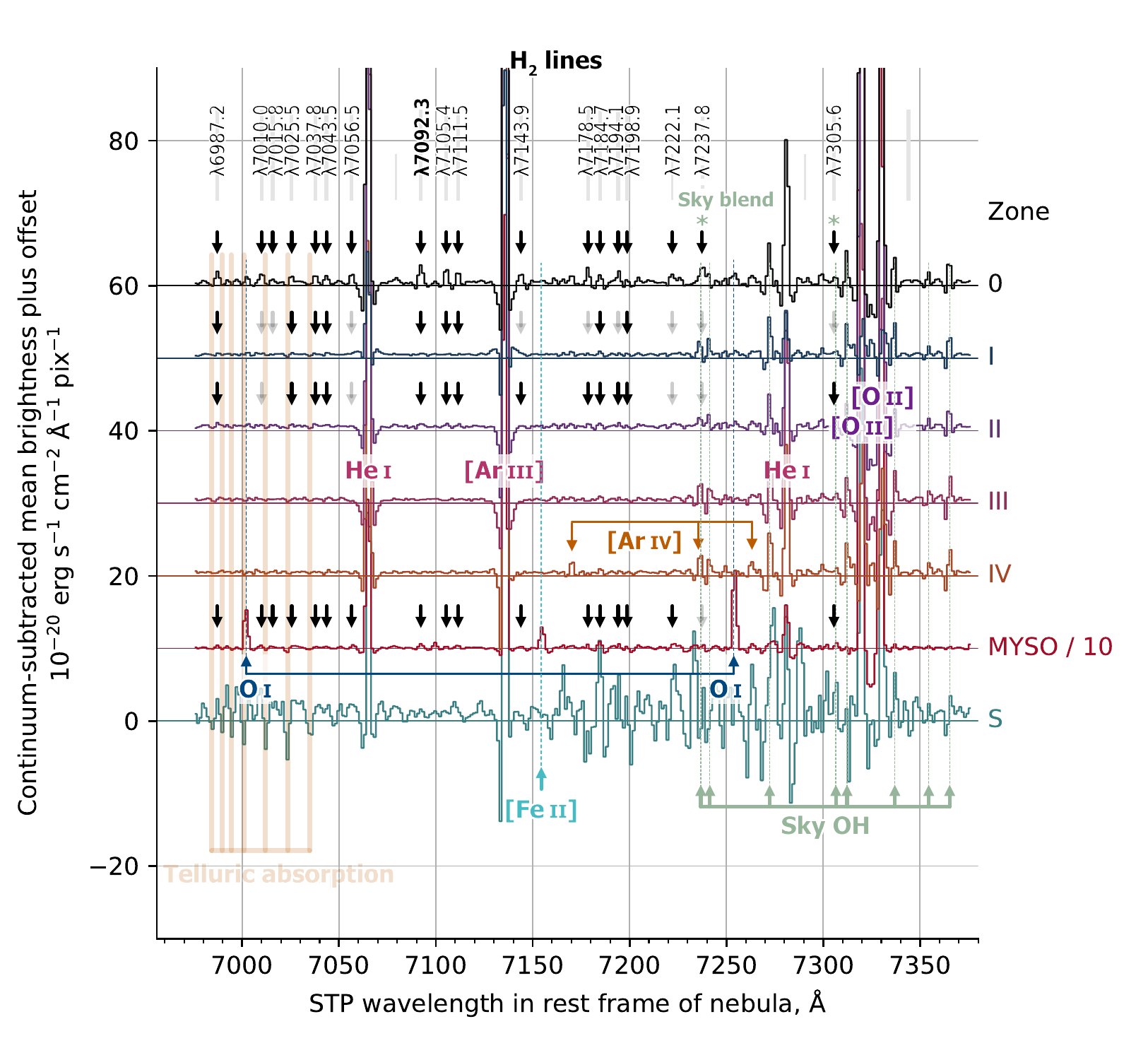}
  \caption{
    As Figure~\ref{fig:spectrum-8xxx-B} but for the region around
    the red [\ion{Ar}{3}] and [\ion{O}{2}] nebular lines.
  }
  \label{fig:spectrum-7xxx-A}
\end{figure*}

\begin{figure*}
  \centering
  \includegraphics[width=\linewidth]{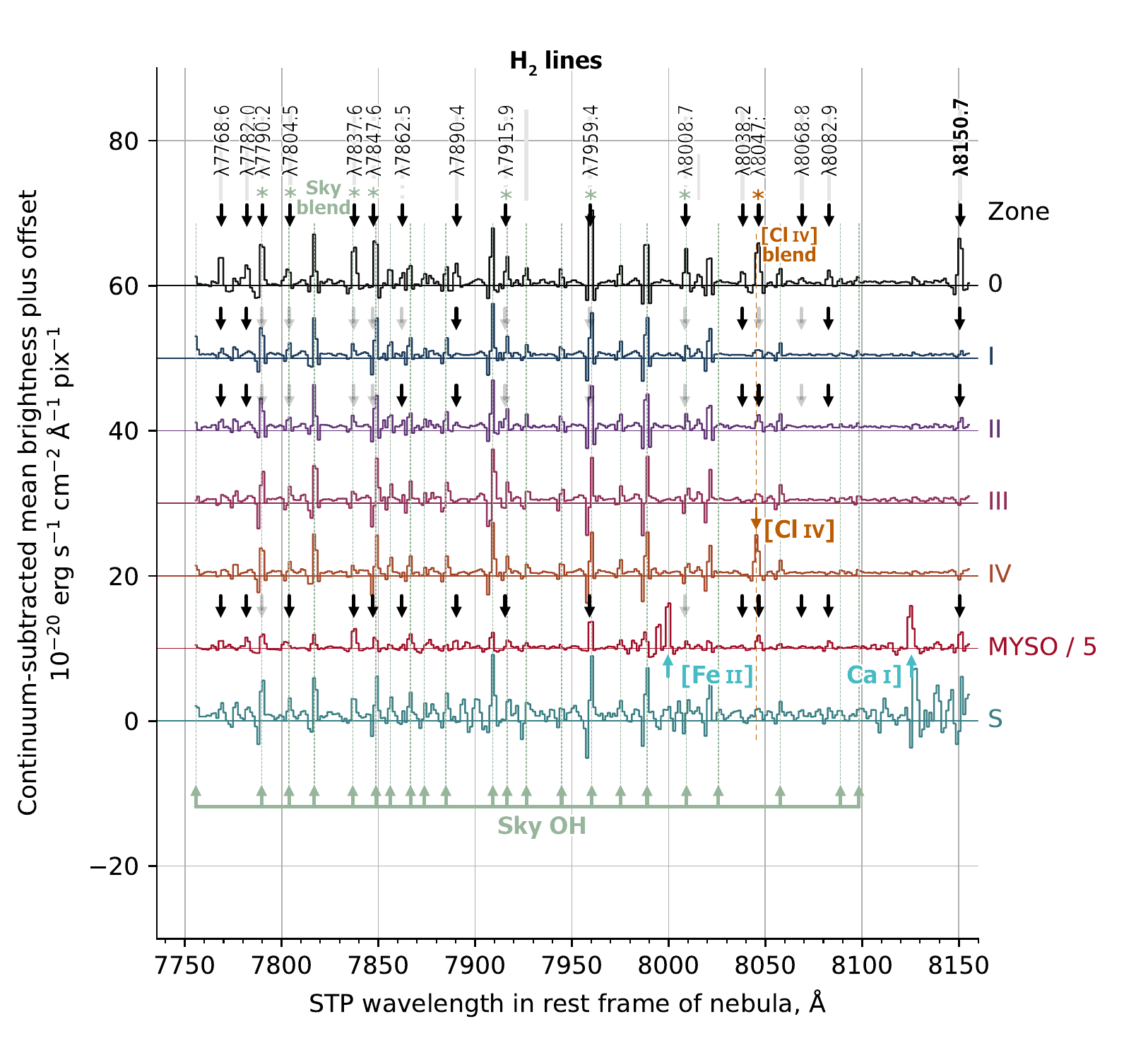}
  \caption{
    As Figure~\ref{fig:spectrum-8xxx-B} but for a far-red region of the spectrum
    that is devoid of strong lines.
  }
  \label{fig:spectrum-7xxx-B}
\end{figure*}

\begin{figure*}
  \centering
  \includegraphics[width=\linewidth]{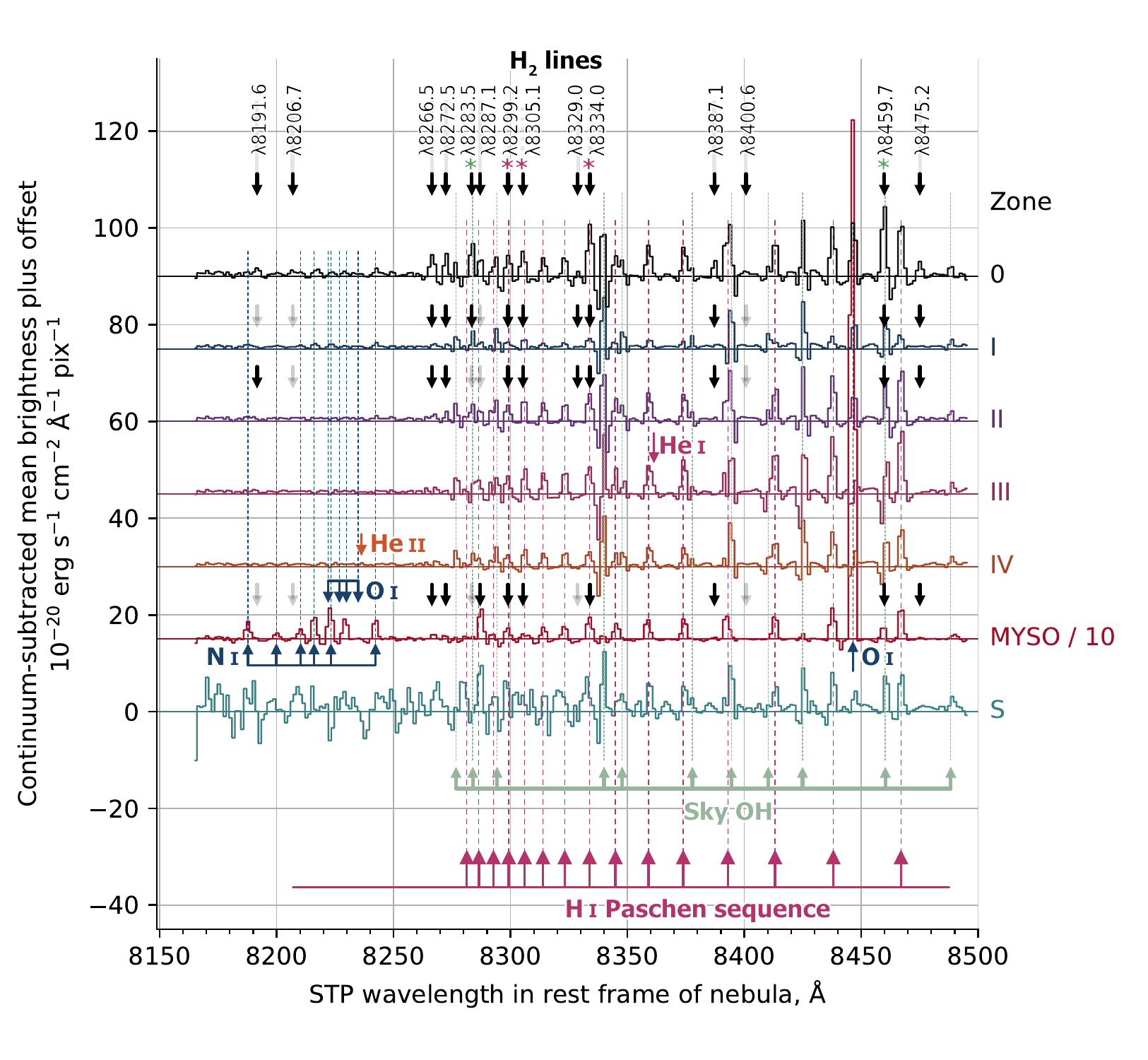}
  \caption{
    As Figure~\ref{fig:spectrum-8xxx-B} but for the region around
    the Paschen  jump.
  }
  \label{fig:spectrum-8xxx-A}
\end{figure*}

\begin{figure*}
  \centering
  \includegraphics[width=\linewidth]{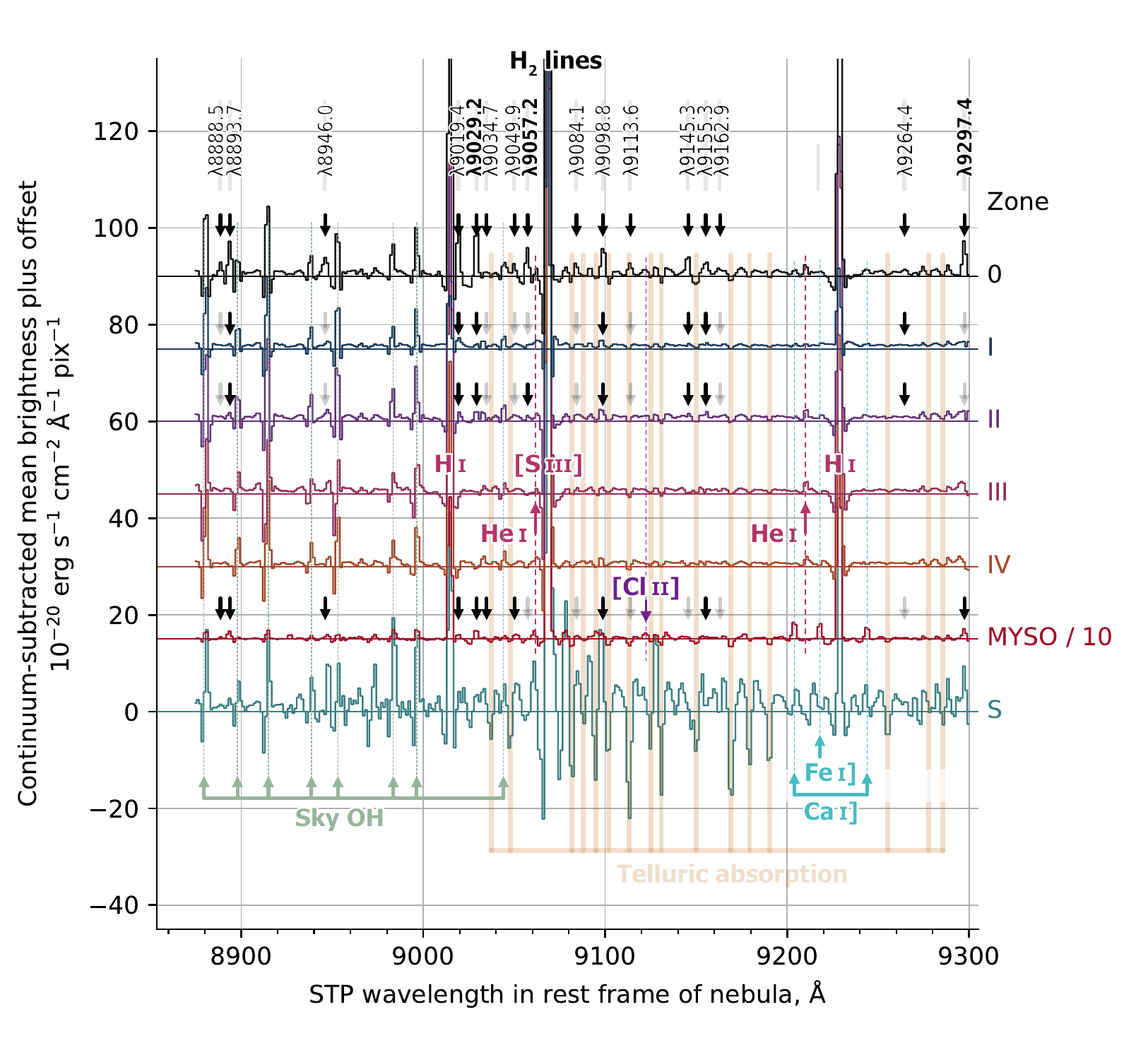}
  \caption{
    As Figure~\ref{fig:spectrum-8xxx-B} but for the farthest red
    region of the spectrum, which contains the [\ion{S}{3}]
    nebular line.
  }
  \label{fig:spectrum-9xxx-A}
\end{figure*}

\startNEW
Figures~\ref{fig:spectrum-6xxx-A} to \ref{fig:spectrum-9xxx-A}
complement Figures~\ref{fig:spectrum-8xxx-B} and \ref{fig:spectrum-6xxx-B}
by showing additional wavelength ranges of the MUSE spectra between
\SI{6000}{\angstrom} and \SI{9500}{\angstrom},
where molecular hydrogen lines are observed.
\stopNEW


\startNEW
\section{The path to identifying the molecular hydrogen lines}
\label{sec:path-ident-molec}
After finding over 100 unknown (to us) optical emission lines
in the MUSE spectra of NGC~346, we spent a lot of effort on searching
for plausible identifications before discovering that they were
in fact molecular hydrogen lines.
The following appendices give the details of this search.
During this phase of the research
we used the term Deep Red Line (DRL) for the as-yet unidentified lines
as a reflection of both their spectral and spatial distribution.
We maintain this terminology in these appendices
but it should be remembered that all of the DRLs
have now been positively identified as \hmol{}. 
\stopNEW

In Appendix~\ref{sec:struct-wavel-distr} we
employ various nearest-neighbor measures
to analyze the clustering of the emission lines in wavelength.
\startNEW
Before making the identification of the Deep Red Lines
with molecular hydrogen, 
the results of this analysis (see Figure~\ref{fig:nearest-neighbor})
were as follows:
\stopNEW
\begin{enumerate}[1.]
\item The Deep Red Lines show a greater regularity in their spacing
  than would arise from a purely random distribution
  (Figure~\ref{fig:nearest-neighbor}b).
  This is similar to what is seen in the spacing of the \NEW{atomic}
  emission lines.
\item Deep Red Lines in the wavelength ranges
  \SIrange{7000}{7500}{\angstrom}
  and \SIrange{8000}{8500}{\angstrom} show a marked
  preference for doublet patterns in the spacings
  (Figure~\ref{fig:nearest-neighbor}c),
  which is also apparent by eye in Figure~\ref{fig:bar-code}.
  Deep Red Lines around \SI{6500}{\angstrom},
  on the other hand, show a preference for asymptotic sequences
  (line spacing systematically increasing or decreasing
  with wavelength).
  For comparison, the known lines show a preference
  for doublet patterns at shorter wavelengths,
  but a marked preference for asymptotic sequences at wavelengths
  \(> \SI{8000}{\angstrom}\), which is mainly due to the \ion{H}{1} Paschen sequence. 
\item The doublet separation of the Deep Red Lines
  with \(\lambda < \SI{8000}{\angstrom}\) is typically larger
  (\SIrange{12}{15}{\angstrom})
  than for the lines with \(\lambda > \SI{8000}{\angstrom}\)
  (\SIrange{5}{10}{\angstrom}). 
\item \label{item:cross} Cross comparison  of the separations between
  the identified and the unidentified lines is consistent with
  pure randomness, which implies that Deep Red Lines primarily
  arise from different species than any of the known lines.
\end{enumerate}
\startNEW
In the light of the identification of the DRLs with \hmol{},
it is interesting to critically evaluate the usefulness of these inferences.
It turns out that the supposed ``doublet'' patterns were not true doublets,
but were caused by the interleaving of independent overlapping rotational ladders.
For instance, the 8-4 and 4-1 S~branches overlap
in the range \SIrange{8200}{8500}{\angstrom},
while the 9-4 S, 6-2 S, and 5-1 Q~branches overlap
in the range \SIrange{7000}{7200}{\angstrom}.
On the other hand, the asymptotic sequence identified around \SI{6500}{\angstrom}
does indeed turn out to be due to the band head of the 8-3 S~branch,
which does not overlap with other sequences. 
Furthermore, the lack of any correlations with the atomic lines
(item~\ref{item:cross} above) was a key piece of evidence
that led us to concentrate on molecular candidates,
resulting in eventual successful identification.

The remainder of the current appendix describes how we systematically
ruled out atomic line candidates for the then-unidentified Deep Red Lines,
together with more exotic possibilities such as Diffuse Interstellar Bands
and Raman scattering.
\stopNEW

% \section{Candidate line identifications}
% \label{sec:cand-line-ident}

% We have been unable to find convincing identifications for
% any of the Deep Red Lines.
% We have concentrated most of our search effort on neutral atomic lines,
% but also briefly discuss the possibility of a molecular origin for the lines. 

\subsection{\NEW{Ruling out} neutral atoms}
\label{sec:neutral-atoms}
The most pedestrian explanation
for the unidentified lines would be
fluorescent excitation of neutral atoms by starlight
that penetrates into the PDR.
Fluorescent lines typically form at an
optical depth of order unity in the resonant
pumping line (at UV or optical wavelengths). 
The fact that the lines are found much deeper in
the PDR than the known \ion{O}{1} and \ion{N}{1}
fluorescent lines
then implies that either the pumping transitions are weak
(for example, a semi-forbidden intercombination transition)
or that the gas phase abundance of the neutral species is low.
The second possibility seems more plausible and could be due
to a combination of three factors:
the intrinsic chemical abundance of the element,
depletion onto dust grains,
and the degree of ionization.
For example, atoms with low first ionization energies
(e.g., C, Si, S)
are predominantly singly ionized in the translucent outer layers of the PDR,
so a significant column of the neutral atoms occurs only at greater depths. 

However, one argument against a fluorescent origin for these lines
is that they do not show any particular enhancement in
the MYSO zone, whereas most known fluorescent lines
are greatly enhanced there, which is
possibly due to a relatively softer illuminating UV spectrum.

Apart from fluorescence, other possible line excitation
mechanisms are collisions with atoms or free electrons,
recombination, and dissociation pumping.
Collisional excitation would require the PDR to be very warm,
with gas temperature \( \sim \SI{5000}{K}\),
since lower limits on the line excitation temperature
(from assuming a transition to the ground state)
are of order \SI{20000}{K},
and collisional rates are exponentially suppressed for
gas temperatures much smaller than this.
Collisional excitation by electrons is the dominant excitation
mechanism for the far-red [\ion{C}{1}] lines in
some planetary nebulae \citep{Liu:1995a},
with an inferred electron temperature of order \SI{8000}{K}.
In those cases, the illumination is from a very hot star
(with effective temperature typically
\(T_\mathrm{eff} > \SI{e5}{K}\)).
In contrast, in classical photodissociation regions
that are illuminated by OB stars
\(T_\mathrm{eff} < \SI{50000}{K}\))
the temperature in the [\ion{C}{1}] emitting regions
is expected to be below \SI{1000}{K}
\citep{Rollig:2007a}.
In this case, radiative recombination is 
the dominant excitation mechanism of the forbidden [\ion{C}{1}] lines
\citep{Escalante:1991a}.
However, the weak temperature dependence of the radiative cooling rate in neutral gas
can lead to thermal bistability
and the coexistence of cold and warm phases in pressure equilibrium,
meaning that both excitation mechanisms might be acting in the same PDR
(collisions in the warm phase and recombination in the cold phase). 

Photodissociation of \chem{OH} that leaves \chem{O} atoms in the
excited \chem{^1D} level has been proposed as an excitation mechanism
for the [\ion{O}{1}] \SI{6300}{\angstrom} line in
circumstellar environments, such as photoevaporating
protoplanetary disks \citep{Storzer:1998f, Gorti:2011a, Ballabio:2023a},
but to our knowledge
this dissociation pumping mechanism has not been detected in
the larger scale PDRs associated with molecular clouds.
 
\begin{table*}
  \caption{Results of semi-automated identification for the brightest Deep Red Lines}
  \label{tab:possible-ids}
  \centering
  \begin{tabular}{S r@{\ \ }S S@{}S lll}
    \toprule
    {\(\lambda(\text{obs})\)}
    & Ion & {\(\lambda(\text{lab})\)}
    & {\(E(\text{lower})\)} & {\(E(\text{upper})\)}
    & Pro & Contra & Confidence level\\
    {\AA} & & {\AA} & {\si{cm^{-1}}} & {\si{cm^{-1}}} & & & \\
    {(1)} & {(2)} & {(3)} & {(4)} & {(5)} & {(6)} & {(7)} & {(8)}\\
    \midrule
    6221.82 \pm 0.53 & \ion{Si}{1} & 6221.70 & 49399.67 & 65468.00 &  & Mult, BR, FP & Very low\\
            & \ion{Fe}{1}] & 6221.6722 & 6928.27 & 22996.67 &  & Mult & Low\\
    \addlinespace{}
    6636.76 \pm 0.57 & \ion{N}{1} & 6636.938 & 94793.49 & 109856.52 & Mult & LC, EP, BR, Casc & Very low\\
            & [\ion{Co}{1}] & 6636.6227 & 1406.85 & 16470.60 & Ground, Mult? & Abun, BR & Low\\
    \addlinespace{}
    6776.75 \pm 0.58 & \ion{Fe}{1}] & 6776.689 & 29798.93 & 44551.33 &  & BR & Very low\\
            & \ion{Ca}{1} & 6776.50 & 45049.07 & 59802.21 &  & BR & Very low\\
    \addlinespace{}
    7092.29 \pm 0.21 & \ion{C}{1}] & 7092.51 & 75255.27 & 89350.10 &  & BR & Very low\\
            & \ion{Fe}{1} & 7092.077 & 43137.48 & 57233.84 &  & BR, Mult, FP & Very low\\
    \addlinespace{}
    8150.67 \pm 0.24 & [\ion{Fe}{1}] & 8151.3424 & 704.01 & 12968.55 & Ground, BR & Wav, Mult? & Low \\
            & \ion{Fe}{1} & 8150.462 & 47831.15 & 60097.02 &  & FP, Mult, BR & Very low\\
    \addlinespace{}
    8550.06 \pm 0.25 & \ion{Si}{1} & 8550.34 & 50189.39 & 61881.6 & BR? & Mult & Low \\
            & \ion{Cl}{1} & 8550.441 & 74225.85 & 85917.94 & BR & Abun & Low \\
    \addlinespace{}
    8789.18 \pm 0.26 & \ion{Fe}{1} & 8789.660 & 46744.99 & 58118.87 &  & Mult, FP, BR & Very low\\
    \addlinespace{}
    8850.65 \pm 0.26 & \ion{Fe}{1} & 8850.1114 & 26224.97 & 37521.16 &  & Mult, BR & Very low\\
    \addlinespace{}
    9029.22 \pm 0.27 & \ion{N}{1} & 9028.922 & 93581.55 & 104654.03 & BR & LC, Mult & Very low\\
    \addlinespace{}
    9057.17 \pm 0.27 & \ion{O}{1} & 9057.02 & 113204.45 & 124242.58 &  & AI & Nil\\
            & \ion{Fe}{1} & 9056.759 & 48475.68 & 59514.13 & Mult? & FP & Very low \\
    \addlinespace{}
    9297.36 \pm 0.27 & \ion{Fe}{1} & 9297.61 & 41018.05  & 51770.55 & Mult? & BR & Very low\\
    \bottomrule
    \addlinespace
    \multicolumn{8}{p{0.9\linewidth}}{%
    Columns:
    (1)~Observed rest wavelength (air) in \si{\angstrom} from Table~\ref{tab:all-lines}. 
    (2)~Ion species of candidate line ID.
    (3)~Laboratory wavelength of candidate line ID in \si{\angstrom}.
    (4, 5)~Lower and upper energy levels of candidate line ID on wavenumber scale (\si{cm^{-1}}).
    (6, 7)~Arguments pro or contra the candidate line ID. Mult: multiplet; BR: branching ratio;
    Casc: cascade;
    FP: false positive; LC: line class; EP: EUV pumping; Ground: transition to ground configuration; 
    Abun: elemental abundance; AI: Autoionizing state; Wav: wavelength difference.
    Question mark indicates weak or mixed evidence.
    See text for details. 
    (8)~Subjective level of confidence in the ID.
    Note that we do not have high confidence in \emph{any} of these identifications. 
    }
  \end{tabular}
\end{table*}

\subsection{Semi-automated identification of atomic lines}
\label{sec:spec-cand-ident}
For the selected Deep Red Lines that are illustrated in Figure~\ref{fig:uil-stamps},
we carried out a semi-automated search for identifications using the
EMILI code \citep{Sharpee:2003a},
based on the Atomic Line List database v2.04 of \citet{van-Hoof:1999a}.
This uses a set of identification criteria based on wavelength agreement,
relative intensities, and detection of other lines of the same multiplet. Details of the parameters used are given in Appendix~\ref{sec:deta-semi-autom}.
In Table~\ref{tab:possible-ids} we list the highest ranked identifications
found by EMILI in which the line arises in a neutral atom.\footnote{%
  The spatial distribution of the DRL emission (section~\ref{sec:interz-diagn-rati})
  argues strongly against an origin within the \hii{} region or ionization front.
  Nonetheless, singly ionized species of low ionization potential
  (e.g., \ion{C}{2}) exist in the PDR and
  cannot be conclusively ruled out, so we have checked for these
  also, but find no plausible candidates.
}
We have supplemented the EMILI results with updated data from the latest beta version v3.00b4
of the Atomic Line List database \citep{van-Hoof:2018a, van-Hoof:2021a}.

\subsubsection{Pro and contra arguments for specific IDs}
\label{sec:arguments-pro-contra}

Columns 6 and 7 of Table~\ref{tab:possible-ids} list different arguments
pro and contra each identification, which can be classified as follows.

\subparagraph{Wavelength difference}
We consider a wavelength agreement within \(2\sigma\) to be acceptable and
nearly all of our candidates fall within this limit, which is not surprising
since it is the principle criterion used by EMILI.
The exception is [\ion{Fe}{1}] \wav{8151.3424} as a potential ID for DRL \wav{8150.67}
with a wavelength discrepancy of nearly \(3\sigma\),
which argues against this otherwise promising identification.

\subparagraph{Multiplet}
How many lines do we detect that arise from the same multiplet
(that is, sharing the same upper and lower spectroscopic term)?
For heavy atoms, the multiplicity is often high and there may be 10
or more potential members of the multiplet,
with typically half of these falling in spectral regions that
are both within the MUSE range and unaffected by blending with
other nebular or sky lines. 
In most cases, no other members are detected at a level \(> 5\%\)
of the observed DRL, which argues strongly against
the reality of the identification.

In some cases, we do find partial matches, with the one promising case
being \ion{Fe}{1} \wav{9297.61}
(\Term{g}{5}{D} -- \Term{x}{5}{F^o}, \(J = 3\)--2)
as a possible ID for DRL \wav{9297.36}.
There are 12 members of the multiplet, of which 3 coincide with observed DRLs.
In addition to DRL \wav{9297.36},
these are DRL \wav{9155.28} (with 2--1 \wav{9155.6472})
and DRL \wav{9019.36} (with 1--1 \wav{9019.7445}).
A fourth component (4--3 \wav{8945.1893}) could plausibly contribute to the blue wing
of DRL \wav{8945.95}. 
Of the remaining 8 members, only one (3--2 \wav{8920.0129}) has a \(2\sigma\) upper limit
less than 5\% of the intensity of DRL \wav{9297.36},
with the remainder all suffering from blends or outside the MUSE observed range.
However, even in this case, we consider the multiplet evidence to be only weakly in favor
of the identification.

Another case of partial multiplet matches is \ion{N}{1} \wav{6636.938},
discussed in detail in Appendix~\ref{sec:possible-id-ni},
where we show that the countervailing evidence against this identification is overwhelming. 

\subparagraph{Branching ratio}
Does the candidate line represent a significant fraction of all the downward
radiative transitions from its upper level? Or are there other observable transitions that should be much stronger? We compare the predicted radiative transition probabilities
(Einstein \(A\) values)
for all relevant lines in order to address this question. For example, for the
same \ion{Fe}{1} \wav{9297.61} line mentioned above,
we find that the branching ratio is only \(0.003\) and that there are many lines
from the same upper state that fall within the MUSE range and should be more than
10 times brighter, such as \wav{4885.4304}, \wav{5473.9008}, and \wav{6841.3387},
none of which are observed.
In fact, even members of the same multiplet,
whose possible detection was counted above as pro evidence,
are predicted to be much brighter than \wav{9297.61}.
For instance, the predicted transition probability ratio of
\(\wav{8945.1893} /  \wav{9297.61}\) is about 50 \citep{Fuhr:2006a},
whereas the observed intensity of the blue wing of DRL \wav{8945.95} is only
about \(0.2\) times DRL \wav{9297.36}.
This is strong evidence against the validity of the identification.

In the case of some semi-forbidden lines, the situation is even more extreme.
For instance, \ion{Fe}{1}] \wav{6776.689} (candidate ID for DRL \wav{6776.75})
has a branching ratio of \(< \num{e-8}\).
Even though the strongest transitions down from its upper level are in the NUV,
and thus unobservable with MUSE,
it is extremely unlikely that \wav{6776.689} could ever be emitted with any significant intensity.

In other cases, the branching ratio argument works in favor of the proposed ID.
For example \ion{Cl}{1} \wav{8550.441} (candidate ID for DRL \wav{8550.06})
is the dominant transition down from its upper level,
with a branching ratio of \num{0.998}.

\subparagraph{Cascade}
Do we (or should we) observe other lines that cascade down from the lower level
of the candidate transition?
If so, then the summed intensity of these cascade lines (in photon units)
must be at least as large as the intensity of the candidate line
if the identification is correct
(at least in the case of permitted lines, for which collisional deexcitation can be ruled out).
We apply this criterion in detail in Appendix~\ref{sec:possible-id-ni}
to the case of \ion{N}{1} \wav{6636.938} as a candidate ID for DRL \wav{6636.76},
finding that the relative weakness of the \ion{N}{1} V~1 \wavv{8700} multiplet
in the spectrum of Zone~0 is strong evidence against this identification.

\subparagraph{False positive}
If lines of a given type are so numerous that their mean spacing is
smaller than the uncertainty in the observed DRL wavelengths
(\num{0.3} to \SI{1.0}{\angstrom})
then the a priori probability of a spurious false positive identification becomes high.
This is particularly an issue for \ion{Fe}{1}, which has
a very large number of bound levels \citep{Nave:1994a}.
The mean line spacing is a strong function of the energy of the upper level,
being \(\approx \SI{50}{\angstrom}\) for \(E < \SI{30000}{cm^{-1}}\),
\(\approx \SI{5}{\angstrom}\) for \(E < \SI{45000}{cm^{-1}}\),
and \(\approx \SI{0.5}{\angstrom}\) for \(E < \SI{60000}{cm^{-1}}\).
Similar results are found for other heavy atoms, such as \ion{Si}{1}.
One symptom of this is that EMILI flags multiple unrelated lines
of the same atom as potential matches for a particular line.
Such is the case for DRL \wav{7092.29}: in addition to
the candidate \ion{Fe}{1} \wav{7092.077} \Level{4s(^6D)5d}{5}{G}{3} -- \Level{y}{5}{G^o}{2}
listed in Table~\ref{tab:possible-ids},
there is also \ion{Fe}{1} \wav{7092.080} \Level{4s(^4D)4d}{5}{F}{4} -- \Level{v}{5}{D^o}{4}.

\subparagraph{Line class}

Although we do not know the exact mechanism responsible for exciting the DRLs,
their particular spatial distribution means
we can be sure there must be some significant difference in the required
physical conditions with respect to all the other line classes listed in Table~\ref{tab:ion-class}.
It therefore counts as an argument contra a proposed ID if
the line seems similar to others that are emitted predominantly from zones other than Zone~0.
Such is the case for permitted lines of light metal atoms,
such as \ion{N}{1} and \ion{O}{1}, which are relatively strongest in Zones~I and MYSO.
A clear example is \ion{N}{1} \wav{9028.92} (candidate ID for DRL \wav{9029.22}),
which is predicted to be produced by the same fluorescent pumping mechanism
as the [\ion{N}{1}] \wav{5199} doublet \citetext{see Table~9 of \citealt{Ferland:2012a}}.

\subparagraph{EUV pumping} If the energy of the upper level is greater than
\(\SI{1}{Rydberg} = \SI{109678}{cm^{-1}}\),
then any radiation capable of pumping the transition would also be capable of
ionizing neutral H, and therefore would be strongly attenuated in the PDR.
This is an argument contra \ion{N}{1} \wav{6636.938} as candidate ID
for DRL \wav{6636.76}, see Appendix~\ref{sec:possible-id-ni}.
The argument would not apply in the case of a recombination line,
but for \ion{N}{1} that would require \chem{N^+}, which is not present in the PDR. 

\subparagraph{Transition to ground configuration}
Any candidate transition in which the lower level is in the ground configuration of the atom
is favored for two reasons. First, there are relatively few such transitions, making each
one ``special'' and lowering the risk of a false positive ID. Second, the upper level will
be of relatively low energy and therefore easier to excite in general,
be it by collisions or radiation. They will also be favored in the case of recombination lines
since a relatively high fraction of the total recombinations will pass through the transition. 

Only two of our candidate IDs have this property:
[\ion{Co}{1}] \wav{6636.620} and [\ion{Fe}{1}] \wav{8151.3424}.

\subparagraph{Elemental abundance}
All else being equal, more abundant elements will give rise to stronger lines.
We have considered this to be an argument contra any ID where the
estimated gas-phase abundance is
less than \num{e-6} with respect to H by number, which includes
[\ion{Co}{1}] \wav{6636.620}, \ion{Ca}{1} \wav{6776.50}, and \ion{Cl}{1} \wav{8550.44}.

\subparagraph{Autoionizing state}

If two electrons in the atom are excited, then the resultant energy level
can lie above the ground state ionization potential.
Such levels have extremely short lifetimes to the non-radiative
autoionization process.
The resultant natural broadening means that any bound-bound transitions between
such levels will produce very broad lines.
They are therefore conclusively ruled out as candidate IDs for the DRLs,
which applies to \ion{O}{1} \wav{9057.02}.

\subsubsection{(Lack of) confidence in the \NEW{atomic} IDs}
\label{sec:lack-of-confidence}

It can be seen from Table~\ref{tab:possible-ids} that for nearly all the
proposed IDs, there are more contra arguments than pro arguments,
and the contra arguments tend to be stronger.
As a result,
we do not consider \emph{any} of these candidate identifications
of neutral atomic lines to be very probable.

One can also raise more general objections to the proposed IDs,
fully two-thirds of which are for refractory elements
(Fe, Si, Co, Ca).
The ionized emission lines from these elements show
a marked increase by a factor of \(\approx 10\) in Zone~MYSO
with respect to Zone~0 (see section~\ref{sec:interz-diagn-rati}),
whereas the DRLs show no such increase,
which would argue against the DRLs being due to these elements.
Also, the particular patterns that we identify in the
wavelength distribution of the DRLs
(section \ref{sec:struct-wavel-distr})
are not explained by these IDs. 

\subsection{\NEW{Ruling out heavier} diatomic molecules and molecular ions}
\label{sec:diat-molec-molec}

Given that the none of the proposed neutral atom IDs for the DRLs are satisfactory, it is worth considering other possibilities.
In the translucent outer layers of the PDR, neutral hydrogen
may coexist with simple molecular species, primarily
diatomic or triatomic molecules such as
\chem{CN}, \chem{C_2}, \chem{CH}, \chem{NH}, \chem{C_3},
together with molecular ions such as \chem{CH^+} and \chem{OH^+}
\citetext{see \citealp{2006ARA&A..44..367S} and references therein}.
Many of these molecules have lines in the optical and near-IR
that could potentially be seen in emission from the PDR.
The excitation mechanisms for these lines include
those considered above for atomic lines,
such as fluorescence and collisional excitation,
plus others specific to molecules
such as chemical formation into excited states.
For the wavelength range of the observed Deep Red Lines
(\SI{0.6}{\micro\meter} to \SI{0.93}{\micro\meter}),
transitions with upper level energy greater than \SI{10000}{cm^{-1}}
must be involved.
These may be electronically excited states,
or alternatively highly vibrationally excited levels
of the ground electronic state.
In either case, they would be most easily excited by fluorescence
unless the PDR temperature is high enough ($\gtrsim \SI{3000}{K}$)
to thermally populate the upper levels by collisional excitation.

For example, several of the \chem{C_2} Phillips bands
\citep{van-Dishoeck:1982r}
overlap with spectral regions where we find the Deep Red  Lines.
% such as the (2--0) band at \(\lambda >  \SI{8750}{\angstrom}\)
% and the (3--0) band at \(\lambda > \SI{7715}{\angstrom}\).
However, comparing the exact wavelengths, we find few convincing matches. We observe no DRLs near the (2--0) band head
at \SI{8750}{\angstrom} and the closest match is for
DRL \wav{8809.89} with (2--0) \chem{Q(16)} \wav{8809.841}.
It seems most likely that this is just a chance coincidence,
since there seems no particular reason why this highly rotationally
excited member of the band should be the only one observed.

On the other hand, there are many other bands, even within
the \chem{C_2} molecule, and many other molecules,
so this seems a promising avenue for further investigation.
Diatomic molecular bands from UV to IR wavelengths
are typically observed in absorption in the ISM
\citep{2006ARA&A..44..367S}, but they have also been seen in emission
in comets \citep{Nelson:2019a}
and in the Red Rectangle nebula \citep{Wehres:2010g}.
Given the evidence we find for doublet structure in the distribution
of the DRLs with \(\lambda > \SI{7000}{\angstrom}\)
(see section \ref{sec:struct-wavel-distr}),
\chem{C_2} is not a promising candidate,
and we should instead look for molecules that show a similar
structure to their spectra.
Doublet splitting of optical emission lines
is present in molecules such as \chem{OH}
and \chem{O_2} that are seen in the terrestrial night sky spectrum
\citep{Osterbrock:1996a},
although the splitting is typically \SI{1}{\angstrom} or less,
whereas we see splitting of order \SI{10}{\angstrom} with the DRLs.

It is even possible to imagine the presence of a bandhead in
our DRL spectrum at \SI{8266}{\angstrom}
(see Figures~\ref{fig:bar-code} and \ref{fig:spectrum-8xxx-A}),
which would correspond to \SI{12100}{cm^{-1}}.
The doublet splitting is about \SI{7}{cm^{-1}}
close to the bandhead, rising to \SI{14}{cm^{-1}}
for \(\lambda > \SI{8500}{\angstrom}\), then remaining constant
until the doublet structure disappears for \(\lambda > \SI{9000}{\angstrom}\).
At the same time, the gap between doublets,
which in this interpretation corresponds to
rungs of a rotational ladder,
increases from about \SI{25}{cm^{-1}} at the bandhead
to about \SI{75}{cm^{-1}} at longer wavelengths.
In principle, it would be possible to search
in databases
\citep{McKemmish:2021z}
for molecular states with diatomic constants that match these spacings,
but we leave that for future work.

\subsection{\NEW{Ruling out} Diffuse Interstellar Bands}
\label{sec:carr-diff-interst}
The Diffuse Interstellar Bands (DIBs, \citealp{Heger:1922a}) are visible and near-infrared absorption features
\citetext{see \citealp{Sarre:2006a} and the symposium volume \citealp{Cami:2014a}}
that are ubiquitous in the ISM of the Milky Way and other galaxies,
including the SMC \citep{Bailey:2015u}.
The most accepted explanation for the DIBs is that they are due to
large carbon-based molecules associated with neutral gas
\citep{Fulara:2000a},
and that they are related to the mid-infrared emission features
that are commonly attributed to polycyclic aromatic hydrocarbons
(PAHs, \citealp{Salama:2014a}).
Optical emission that might be related to the DIBs
is relatively less common,
but unidentified emission features
observed in the Red Rectangle pre-planetary nebula
\citep{Wehres:2011a}
and in the circumstellar dust shells of
R~Coronae Borealis stars \citep{Oostrum:2018a}
have been tentatively ascribed to the DIB carriers.
It is therefore worth investigating whether our Deep Red Lines
might be related to these.

However, a detailed comparison suggests that this is unlikely.
The putative DIB emission features in the Red Rectangle
extend to shorter wavelengths than the DRLs, with the strongest
features
\citetext{Table~2 of \citealp{Wehres:2011a}}
appearing in the range \num{5800}--\SI{5900}{\angstrom},
where we see no DRLs.
There are some coincidences of the weaker features,
such as the \SI{6615}{\angstrom} feature
with DRL \wav{6615.00},
but in many other cases, the wavelengths do not match.\footnote{%
  Although the comparison is complicated by the fact that
  the Red Rectangle features shift to shorter wavelengths
  with increasing distance from the central star.
}
Furthermore, even the narrowest of the Red Rectangle features
have a FWHM of \SI{3}{\angstrom},
which should be easily resolvable by MUSE,
whereas we find that the DRLs, like the known nebular lines,
have widths \(< \SI{1}{\angstrom}\). 

\subsection{\NEW{Ruling out} Raman-scattered ultraviolet lines}
\label{sec:raman-scatt-ultr}

A more exotic possibility is that
some of the Deep Red Lines may be due to
Raman scattering of ultraviolet lines into the optical range
by neutral hydrogen,
as is seen in some symbiotic stars and planetary nebulae
\citep{Schmid:1989a, Lee:2006a}.
The column density of neutral hydrogen in PDRs is much
lower than in the circumstellar environment of symbiotic stars,
meaning that the scattering would only be efficient
for final wavelengths within
\(\pm \SI{150}{\angstrom}\) of the H lines \citep{Henney:2021b}.
We do find an apparent concentration of DRLs around
the \ha{} line, but there are several reasons to be sceptical
of this interpretation.
First, we definitely detect Raman scattering of the continuum
and superimposed absorption lines near \ha{}
(see Figure~\ref{fig:spectrum-6xxx-B})
but this is more than ten times stronger in Zone~MYSO than in Zone~0,
whereas the DRLs all show \(\Rat{MYSO}{0} < 2\) (Figure~\ref{fig:ratios}b).
Second, Raman-scattered lines are naturally broad,
whereas, as discussed in the previous section, the DRLs are narrow.

\section{Rest wavelength accuracies of the unidentified lines}
\label{sec:rest-wavel-accur}

\begin{figure*}
  \centering
  \includegraphics[width=\linewidth]{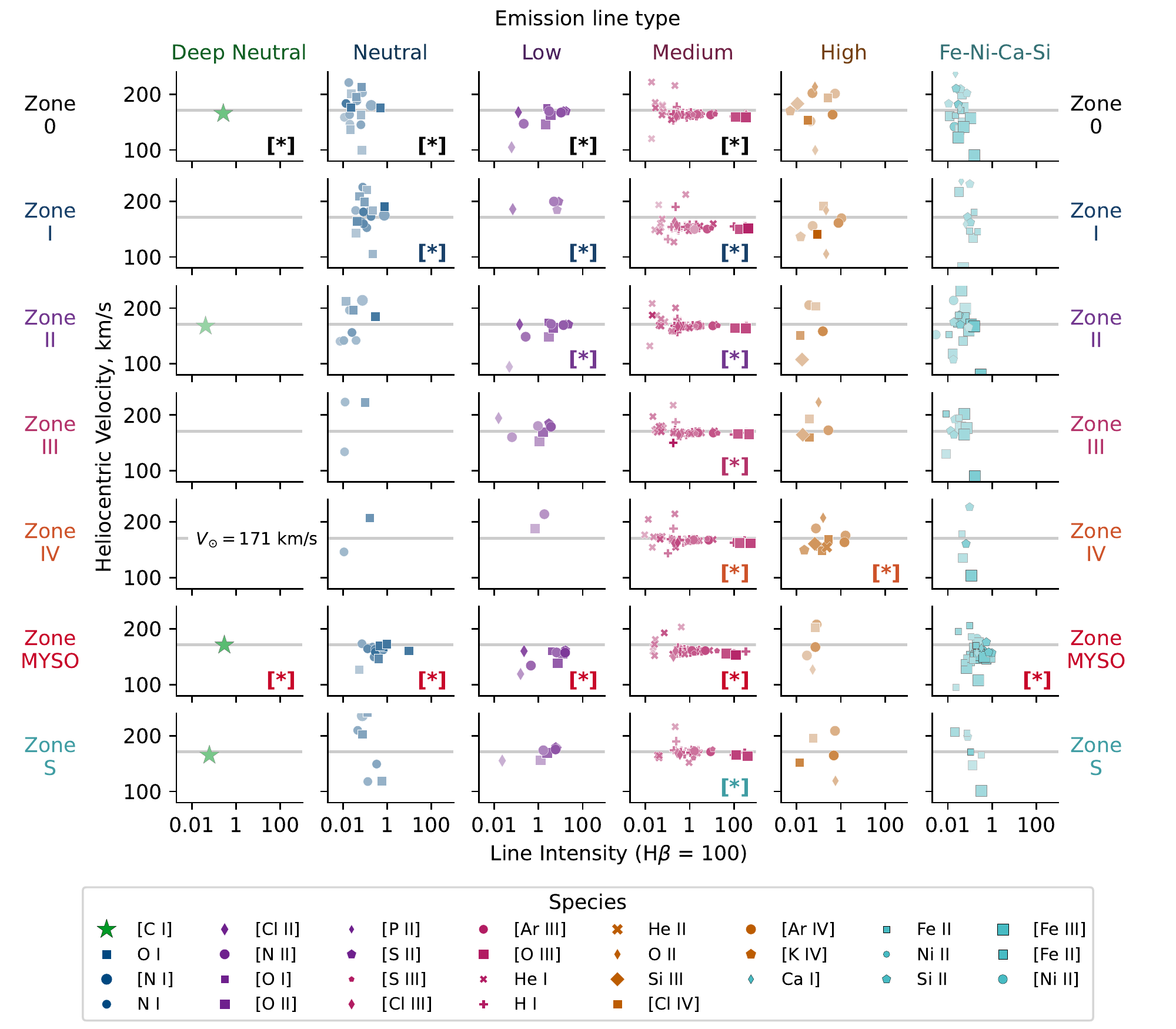}
  \caption{
    Centroid velocity versus line intensity (normalized to \(\hb = 100\))
    for different emission line types (panel columns, labeled at top)
    and different spatial zones in the nebula (panel rows, labeled on left and right).
    Different species are indicated by symbol type, as indicated in the key at bottom.
    Darkness of the symbols indicates the signal-to-noise (S/N) of each line measurement,
    with the darkest symbols having \(\text{S/N} > 100\).
    Measurements with \(\text{S/N} < 3\) are omitted.
    Gray horizontal bars indicate the previously determined
    \citep{Zeidler:2022a}
    ``group 2'' gas velocity of \(\SI{171.1 \pm 2.7}{km.s^{-1}}\).
    Asterisks \([*]\) indicate the zones with the highest quality measurements
    for each emission line type.
    Note that the unidentified lines do not appear in this figure because,
    without a known rest wavelength, the Doppler shift cannot be determined.
  }
  \label{fig:velocities}
\end{figure*}

We estimate the accuracy with which we can determine the rest wavelengths
of the unidentified lines in two ways:
(1)~from the dispersion in radial velocities of known lines within each spatial zone,
and (2)~from the dispersion in wavelength differences of the unidentified lines
between the different zones.
The first will be an upper limit since it includes additional effects,
such as uncertainty in the rest wavelengths of the known lines and the
possibility of correlated variations in velocity and ionization along the line of sight,
although the latter can be minimised by analysing different emission line types separately.
The second will be a lower limit since it does not include any systematic uncertainties
in the wavelength calibration, but these are predicted to be very small
(less than \SI{0.03}{\angstrom} or \SI{1}{km.s^{-1}},
see section~6.2 of \citealp{Weilbacher:2020a}).

Results from the first approach are shown in Figure~\ref{fig:velocities}
as an array of panels in which each panel shows
the centroid velocities for all emission lines
of a single line type (columns) and from a single spatial zone (rows).
As a reference, we show as a gray line in each panel
the mean velocity of the dominant gas component
\(V_\odot = \SI{171}{km.s^{-1}}\) derived by \citet{Zeidler:2022a}
from an independent reduction of the same observations
using lines of [\ion{S}{2}], [\ion{N}{2}], and \ha{}.
Variations within each individual panel are probably dominated
by inaccuracies in the individual wavelength measurements.
Panels with a large number of lines all show a similar behavior:
a large dispersion in velocities for the weakest lines
that rapidly converges on a much narrower distribution around
a constant value for the stronger lines.
A small number of outliers may be the result of misidentification,
unresolved blends, or errors in the rest wavelengths.
There is no evidence for any systematic trend of velocity with brightness,
except in the case of the medium ionization lines, where the faint lines
tend to have a more positive velocity than bright lines,
especially in the case of the MYSO and S regions.
This is probably a result of underlying photospheric absorption from
the cluster stars
since these are the two regions with the largest starlight contribution.
The median stellar velocity is blueshifted by \SI{8}{km.s^{-1}}
with respect to the gas \citep{Zeidler:2022a},
which is consistent with the sign of the observed trend if the
weaker emission lines are more affected by photospheric absorption. 

Variation across the plane of the sky of the line of sight velocity
will be seen as a systematic difference between the rows,
whereas variation along the line of sight will be seen
as difference between the columns,
so long as the velocity variation is correlated with changes in ionization.
An example of the former is that the Medium Ionization lines
from Zone~I are blueshifted (lower velocities) with respect
to Zones~0 and II.\@
The clearest example of the latter is again with Zone~I,
where the Low Ionization lines are redshifted
with respect to the Medium Ionization lines.
In both cases, the velocity differences are of order \SI{30}{km.s^{-1}}
or less, which is consistent with expectations for internal velocity structure
within an \hii{} region.

In the deep neutral category, we have only one known line, \ion{C}{1} \wav{8727},
but there is no evidence for any significant velocity differences
between this and the other line types, as can be seen from the first column
of Figure~\ref{fig:velocities}. 

\begin{figure}
  \centering
  \includegraphics[width=\linewidth]{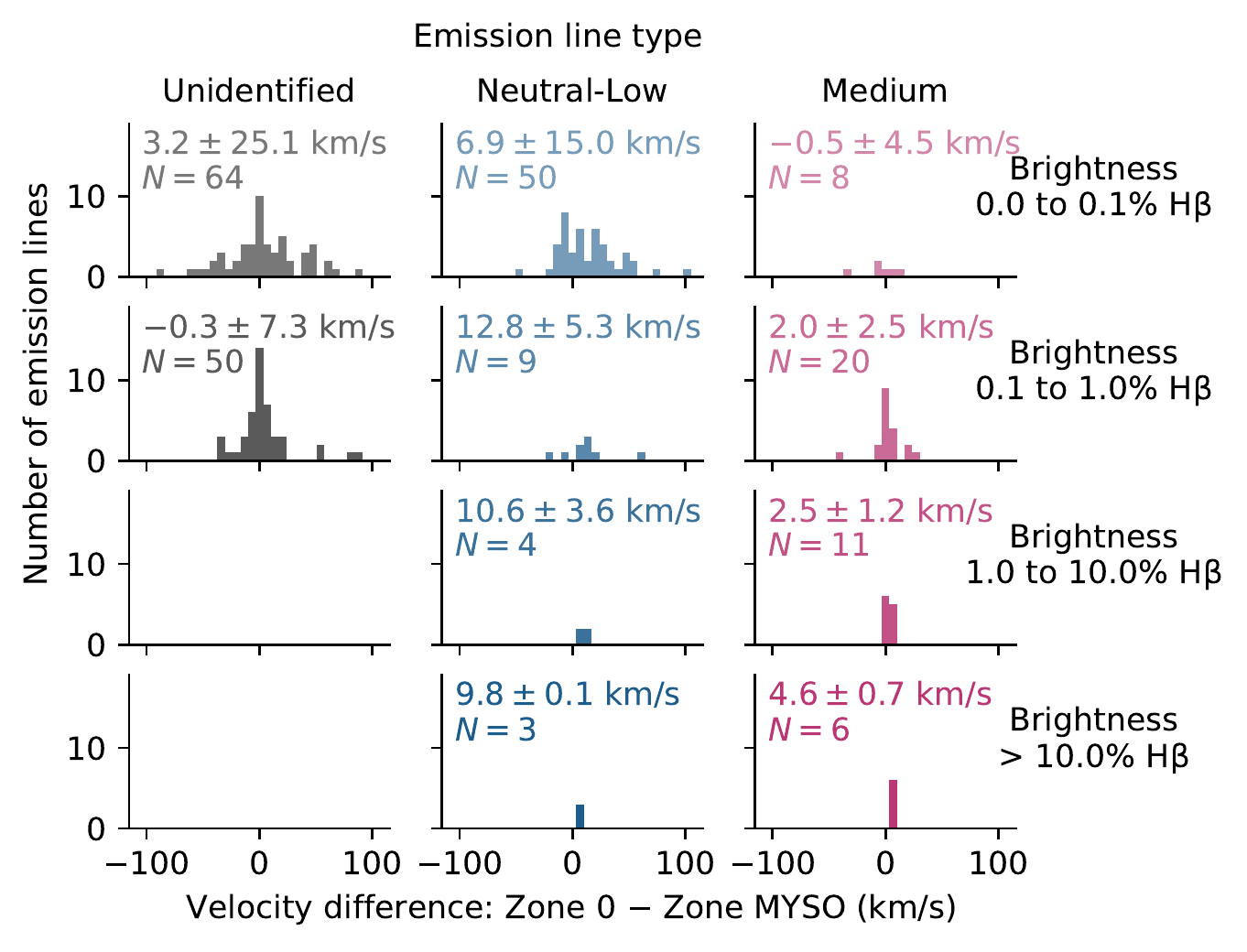}
  \caption{
    Variation in the precision of wavelength measurements
    as a function of line brightness
    for different emission line types (panel columns, labeled at the top).
    The Neutral, Low Ionization, and \feni{} lines are combined into a single category
    (Neutral-Low), while the High Ionization lines are omitted.
    Each row of panels corresponds to a brightness category
    (measured in Zone~0)
    from faintest at the top to brightest at the bottom, as labeled on the right.
    Each panel shows a histogram of measured velocity differences between
    Zone~0 and Zone~MYSO for emission lines of a particular type
    and brightness.
    The sigma-clipped (\(2\sigma\)) mean and standard deviation are marked,
    together with the number of lines that contribute to each panel.
    Assuming that the true dispersion of values within each panel is very small,
    the standard deviation is an estimate of the precision of the rest wavelength measurements
    for the unidentified lines.
  }
  \label{fig:vel-diff-histograms}
\end{figure}

Results from the second approach are presented in Figure~\ref{fig:vel-diff-histograms},
which shows histograms of the velocity differences between Zone~0 and
Zone~MYSO for different line types (columns) and different line strengths (rows).
Although we cannot measure absolute heliocentric velocities for
the unidentified lines (because we have no independent determination of the rest wavelength),
we can measure velocity \emph{differences} between spatial zones
by using a guess for the rest wavelength.
Even an error as large as \SI{1}{\angstrom} in this guess would contribute a relative
error less than 0.2\% in the derived velocity difference,
which is completely negligible.
As with the absolute velocities considered above,
we see a similar behavior: the dispersion in velocity differences
becomes smaller as the line brightness increases,
which is consistent with random noise being the limiting factor
in the accuracy of the wavelength measurements.
Furthermore, the unidentified lines behave in a similar way to the known lines,
albeit with slightly higher dispersion at each brightness level.

\section{Nearest-neighbor analysis of the spectral distribution of emission lines}
\label{sec:struct-wavel-distr}
\newcommand\nn{\ensuremath{_{\mathrm{nn}}}}
\newcommand\nnratio{\ensuremath{\langle \Delta\lambda\nn \rangle / \langle \Delta\lambda \rangle}}
\newcommand\nnnn{\ensuremath{_{\textrm{nn-nn}}}}
\begin{figure*}
  \centering
  \includegraphics[width=\linewidth]{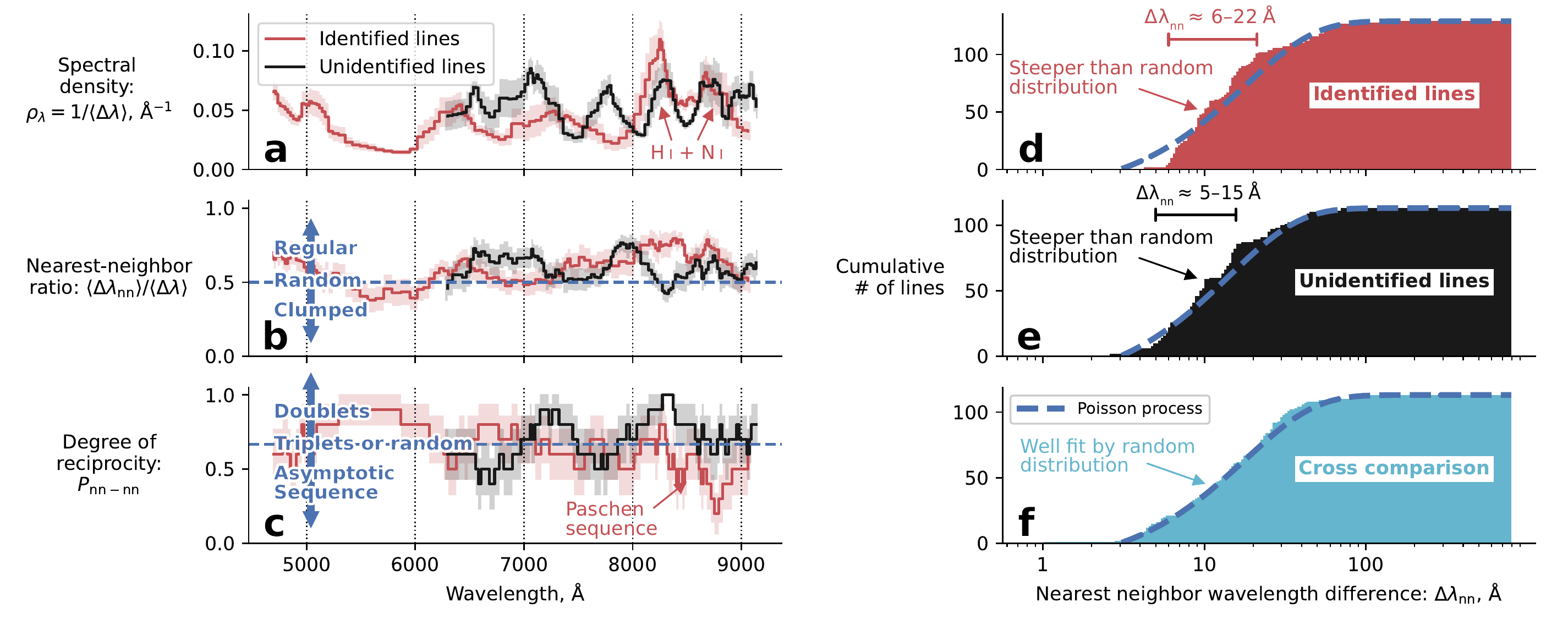}
  \caption{
    Clustering properties of the emission lines.
    (a)~Spectral density, \(\rho_\lambda  = 1 / \langle \Delta\lambda \rangle\),
    of identified (red) and unidentified (black) emission
    lines across the observed wavelength range.
    This is essentially a smoothed version of the
    distributions shown in Figure~\ref{fig:bar-code}.
    In this and the following two panels,
    solid lines shows running means, calculated in a
    10-point window, while translucent bands show plus/minus
    one standard error of the mean.
    All detected lines are treated identically,
    irrespective of their relative intensities.
    (b)~Ratio of nearest-neighbor distance to mean spacing,
    \nnratio.
    (c)~Degree of reciprocity, \(P\nnnn\).
    Note that, since this is the fraction calculated
    over a 10-point window, the only allowed values are integer
    multiples of \(0.1\).
    (d)~Cumulative Distribution Function (CDF) of
    nearest-neighbor separations for identified lines
    (e)~Same as (d) for unidentified lines.
    (f)~Same as (d) but for a cross-comparison
    between the two types of lines:
    the nearest neighbor among the identified lines of each
    \emph{unidentified} line. 
    Blue dashed lines show the predicted exponential CDF of
    a homogeneous Poisson process with the mean observed
    spectral density,
    truncated at \(\Delta\lambda\nn < \SI{3}{\angstrom}\) to account
    for the finite spectral resolution of the observations. 
  }
  \label{fig:nearest-neighbor}
\end{figure*}

Clues about the nature of the unidentified emission lines
can be obtained by studying the structure of their
wavelength distribution.
As mentioned above, there seems to be a clear preference for
closely spaced doublets,
but it is important to objectively quantify this.
We therefore calculate various clustering measures related to the nearest
neighbor (in wavelength) of each line:
\begin{enumerate}[1.]
\item The ratio between the mean nearest-neighbor distance
  and the harmonic mean separation between lines: \nnratio.
\item The degree of reciprocity, \(P\nnnn\),
  which is the fraction of lines that
  are the mutual nearest neighbor of their own nearest neighbor.
\item The probability distribution of nearest-neighbor distances,
  \(\Delta\lambda\nn\).
\end{enumerate}

If the emission line rest wavelengths were
randomly and independently sampled
from a smooth probability distribution
in one dimension (not necessarily uniform),
then one would expect a ratio \( \nnratio = 1/2\)
\citep{Boots:1988a},
whereas completely regular uniform spacing
would yield \(\nnratio = 1\),
and values smaller than \(1/2\) are also possible if the
distribution is highly clumped. 
In a similar way, the expected degree of reciprocity from a
random distribution is \(P\nnnn = 2/3\)
\citep{Schwarz:1980a}, while
different kinds of systematic patterns in the line spacing
can produce smaller or larger values than this.
For instance, if all lines are closely spaced non-overlapping doublets
then the reciprocity is perfect, so \(P\nnnn = 1\).
On the other hand, non-overlapping hierarchical triplets would have
\(P\nnnn = 2/3\), the same as in the random case.
Hierarchical quartets (doublets of doublets) would again have \(P\nnnn = 1\),
and the same holds for any even multiplet,
so long as the inner doublets do not overlap.
A very different result holds for a limiting series of lines,
such as the Paschen series of \chem{H^0}.
In that case, each line's nearest neighbor is on its blue side,
so there is no reciprocity at all, leading to \(P\nnnn = 0\).
Molecular band heads are more complicated because of overlap
between the different branches.

For a given collection of observed emission lines, these measures
will vary across the spectrum, according to the nature of the
lines that predominate in each wavelength range.
In spectral regions where lines of many different species are interleaved,
one expects that both the nearest-neighbor ratio and degree of reciprocity
should tend towards their fully random values:
\( \langle \Delta\lambda\nn \rangle / \langle \Delta\lambda \rangle = 1/2\) and \(P\nnnn = 2/3\).

In Figure~\ref{fig:nearest-neighbor}a we show the variation across
the entire observed spectral range of the
10-point running arithmetic mean of the spectral density 
(reciprocal of the harmonic mean spacing) for both
the identified and unidentified lines,
while panels b and c show the same for the clumping measures.
At this level of smoothing, the variation in spectral density is
roughly a factor of 2 for the identified lines
and 4 for the unidentified lines,
reaching peaks of \(\rho_\lambda = 0.05\) to \SI{0.1}{\angstrom^{-1}},
which correspond to the densest concentrations of lines visible in Figure~\ref{fig:bar-code}.

The nearest neighbor ratio (Figure~\ref{fig:nearest-neighbor}b)
shows deviations both above and below the pure-random value of \(1/2\).
In general, \nnratio{} is positively correlated with \(\rho_\lambda\),
indicating a more regular spacing in spectral regions where
the lines are closer together, possibly because such regions tend
to be dominated by lines from only one or two species.
For instance, the twin concentrations of identified lines
around \SI{8250}{\angstrom} and \SI{8700}{\angstrom}
consist largely of \ion{H}{1} and \ion{N}{1} transitions.
A clear exception to this behavior, however, is the concentration
of unidentified lines also at \SI{8250}{\angstrom},
which corresponds to a pronounced dip in \nnratio{}.

A clearer picture emerges from the graph of degree of reciprocity
(Figure~\ref{fig:nearest-neighbor}c).
For the identified lines, high reciprocity values
(\(P\nnnn > 2/3\)) are found primarily in
the sparsely populated region from \num{5000} to \SI{6000}{\angstrom}.
From inspection of Figure~\ref{fig:bar-code}, it is apparent that
many nearest-neighbor pairs in this region
are doublets from the same species, such as [\ion{Cl}{3}]
(the relatively low spectral density at these wavelengths helps ensure
that no unrelated lines from other species fall nearby).
The reciprocity for identified lines
falls to its lowest values (0.2 to 0.5) in
the range \SI{8300}{\angstrom} to \SI{9000}{\angstrom}, where
a large fraction of the lines are from the \ion{H}{1} Paschen sequence.
This can be understood as the effect of the regular decrease in
line separation as one approaches the Paschen limit, which drives
the reciprocity towards zero.

For the unidentified lines, low reciprocity is seen in the
range \SI{6300}{\angstrom} to \SI{7000}{\angstrom} that
flanks the \ha{} \wav{6563} line, suggesting that
one or more limit sequences may be present in this region.
For most other spectral ranges the reciprocity is high,
consistent with doublets being the dominant pattern.
The only exception to this is the concentration around \SI{7700}{\angstrom},
where the reciprocity hovers around the random value (\(P\nnnn \approx 2/3\)).
Inspection of Figure~\ref{fig:bar-code} suggests that doublets
may still be present in this region, but with separations
\(> \SI{10}{\angstrom}\),
which is comparable with the mean line spacing, so that
the doublets will tend to overlap, lowering the reciprocity. 
In contrast, the concentration of unidentified lines
around \SI{8300}{\angstrom} has much closer doublets
(separations \(\sim \SI{5}{\angstrom}\))
so that a high reciprocity is maintained, despite the high spectral density.

Panels d and e of Figure~\ref{fig:nearest-neighbor} show the
cumulative distribution functions (CDFs)
of the nearest-neighbor separations.
For a homogeneous Poisson point process
(independent random wavelength samples with constant \(\rho_\lambda\)),
the predicted normalized CDF of \(\Delta\lambda\nn\) is exponential
\citep{Boots:1988a}:
\begin{equation}
  \label{eq:CDF}
  \text{CDF}(\Delta\lambda\nn) / N = 1 - e^{-2\, \rho_\lambda\, \Delta\lambda\nn}, 
\end{equation}
where \(N\) is the total number of samples (emission lines).
However, this needs to be modified to account for the fact
that the finite width of the lines
(both intrinsic and due to the spectrograph resolution)
means that no neighbors will be detected with separations
below a certain threshold, \(\Delta\lambda_{\mathrm{min}}\).
We approximate this effect by plotting the interior-truncated
distribution:
\(\text{CDF}(\Delta\lambda\nn) - \text{CDF}(\Delta\lambda_{\mathrm{min}})\),
where \(\Delta\lambda_{\mathrm{min}} = \SI{3}{\angstrom}\) is assumed.
It can be seen that for both sets of lines,
identified and unidentified, the CDF is not well represented
by the Poisson curve,
but instead shows a significantly steeper slope for nearest-neighbor
separations of order \SI{10}{\angstrom}.
It is clear that this steepness cannot be due to the breakdown of the homogeneity assumption,
since the effect of variations in the spectral density is to
convolve the predicted CDF by the distribution of the mean spacing \(\langle\Delta\lambda\rangle\),
which can only serve to broaden it and make it shallower not steeper.

It is more reasonable to interpret the steepness as the signature
of the typical separation among lines \emph{of the same species}.
In the case of the identified lines, we know (see above) that this is
due primarily to the \ion{H}{1} Paschen sequence and
to fine-structure doublets of metal lines, such as
[\ion{Cl}{3}] \wavv{5518, 5538},
[\ion{S}{2}] \wavv{6716, 6731}, and
[\ion{O}{2}] \wavv{7320, 7331}.
Note, however, that the apparent upper bound of \SI{22}{\angstrom}
for the steep section of the CDF (Figure~\ref{fig:nearest-neighbor}d)
is simply due to having exceeded the average spacing. 
It does not mean that there are no doublets with larger separations,
but merely that they are unlikely to be nearest neighbors because of
overlap with unrelated lines from other species.
Prominent examples are
[\ion{O}{3}] \wavv{4959, 5007} (interrupted by [\ion{Fe}{3}] \wav{4987}),
[\ion{O}{1}] \wavv{6300, 6363} (interrupted by [\ion{S}{3}] \wav{6312}), and
[\ion{N}{2}] \wavv{6548, 6583} (interrupted by \ha{} \wav{6563}).

In order to further test this interpretation,
we have calculated a cross-comparison CDF of the nearest-neighbor
separations \emph{between} the identified and unidentified lines.
That is, redefining \(\lambda\nn\) as the wavelength difference
between each unidentified line and its nearest identified line.
Results are shown in panel~f of Figure~\ref{fig:nearest-neighbor},
where it can be seen that in this case the Poisson process
\emph{does} provide a reasonable fit to the observed distribution,
apart from some small deviations that are probably due to the fact that
the spectral density is not homogeneous, see above.
The good agreement with the fully random prediction for this cross comparison
implies that the wavelengths of the
unidentified lines are completely uncorrelated with the identified lines,
which is exactly what would be expected if they arise from different
atomic or molecular species.
It also strengthens the case that the excess of pairs 
within each emission line type
for separations  \(\Delta\lambda\nn \sim \SI{10}{\angstrom}\)
is due to departure from randomness.

Given the general similarity between the CDFs of the identified and unidentified lines
(Figure~\ref{fig:nearest-neighbor}d and e),
it is likely that we are seeing the signature of doublet structure
in the unidentified lines too.
The horizontal ``ledge'' seen in the CDF at \(\Delta\lambda\nn = \SI{10}{\angstrom}\)
is evidence that the distribution of separations may be bimodal,
with two distinct populations:
the first with \(\Delta\lambda\nn = 5\) to \SI{10}{\angstrom} and
the second with \(\Delta\lambda\nn = 12\) to \SI{15}{\angstrom}.
These may even be concentrated in different regions of the spectrum,
see second paragraph of description of Figure~\ref{fig:nearest-neighbor}c above. 

\section{Details of semi-automated line identification with EMILI}
\label{sec:deta-semi-autom}

We ran the EMILI spectral identification code \citep{Sharpee:2003a}
on the integrated spectrum from each spatial zone: 0, I, II, III, IV, MYSO, and S (see Table~\ref{tab:zones}). 

We initialized EMILI with the measured NGC~346 abundances  \citep{Valerdi:2019a} for
He, N, O, Ne, S, Ar, Cl,
assuming \num{0.2} times default Galactic ISM abundances for all other elements.
For the spatial zones corresponding to diffuse emission
(all except MYSO and S)
we included an additional gas-phase depletion factor of 10
for the refractory elements Fe, Ni, Si, Ca, Mg, Ti, and Al. 

EMILI divides lines into bins \(k = 1\) to \(5\)
based on the degree of ionization
of the parent ion, in a similar way to our own classification
(see Table~\ref{tab:ion-class}).
Bin~1 corresponds to our Deep Neutral lines,
bins~2 and 3 correspond to our Medium Ionization lines,
and bin~4 corresponds to our High Ionization lines.
Bin~5 corresponds to very highly ionized lines,
none of which are detected in our spectra. 
Although the code can automatically calculate fractional ionic
abundances \(x_k\)  for each bin,
we found it was more reliable to specify these
by hand,
which we did using equations~(1--3) of \citet{Sharpee:2003a}
with our observed ratios of \ion{He}{1} \wav{5876},
\ion{He}{2} \wav{4686}, and [\ion{C}{1}] \wav{8728}
to \hb. 

\section{\boldmath \ion{N}{1} multiplet V~20 as candidate ID for 4 Deep Red Lines}
\label{sec:possible-id-ni}
\begin{figure}
  \centering
  \includegraphics[width=\linewidth]{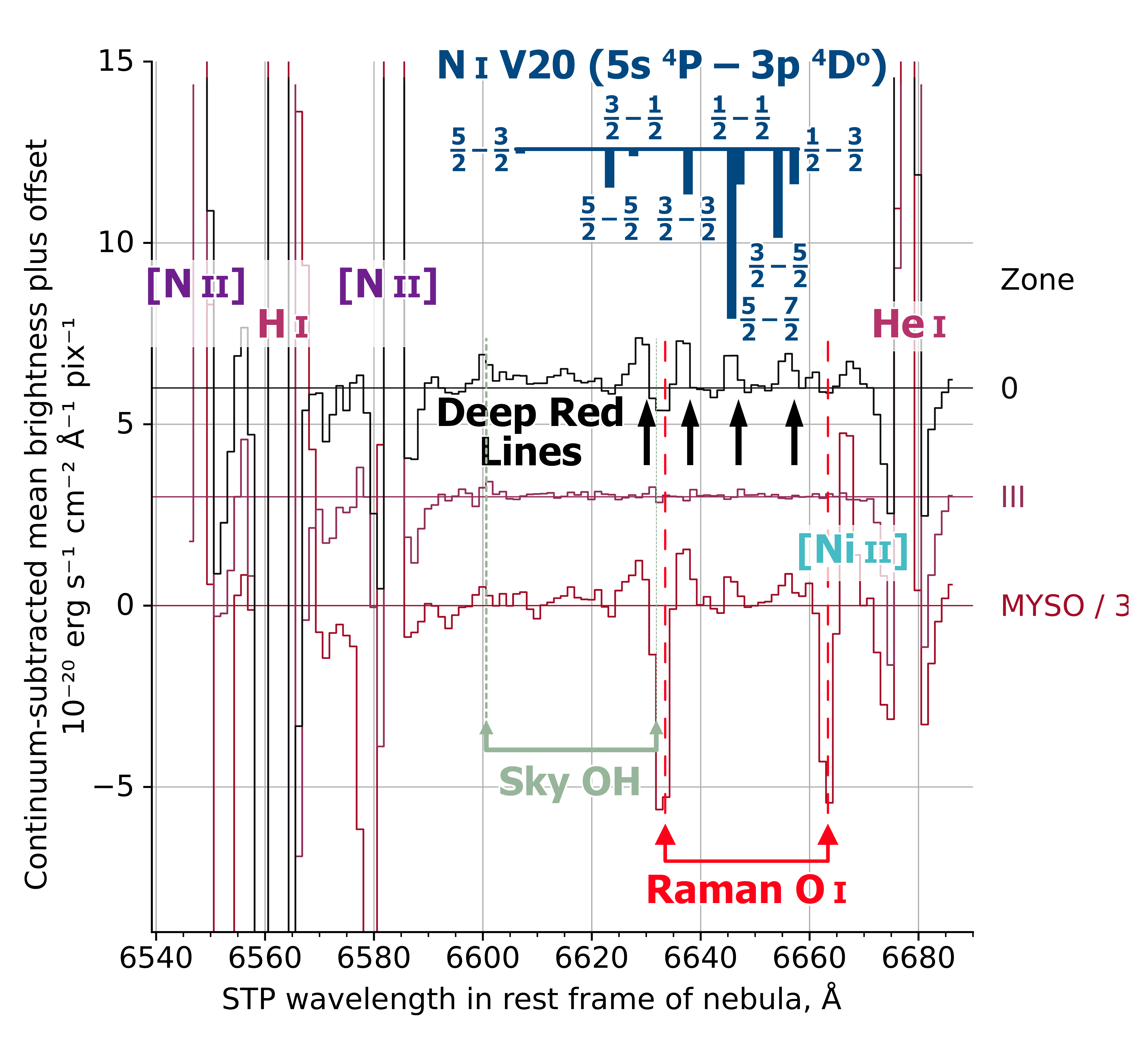}
  \caption{Expanded view of
    the observed spectrum immediately redward of the
    \ha{} line (see Figure~\ref{fig:spectrum-6xxx-B}),
    showing Deep Red Lines around \SI{6640}{\angstrom}
    from Zone~0 and Zone~MYSO
    (indicated by black arrows)
    and potential identification with the \ion{N}{1} V~20 multiplet.
    The predicted wavelengths and LTE intensities of
    the 8 components of the multiplet are indicated by
    thick blue lines, labelled by the angular momenta
    of their upper and lower states, \(J_k\)--\(J_i\).
    Despite a good correspondence in wavelength,
    the observed relative intensities of the components
    differs markedly from the theoretical predictions.
  }
  \label{fig:ni-spectrum}
\end{figure}

\begin{figure}
  \centering
  \includegraphics[width=\linewidth]{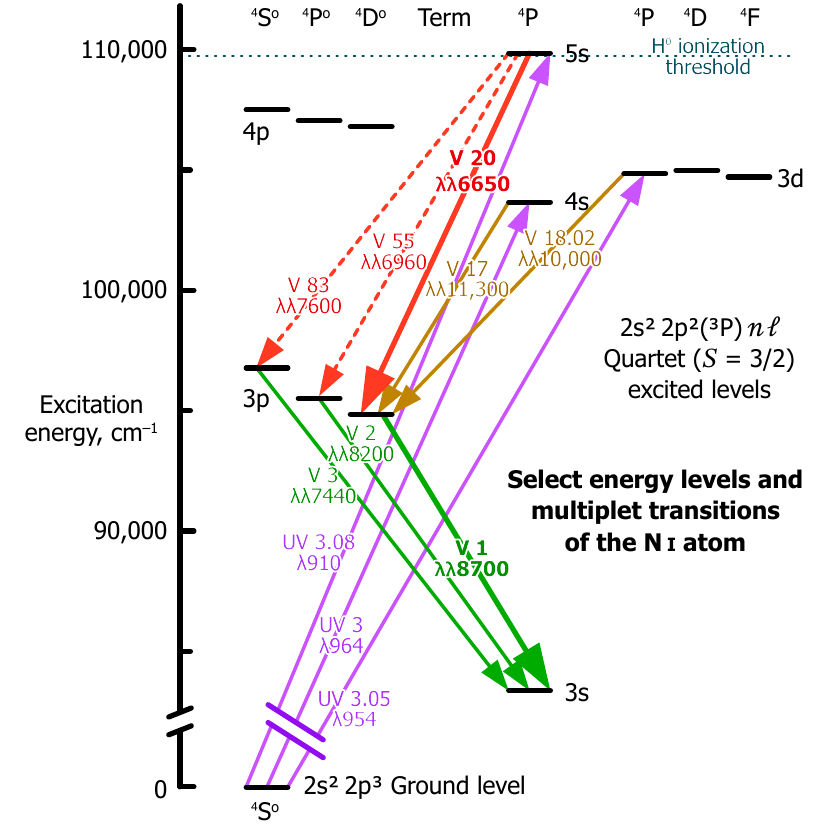}
  \caption{Partial Grotrian diagram of \ion{N}{1} showing
    cascades from the \chem{^4P} levels
    that can be directly pumped via permitted radiative transitions
    from the ground level.
    Observed optical multiplets are shown in green,
    while ultraviolet and infrared transitions are shown in 
    purple and brown, respectively.
    The V~20 multiplet (solid red arrow) is a potential ID
    for some of the Deep Red Lines, but the relative weakness of the
    V~1 multiplet from Zone~0 together with the non-detection of
    V~55 and V~83 argue against this. 
  }
  \label{fig:ni-grotrian}
\end{figure}

A tantalising coincidence is found between the neutral Nitrogen
multiplet V~20
(\(\Term{5s}{4}{P} \to \Term{3p}{4}{D^o}\), \citealp{Moore:1975a})
and a group of four Deep Red Lines close to \ha:
\wav{6629:}, \wav{6636.76}, \wav{6645.64}, and \wav{6655.89}.
This is illustrated in Figure~\ref{fig:ni-spectrum},
where it can be seen that nearly all the 8 members of the multiplet
coincide with emission features that are present
in Zone~0 (neutral filaments)
and Zone~MYSO, but which are absent from Zone~III (\hii{} region).
These lines are relatively more prominent in Zone~MYSO than is typical
for the DRLs, yielding \(\Rat{MYSO}{0} = 1.6 \pm 1.0\) when averaged over the four lines,
as compared with the median value over all 114 DRLs of \(\Rat{MYSO}{0} \approx 0.3\)
(see Figure~\ref{fig:ratios}).\footnote{%
  Recall that these are ratios between
  line intensities in units of \hb{}
  from different spatial ones of the nebula
  (section \ref{sec:interz-diagn-rati}).
}
On the other hand, their other interzone ratios are
\(\Rat{I}{0} \approx 0.6\) and \(\Rat{II}{0} \approx 0.3\),
which are typical of the bulk of the DRLs (Figure~\ref{fig:ratios}a).

Figure~\ref{fig:ni-grotrian} shows relevant energy levels
in the quartet term diagram of \ion{N}{1},
which can all be populated via sequences of permitted
electric dipole transitions starting from the ground level.
The far-red multiplets V~1 (Figure~\ref{fig:spectrum-8xxx-B}),
V~2 (Figure~\ref{fig:spectrum-8xxx-A}),
and V~3 (not illustrated) are all detected in our spectra.
The three multiplets (green arrows in Figure~\ref{fig:ni-grotrian})
all arise from upper levels in the
\Config{3p} configuration, which can be fluorescently pumped
via FUV absorption lines \wavv{954,964} (purple arrows in the figure)
that excite \Config{4s} and \Config{3d},
followed by infrared transitions to \Config{3p} (brown arrows).
In NGC~346, although the V~1, 2, 3 lines are weakly detected in Zone~0,
they are much more prominent in Zones~I and MYSO,
with typical interzone ratios of
\(\Rat{I}{0} \approx 3\), \(\Rat{MYSO}{0} \approx 10\) and \(\Rat{II}{0} \approx 0.5\).
In principle, the upper \Config{5s} level of the V~20 multiplet could be
directly radiatively excited from the ground level,
but an important difference from the other multiplets is that
the driving line \wav{910} is in the EUV, just short of the Lyman limit,
and therefore subject to strong competing continuum absorption
from neutral hydrogen.

We can apply some of the criteria introduced in section~\ref{sec:arguments-pro-contra}
to evaluate the feasibility of this identification. The Multiplet criterion
is well satisfied with respect to the wavelengths of the components,
with only minor discrepancies, which can be explained by blending
with the Raman \ion{O}{1} absorption features. 
However, there is a large difference in relative intensities between
the observed lines and the theoretical prediction if one assumes that the
angular momentum states \(J_k\) of the upper level are populated according to
Boltzmann statistics (in proportion to their statistical weights, \(2 J_k + 1\)).
Even if this assumption is not valid, the relative intensities between
components that share the same \(J_k\) should be in proportion to their
transition probabilities, which also gives rise to serious discrepancies.
For instance, the observed equal intensities of DRL \wav{6645.64} and \wav{6655.89}
could be explained if \(J_k = 1/2\) were preferentially populated
with respect to \(3/2\) and \(5/2\). However, this would also imply that
DRL \wav{6636.76} should be much weaker than \wav{6655.89},
whereas it is observed in fact to be stronger.

For the Branching Ratio criterion, we find the following branching ratios
from the \Term{5s}{4}{P} upper level:
74\% to return to the ground level,
10\% to transition to \Config{4p} levels via various infrared lines
between \num{3} and \SI{4}{\micro\meter} (outside the MUSE observed range),
9\% to transition to \Term{3p}{4}{D^o} via the V~20 multiplet
(our candidate ID),
and 7\% to transition to other \Config{3p} levels
via the V~83 and V~55 multiplets around \wavv{7600} and \wavv{6960}, respectively
(see Figure~\ref{fig:ni-grotrian}).
In addition, there are a number of semi-forbidden intercombination transitions to
doublet levels (not illustrated), but these have very small branching ratios.
If the identification is correct, one would therefore expect to observe
lines from the V~83 and V~55 multiplets with an intensity that is only
slightly less than the V~20 multiplet.
It is conceivable that the observed DRL \wav{7628.73} may correspond
to one of the V~83 components at \wav{7628.180}, but no other component
of either multiplet is detected.
However, both multiplets lie
in regions of the spectrum that are strongly affected by telluric absorption,
so the evidence is not conclusive.

For the Cascade criterion, we concentrate on the V~1 multiplet:
\Term{3p}{4}{D^o} -- \Term{3s}{4}{P}.
Since \Term{3s}{4}{P} is the only even parity quartet state
with lower energy than \Term{3p}{4}{D^o},
the V~1 lines are the only permitted transitions that can follow on
from the V~20 multiplet.
A test of the proposed identification is therefore to compare
the relative intensities in photon units by calculating the ratio
\begin{equation*}
  R = \frac{
    \displaystyle
    \sum_{\lambda \in \text{V1}} \lambda\,I(\lambda)
  }{
    \displaystyle
    \sum_{\lambda \in \text{DRL}} \lambda\,I(\lambda)
  }
  \quad
  \text{where}
  \quad
  \begin{array}{r}
    \text{V1} = \{8680, 8703, 8712, 8719\} \\[\medskipamount]
    \text{DRL} = \{6629, 6637, 6646, 6656\}
  \end{array}
\end{equation*}
If the identification is correct then we expect \(R \ge 1\),
but we find \(R = 0.4 \pm 0.2\) for Zone~0, which is a very
strong argument against the validity of the identification.

% Don't change these lines
\bsp	% typesetting comment
\label{lastpage}

\end{document}